\documentclass[12pt,a4paper,twoside]{article}

\usepackage{latexsym}
\usepackage{epsfig}
\usepackage{psfig}
\usepackage{cite}

\newlength{\dinwidth} 
\newlength{\dinmargin}
\newcommand{\lap}{\ensuremath{\stackrel{_{\scriptstyle <}}{_{\scriptstyle\sim}}}}
\newcommand{\gap}{\ensuremath{\stackrel{_{\scriptstyle >}}{_{\scriptstyle\sim}}}}

\newcommand{\xbj}{\ensuremath{x_{\mbox{\rm\tiny BJ}}\/\/}}

\newcommand{\xl}{\ensuremath{x_{\mbox{\it \tiny L}}\/\/ }}
\newcommand{\pt}{\ensuremath{p_{\mbox{\it \tiny T}}\/\/ }}
\newcommand{\tmin}{\ensuremath{t_{\mbox{\it \tiny 0}}\/\/ }}

\newcommand{\etamax}{\ensuremath{\eta_{\mbox{\rm\tiny max}}\/\/ }}
\newcommand{\ptmax}{\ensuremath{p_{\mbox{\it\tiny T}}^{\mbox{\rm\tiny max}}\/\/ }}
\newcommand{\ptsq}{\ensuremath{p_{\mbox{\it \tiny T}}^2\/\/ }}
\newcommand{\Rabs}{\ensuremath{R_{\mbox{\rm\tiny abs}}\/\/ }}

\newcommand{\pom}{\ensuremath{{I\!\!P}\/\/ }}

\newcommand{\ff}{\ensuremath{F_2(x, Q^2)}}
\newcommand{\feff}{\ensuremath{f_{\rm eff}}}

\newcommand{\emcap}{}

\newcommand{\Journal}[4]{{#1}{#2} (#3) #4}

\newcommand{\CPC}{Comp.\ Phys.\ Comm.\ }
\newcommand{\EPA}{Eur.\ Phys.\ J.\ A}
\newcommand{\EPC}{Eur.\ Phys.\ J.\ C}

\newcommand{\NCA}{Nuovo Cimento A}

\newcommand{\NIMA}{Nucl.\ Instr.\ and Meth.\ A}
\newcommand{\NP}{Nucl.\ Phys.\ }

\newcommand{\NPB}{Nucl.\ Phys.\ B}
\newcommand{\NPPS}{Nucl.\ Phys.\ Proc.\ Suppl.\ }

\newcommand{\PLB}{Phys.\ Lett.\ B}
\newcommand{\PR}{Phys.\ Rev.\ }
\newcommand{\PRep}{Phys.\ Rep.\ }
\newcommand{\PRL}{Phys.\ Rev.\ Lett.\ }

\newcommand{\PRD}{Phys.\ Rev.\ D}
\newcommand{\ProgPNP}{Prog.\ Part.\ Nucl.\ Phys.\ }

\newcommand{\SNP}{Sov.\ J.\ Nucl.\ Phys.\ }
\newcommand{\ZPA}{Z.\ Phys.\ A}
\newcommand{\ZPC}{Z.\ Phys.\ C}

\begin{document}
\thispagestyle{empty}
%
\title {
\begin{flushright}{\large DESY--02--039}  \end{flushright}
\hspace*{-5mm}\large\rm
\rm\bf\LARGE  
Leading neutron production\\
in $e^+p$ collisions at HERA
}
           
\author{ZEUS Collaboration}
\date{}
\maketitle

\begin{abstract}
\noindent

The production of neutrons carrying at least 20\% of the proton beam energy
($\xl > \, 0.2$) in $e^+p$  
collisions has been studied with the ZEUS detector at HERA for a wide
range of $Q^2$, the photon
virtuality, from  photoproduction to deep inelastic scattering.
The neutron-tagged cross
section, $e p\rightarrow e' X n$, is measured relative 
to the inclusive cross section, $e p\rightarrow e' X$, 
thereby reducing the systematic uncertainties.
For $\xl >$ 0.3,
the rate of neutrons in photoproduction is about half of that measured in
hadroproduction, which constitutes a clear breaking of factorisation. 
There is about a 20\% rise in the neutron rate
between photoproduction and deep inelastic scattering, which may be
attributed to absorptive rescattering in the $\gamma p$ system.
or $0.64 < \xl < 0.82$, the rate
of neutrons is almost independent of the Bjorken scaling variable $x$ 
and $Q^2$. However, at lower and
higher $\xl$ values, there is a clear but weak dependence on these variables,
thus demonstrating the breaking of limiting fragmentation.
The neutron-tagged structure function,
${{F}^{\mbox{\rm\tiny LN(3)}}_2}(x,Q^2,\xl)$, rises at
low values of $x$ in a way similar to that of the inclusive 
\ff\ of the proton.
The total $\gamma \pi$ cross section 
and the structure function of the pion, 
$F^{\pi}_2(x_\pi,Q^2)$ where $x_\pi = x/(1-\xl)$, have been determined 
using a one-pion-exchange model, up to 
uncertainties in the normalisation due to the poorly understood pion flux.
At fixed $Q^2$, $F^{\pi}_2$ has approximately the same $x$ dependence
as $F_2$ of the proton.
\end{abstract}
\newpage

\topmargin-1.cm                                                                                    
\evensidemargin-0.3cm                                                                              
\oddsidemargin-0.3cm                                                                               
\textwidth 16.cm                                                                                   
\textheight 680pt                                                                                  
\parindent0.cm                                                                                     
\parskip0.3cm plus0.05cm minus0.05cm                                                               
\def\3{\ss}                                                                                        
\newcommand{\address}{ }                                                                           
\pagenumbering{Roman}                                                                              
                                                   %
\begin{center}                                                                                     
{                      \Large  The ZEUS Collaboration              }                               
\end{center}                                                                                       
  S.~Chekanov,                                                                                     
  D.~Krakauer,                                                                                     
  S.~Magill,                                                                                       
  B.~Musgrave,                                                                                     
  A.~Pellegrino,                                                                                   
  J.~Repond,                                                                                       
  R.~Yoshida\\                                                                                     
 {\it Argonne National Laboratory, Argonne, Illinois 60439-4815}~$^{n}$                            
\par \filbreak                                                                                     
  M.C.K.~Mattingly \\                                                                              
 {\it Andrews University, Berrien Springs, Michigan 49104-0380}                                    
\par \filbreak                                                                                     
  P.~Antonioli,                                                                                    
  G.~Bari,                                                                                         
  M.~Basile,                                                                                       
  L.~Bellagamba,                                                                                   
  D.~Boscherini,                                                                                   
  A.~Bruni,                                                                                        
  G.~Bruni,                                                                                        
  G.~Cara~Romeo,                                                                                   
  L.~Cifarelli,                                                                                    
  F.~Cindolo,                                                                                      
  A.~Contin,                                                                                       
  M.~Corradi,                                                                                      
  S.~De~Pasquale,                                                                                  
  P.~Giusti,                                                                                       
  G.~Iacobucci,                                                                                    
  G.~Levi,                                                                                         
  A.~Margotti,                                                                                     
  R.~Nania,                                                                                        
  F.~Palmonari,                                                                                    
  A.~Pesci,                                                                                        
  G.~Sartorelli,                                                                                   
  A.~Zichichi  \\                                                                                  
  {\it University and INFN Bologna, Bologna, Italy}~$^{e}$
% ...
\newpage                                                                                           
% ...
$^{\    3}$ now at Minist\`ere de la Culture, de L'Enseignement                                    
Sup\'erieur et de la Recherche, Luxembourg\\                                                       
% ...
                                                           %
                                                           %
\newpage   
                                                           %
                                                           %
\begin{tabular}[h]{rp{14cm}}                                                                       
$^{a}$ &  supported by the Natural Sciences and Engineering Research                               
          Council of Canada (NSERC) \\                                                             
$^{b}$ &  supported by the German Federal Ministry for Education and                               
          Research (BMBF), under contract numbers HZ1GUA 2, HZ1GUB 0, HZ1PDA 5, HZ1VFA 5\\         
$^{c}$ &  supported by the MINERVA Gesellschaft f\"ur Forschung GmbH, the                          
          Israel Science Foundation, the U.S.-Israel Binational Science                            
          Foundation, the Israel Ministry of Science and the Benozyio Center                       
          for High Energy Physics\\                                                                
$^{d}$ &  supported by the German-Israeli Foundation, the Israel Science                           
          Foundation, and by the Israel Ministry of Science\\                                      
$^{e}$ &  supported by the Italian National Institute for Nuclear Physics (INFN) \\                
$^{f}$ &  supported by the Japanese Ministry of Education, Science and                             
          Culture (the Monbusho) and its grants for Scientific Research\\                          
$^{g}$ &  supported by the Korean Ministry of Education and Korea Science                          
          and Engineering Foundation\\                                                             
$^{h}$ &  supported by the Netherlands Foundation for Research on Matter (FOM)\\                   
$^{i}$ &  supported by the Polish State Committee for Scientific Research,                         
          grant no. 620/E-77/SPUB-M/DESY/P-03/DZ 247/2000-2002\\                                   
$^{j}$ &  partially supported by the German Federal Ministry for Education                         
          and Research (BMBF)\\                                                                    
$^{k}$ &  supported by the Fund for Fundamental Research of Russian Ministry                       
          for Science and Edu\-cation and by the German Federal Ministry for                       
          Education and Research (BMBF)\\                                                          
$^{l}$ &  supported by the Spanish Ministry of Education and Science                               
          through funds provided by CICYT\\                                                        
$^{m}$ &  supported by the Particle Physics and Astronomy Research Council, UK\\                   
$^{n}$ &  supported by the US Department of Energy\\                                               
$^{o}$ &  supported by the US National Science Foundation\\                                        
$^{p}$ &  supported by the Polish State Committee for Scientific Research,                         
          grant no. 112/E-356/SPUB-M/DESY/P-03/DZ 301/2000-2002, 2 P03B 13922\\                    
$^{q}$ &  supported by the Polish State Committee for Scientific Research,                         
          grant no. 115/E-343/SPUB-M/DESY/P-03/DZ 121/2001-2002, 2 P03B 07022\\                    
\end{tabular}                                                                                      
                                                           %
\newpage
\pagenumbering{arabic}
\section{Introduction}

Leading baryon production in a process with a hard scale
provides a probe of the relationship between the quantum 
chromodynamics (QCD) of quarks and gluons and
the strong interaction of hadrons. The leading baryons are produced 
with small transverse
momentum ($\pt $), guaranteeing the presence of a soft process 
with its related long-range correlations.
The hard scale can come
from large photon virtuality ($Q^2$) in deep inelastic 
scattering (DIS), or 
high transverse energy ($E_{\mbox{\it \tiny T}}$) in
photoproduction. 
The observation of events with neutrons or protons carrying a large 
fraction ($\xl$) of the 
incident proton beam
energy in positron-proton ($e^+p$) scattering
at HERA\cite{fnc2,h1f2lb,zeuslndijet,lpsf2d,97-238,01-062}
has led to renewed interest in the 
QCD evolution and factorisation properties of proton fragmentation 
to leading baryons in deep inelastic 
scattering\cite{frankfurt,FF,collins,limiting_frag_a,limiting_frag_b}. 

A non-perturbative approach to the strong interaction is needed to
calculate the production rates of leading baryons. Some 
theoretical models retain the QCD building blocks, quarks and gluons, as
fundamental entities but add non-perturbative elements, such as soft
color interactions\cite{sci}. Another approach interprets the production of forward baryons in terms
of the exchange of virtual
particles\cite{yukawa,sullivan,bishari,zoller,kopeliovich,sns,
field,ganguli,zakharov,neu-pi,pdbcollins},
which accounts for charge-exchange processes ($p\rightarrow n$) in hadronic
interactions\cite{erwin,pickup,
engler,robinson,flauger,hanlon,hartner, eisenberg,blobel,abramowicz}.
In leading neutron production, the large diffractive peak in the cross
section, which is observed in leading proton
production\cite{lpsf2d,97-238}, is absent.
Although leading neutrons can be produced indirectly through the
production and decay of heavy baryons or through the fragmentation of
diffractively dissociating protons, direct production 
dominates~\cite{zeuslndijet}. 
In the language of particle exchange models, isovector exchange ($\pi$, $\rho$
and $a_2$) is
required for direct neutron production, and isoscalar exchange ($f$
and $\omega$) is absent.  The production mechanism of leading neutrons
is therefore simpler than that of protons and, since the cross section
for $\pi$ exchange is much larger than that for $\rho$ or $a_2$ exchange,
it is predominantly the structure of the pion that is probed by the hard
process\cite{neu-pi}.

Comparisons between cross sections
for the production of particles in the fragmentation region of a target
nucleon provide tests of the concepts of vertex factorisation
and limiting fragmentation. Factorisation tests involve comparing 
semi-inclusive rates, normalised to the respective total cross section, 
to study whether particle production from a given target
is independent of the nature of the incident projectile.
Thus, for example, data from $\gamma p$ interactions
may be compared with those from $pp$ collisions. The limiting-fragmentation 
hypothesis~\cite{limiting_frag_a} states that, in
the high-energy limit, the momentum distribution of fragments becomes
independent of the incident projectile energy. This has been 
verified in hadronic interactions~\cite{bellettini}. 
It has not been extensively studied 
for produced baryons in electroproduction. In this case, the fragmentation
spectra should be studied as a function of $Q^2$ and the photon-proton
centre-of-mass energy, $W$~\cite{limiting_frag_b}.

In electroproduction, the size of the incident projectile (the virtual
photon) can be varied. At $Q^2 = 0$, the photon has a typical hadronic
size; as $Q^2$ increases, the photon size decreases.
Hence, absorptive rescattering of the produced
baryon is expected to decrease and become independent of $Q^2$ at high 
$Q^2$~\cite{nszak,alesio}. Such effects, if $\xl$ or $\pt$ dependent,
lead to an apparent failure of factorisation and 
limiting fragmentation for low values of $Q^2$.

This paper reports studies of the production of 
leading neutrons using the
ZEUS detector and its forward neutron calorimeter (FNC). 
Leading neutron production can be measured precisely by determining
the neutron-tagged cross sections relative to the inclusive $ep$ cross
section, which has been measured with high accuracy\cite{zeusf294}.
The study of ratios reduces the systematic uncertainties arising from
both theoretical and experimental sources.
 
The paper is organized as follows:  
the kinematics of leading neutron production is defined in Section 2; the one-pion-exchange model
is reviewed in Section~3; the detector, the reconstruction 
of the kinematic variables and the experimental
conditions are described in Section 4; 
the event selection is discussed in Section~5; 
in Section 6, the advantages of analysing the data using the ratio
of the neutron-tagged to the inclusive cross section are outlined;
systematic uncertainties are discussed in Section~7. The general features of 
the neutron energy spectra and the ratio of leading-neutron to total
cross sections are discussed in Section~8, where
the factorisation and limiting-fragmentation 
properties of the data
are investigated and an estimate of absorptive effects is made.
The neutron-tagged structure function is discussed in Section~9. 
The data are analysed in Section~10 in terms of a
one-pion-exchange ansatz and the pion structure function, $F_2^\pi$, is
extracted up to uncertainties in the normalisation due to the lack of knowledge
of the pion flux;
 theoretical models are compared to $F_2^{\pi}$.
Section 11 summarizes the results.

\section{Kinematics and cross sections}

\label{sec:kinematics}

\begin{figure}[htbp!]
\epsfysize=14cm
\centerline{\epsffile{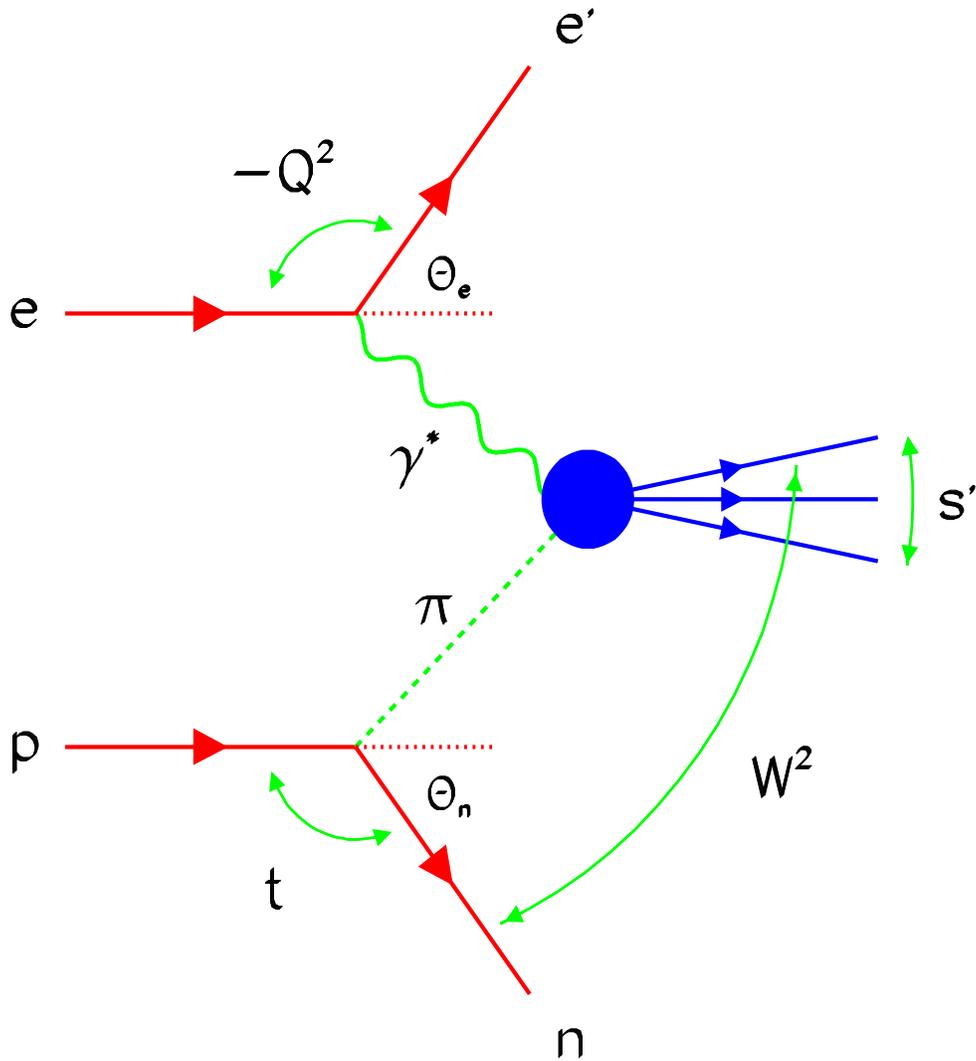}}
\caption{A schematic diagram for the one-pion-exchange model as applied to
$ep \rightarrow e^{\prime}Xn$ showing the Lorentz invariant variables 
$s'$, $Q^2$, $W^2$ and $t$. Also shown is the definition of 
the scattering angle of the positron, $\theta_e$, and
the production angle of the neutron, $\theta_n$.
In the ZEUS coordinate system, the polar angle
of the positron is
$\theta=\pi-\theta_e$. 
}
\label{fig:ope}
\end{figure}

Figure~\ref{fig:ope} shows a Mandelstam diagram of semi-inclusive 
leading neutron production in $ep$ collisions.
Four kinematic variables  are needed to describe the interaction
$ep\rightarrow e' X n$.
They are defined in terms of the four-momenta $k, P, k'$ and $N$, 
respectively, of the incoming and outgoing particles $e, p, e'$ and $n$.
Two variables are chosen
from the Lorentz invariants used in inclusive 
studies:
$Q^2= -q^2 = -(k-k')^2$, the virtuality of the exchanged photon;
$x=Q^2/(2P\cdot q)$, the Bjorken scaling variable;
$y=q\cdot P/(k\cdot P)\simeq Q^2/(sx)$,
the inelasticity; and 
$W^2=(P+k-k')^2=m^2_p+Q^2(1-x)/x$, the 
squared mass of the produced hadronic system. 
In these equations, $m_p$ is
the mass of the proton 
and $\surd s=$ 300 GeV is the $ep$
centre-of-momentum-system (cms) energy.

Two further variables are required to 
describe the leading neutron. They
can be chosen as the laboratory production angle of the neutron, $\theta_n$,
and the energy fraction carried by
the produced neutron, 
\[
\xl = \frac{N \cdot k}{P\cdot k} \simeq \frac {E_n}{E_p}, 
\]
where $E_p$ is the proton beam energy and $E_n$ is
the neutron energy. The transverse momentum of the neutron is 
given by $\pt \simeq \xl  E_p \theta_n$.
In terms of
these variables, the squared four-momentum
transfer from the target proton is
\[
t = (P-N)^2 \simeq -\frac{\ptsq}{\xl }-\tmin ,
\]
where $\tmin$ is the minimum kinematically
allowed value of $|t|$ for a given $\xl $
\[
\tmin(\xl) = \frac{(1-\xl )}{\xl }\left(m^2_n-\xl  m^2_p\right)
          \simeq \frac{(1-\xl )^2}{\xl }m^2_p, 
\]
and $m_n$ is the mass of the neutron.

The differential cross section for inclusive $ep\rightarrow e' X$ 
scattering
is written in terms of the proton structure function,
$F_2(x,Q^2)$, as
\begin{equation}
\frac{d^2\sigma^{ep\rightarrow e^{\prime}X}}{dx \, dQ^2} = 
 {\cal K}
 F_2(x , Q^2)(1+\delta), \label{eq:inc}
\end{equation}
where
\[
 {\cal K}=\frac{4\pi \alpha^2}{x Q^4}
 \left( 1-y+\frac{y^2}{2} \right)
\]
and $\delta$ is the correction 
to the Born cross section for photon radiation, $Z^0$ exchange  
and the longitudinal structure function, $F_L$.
In analogy to this, the differential
cross section for leading neutron production, 
$ep \rightarrow e' X n$, is written as
\begin{equation}
\frac{d^4\sigma^{ep\rightarrow e' Xn}}{dx \,dQ^2 \,d\xl \,d\pt} = 
 {\cal K}
  F^{\mbox{\rm\tiny LN(4)}}_2(x , Q^2, \xl, \pt)
  (1+\delta_{\mbox{\rm\tiny LN}}).   \label{eq:ln}
\end{equation}
The term $\delta_{\mbox{\rm\tiny LN}}$
is analogous to $\delta$, but for the case of leading
neutron production.

In the results presented here, $\theta_n$ (and hence $\ptsq$ and $t$) 
is not measured. 
Integrating Eq. (\ref{eq:ln}) up to the maximum
experimentally accessible angle, $\theta^{\rm max}_n$, corresponding to
a $\ptmax$ which varies with $\xl$, leads to
\begin{eqnarray}
\frac{d^3\sigma^{ep\rightarrow e' Xn}}{dx\, dQ^2 \,d\xl}
\equiv\int_{0}^{\ptmax}
\frac{d^4\sigma^{ep\rightarrow e' Xn}}{dx\, dQ^2 \,d\xl \,d\pt}d\pt \nonumber\\ \nonumber \\
={\cal K}  F^{\mbox{\rm\tiny LN(3)}}_2(x , Q^2, \xl)
(1+\delta_{\mbox{\rm\tiny LN}}) 
\label{eq:ln3}
\end{eqnarray}
where 
\[
F^{\mbox{\rm\tiny LN(3)}}_2(x , Q^2, \xl) \equiv\int_{0}^{\ptmax}  
F^{\mbox{\rm\tiny LN(4)}}_2(x , Q^2, \xl, \pt) d\pt
\]
is the neutron-tagged 
structure function integrated over the measured range in $\theta_n$.

It is sometimes more convenient to 
discuss the results in terms of cross sections rather than 
structure functions.
At low $x$, where  $W^2 \simeq Q^2/x$, the following expressions are valid:
\begin{eqnarray}
\sigma^{\gamma^\ast p}_{\rm tot}(W,Q^2) & = & 
\frac{4\pi^2\alpha}{Q^2}F_2(x,Q^2) \label{eq:sigtot}\\
\frac{d\sigma^{\gamma^\ast p \rightarrow Xn}(W,Q^2)}{d\xl}
 & \equiv & \int_{0}^{\ptmax}
\frac{d^2\sigma^{\gamma^\ast p \rightarrow Xn}(W,Q^2)}{d\xl\, d\pt}d\pt
\nonumber\\
& = & \frac{4\pi^2\alpha}{Q^2}F^{\mbox{\rm\tiny LN(3)}}_2(x , Q^2, \xl), \label{eq:sigln}
\end{eqnarray}
where $\sigma^{\gamma^\ast p}_{\rm tot}$ is the total virtual photon-proton
cross section and ${d^2\sigma^{\gamma^\ast p \rightarrow Xn}(W,Q^2)}/{d\xl d\pt}$
is the differential cross section for leading neutron production
in a virtual photon-proton collision.

\section{One-pion-exchange model}
\label{sec:opeint}

A successful description of the available 
data\cite{erwin,pickup,engler,
robinson,flauger,hanlon,hartner,
eisenberg,blobel,abramowicz}
on charge-exchange processes $(p\rightarrow n)$ 
in hadron-hadron interactions
has been obtained using 
the exchange of virtual particles 
with the quantum numbers of the $\pi$, $\rho$, and $a_2$
mesons\cite{yukawa,sullivan,bishari,field,ganguli,zakharov,zoller}.
In such processes, the pion, due to its small mass,
dominates the $p\rightarrow n$ transition amplitude,
with its relative contribution increasing as 
$|t|$ decreases.
In contrast to the nucleon, whose partonic structure
has been probed directly in deep inelastic scattering,
the structure of mesons has been studied only indirectly
using hadron-hadron collisions.
For the pion, the data from $\pi p$
interactions yielding prompt photons\cite{na24,wa70}, 
high-mass muon pairs\cite{na3,na10,e573,e615}
and dijets\cite{e609} in the final state
have given important information. 
However, these results are mostly sensitive 
to the high-$x$, or valence, structure of the pion:
the sea-quark and gluon content of the pion
at low $x$ have not yet been measured.
The HERA leading neutron data offer an opportunity
to probe this low-$x$ structure via the diagram shown in 
Fig.~\ref{fig:ope}.
In the one-pion-exchange (OPE) approximation, the cross section
for hadroproduction, $hp \rightarrow Xn$, can be written as
\begin{equation}
\frac{d\sigma_{hp \rightarrow Xn}}{d\xl\, dt} = 
f_{\pi /p}(\xl, t) \sigma^{h\pi}_{\rm tot}(s^\prime)
\label{eq:opef}
\end{equation}
where the flux factor, $f_{\pi /p}(\xl, t)$, describes the splitting of a 
proton into a $\pi n$ system, $s^\prime = s(1-\xl)$, where $s^\prime$ and
$s$ are the squares of the cms energy for the $h\pi$ 
and $hp$ systems, respectively. 
Charge-exchange 
processes at the ISR~\cite{engler,flauger} and 
Fermilab~\cite{hanlon,hartner} are well
described by such an expression with~\cite{bishari,field} 
\begin{equation}
f_{\pi /p}(\xl,t)=  \frac{1}{4\pi} \frac{2g_{\pi pp}^2}{4\pi}
                         \frac{-t}{(t-m^2_{\pi})^2}
                          (1-\xl)^{1-2\alpha_{\pi}(t)}
                         \left[ F(\xl ,t) \right]^2,
\label{eq:flux}
\end{equation}
where $g_{\pi pp}^2/(4\pi)=14.5$ is the $\pi^0 pp$ coupling constant and
$\alpha_{\pi}(t)= \alpha' t$ with $\alpha'=1$ GeV$^{-2}$. 
The form-factor $F(\xl ,t)$ parametrises the distribution of the pion cloud in the proton and accounts for final-state rescattering of the 
neutron.

For the hadronic charge-exchange experiments, a good description of most of 
the  $pn \rightarrow Xp$ data~\cite{robinson}
is obtained using the Bishari~\cite{bishari} flux, which corresponds to
 Eq. (\ref{eq:flux}) with $F(\xl ,t)=1$. 
In this case, 
\begin{equation}
f_{\rm eff}(\xl,t)=  \frac{1}{4\pi} \frac{2g_{\pi pp}^2}{4\pi}
                         \frac{-t}{(t-m^2_{\pi})^2}
                          (1-\xl)^{1-2\alpha_{\pi}(t)}.
\label{eq:eff_flux}
\end{equation}
The flux-factor $f_{\rm eff}$ can therefore
be interpreted as an $\it{effective}$ pion flux that takes into account 
processes that occur in hadronic charge-exchange processes,
such as absorption (Section~\ref{sec:comp}), 
non-pion exchange and off-mass-shell effects.

\section{HERA and the ZEUS detector}

During 1995--97, the HERA
collider operated 
at $\surd s \sim$ 300~GeV with
positrons of $E_e$ = 27.5 GeV colliding with protons of 
$E_p$ = 820 GeV. 
In addition to the 175 $ep$ colliding bunches, 
21 unpaired 
bunches of protons or positrons allowed the
determination of beam--related backgrounds.

\subsection{The ZEUS detector}

The ZEUS central detector is described in detail elsewhere
\cite{zeusdet}. Here the main components
used in the present analysis are described briefly.
The central tracking detector (CTD)\cite{ctd}, 
positioned in a 1.43 T solenoidal magnetic field, 
was used to establish an interaction vertex with a typical 
resolution along (transverse to) the beam direction\footnote{
The ZEUS coordinate system is a right-handed Cartesian system, with
the $Z$ axis pointing in the proton beam direction, and the $X$ axis
pointing left towards the centre of HERA. The coordinate origin is at the
nominal interaction point.
The polar angle $\theta$ is defined with
respect to the positive $Z$-direction.
The  pseudorapidity, $\eta$,
is given by $-\ln ( \tan \frac{\theta}{2} ) $.} of
0.4 (0.1) cm.

The high-resolution uranium-scintillator calorimeter (CAL)~\cite{cal} 
consists of three parts:
the forward (FCAL), the barrel (BCAL) and the rear (RCAL) calorimeters.
Each part is subdivided transversely into towers
and longitudinally into one electromagnetic section (EMC)
and either one (in RCAL) or two (in FCAL and BCAL) 
hadronic sections (HAC).
The smallest subdivision of the calorimeter is called a cell.
The CAL relative energy resolutions, as measured under test-beam conditions,
are $\sigma(E)/E=0.18/\sqrt{E}$ for electrons
and $\sigma(E)/E=0.35/\sqrt{E}$ for hadrons ($E$ in GeV).

In addition to the CAL, two small electromagnetic sampling calorimeters,
situated beside the beam pipe,
were used to measure the scattered positron. 
The beam pipe calorimeter (BPC)\cite{zeusf2bpc} 
was located at $Z=-2.9$ m from the interaction 
point, and the LUMI calorimeter\cite{zeussigmatot}, 
which forms part of the ZEUS system for 
monitoring the luminosity, was located at $Z=-35$ m. These additional
calorimeters had relative energy resolutions of $\sigma(E)/E$ = 
$0.18/\sqrt{E}$.

\subsection{The forward neutron calorimeter}

\label{sec:fnc}

\begin{figure}[htbp!]

\epsfysize=3.75cm
\centerline{\epsffile{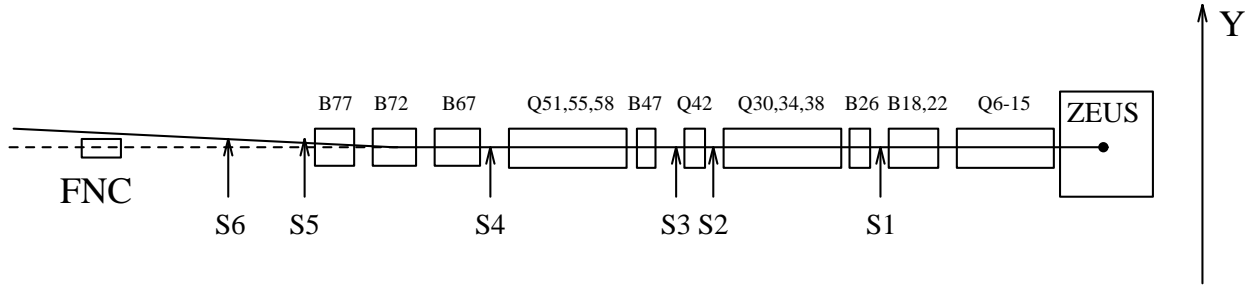}}
\caption{A schematic elevation view of the forward proton beam 
line from the central detector to the FNC at $Z = +106$ m.
Magnets are labelled by B (dipoles) or Q (quadrupoles) 
and a number which is their nominal distance from the interaction point
in meters. 
S1 through S6 label the positions of the six leading
proton spectrometer (LPS) stations. Between 65 and 80 m, the proton beam line
is bent upwards by 6 mrad by three bending magnets which
form a momentum analyzer for stations S5 and S6 of the LPS. The neutron exit
window is located at $Z = +96$ m, immediately following S6.
}
\label{fig:beamline}
\end{figure}

A forward neutron calorimeter (FNC) was
installed in 1995~\cite{fnc3} in the HERA tunnel 
at $Z=+106$ m and at zero degrees
from the incoming proton direction.
Figure~\ref{fig:beamline} shows the layout of the FNC
and that of the six stations
(S1--S6) of the leading proton spectrometer (LPS).

\begin{figure}[hbp!]
\epsfysize=10cm
\centerline{\epsffile{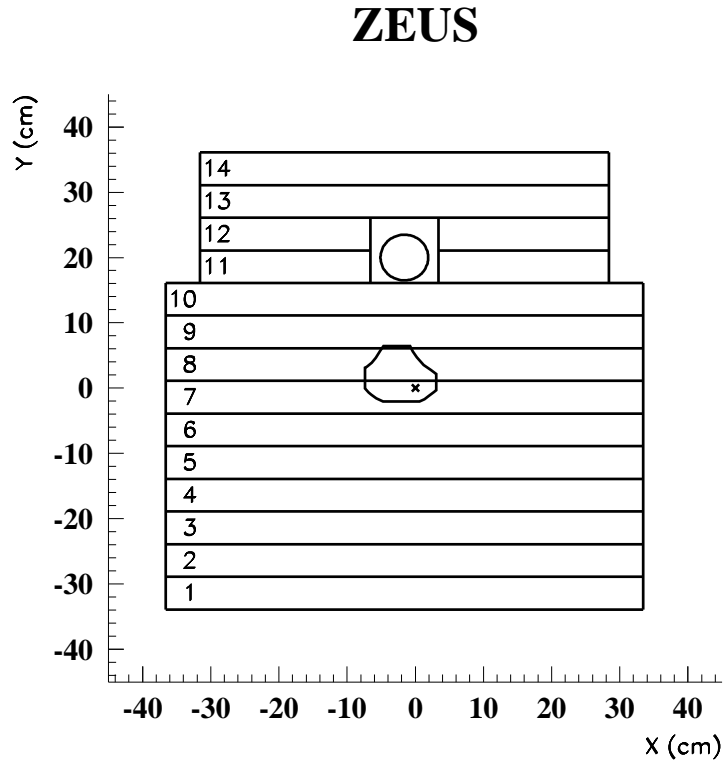}}
\caption{
A schematic view of the FNC, looking towards the interaction region.
The calorimeter is segmented vertically into 14 towers.
The window of geometric acceptance defined by the apertures of the
HERA beam-line elements is outlined in towers 7--9.
The zero-degree point is marked by an $\times$.
Towers 11 and 12 have a $10 \times 10$ cm$^2$ hole to accommodate the 
proton beampipe.
}
\label{fig:beamspot}
\end{figure}

The structure of the calorimeter is shown in 
Fig.~\ref{fig:beamspot}. 
It is a 
finely segmented, compensating,
sampling calorimeter 
with 134 layers of 1.25~cm-thick lead plates
as absorber and 0.26~cm-thick scintillator plates as the active material.
The scintillator is read out on each side with
wavelength-shifting light guides coupled 
to photomultiplier tubes (PMTs). 
It is segmented longitudinally into a front section, seven 
interaction-lengths deep, and a rear section, three interaction-lengths deep.
The front section is divided vertically into 14 towers, each 5~cm high.
The relative energy resolution for hadrons,
as measured in a test beam, 
was $\sigma/E$ = $0.65/\sqrt{E}$.
As seen in Fig.~\ref{fig:beamspot}, 
the FNC completely surrounds the
proton beam, which 
passes through a 10~$\times$10~cm$^2$ hole in towers 11 and 12.
Three planes of scintillation counters, 
each $70\times 50\times\,2$ cm$^3$, 
are located 70, 78, and 199 cm in front
of the calorimeter.  
These counters completely cover the bottom front
face of the calorimeter and are used to identify
charged particles and so reject 
neutrons that interact in front of the FNC 
in inactive material such as magnet support structures,
the beam-pipe wall and the mechanics and
supports of the LPS.

\begin{figure}[htbp!]
\epsfysize=8cm
\centerline{\epsffile{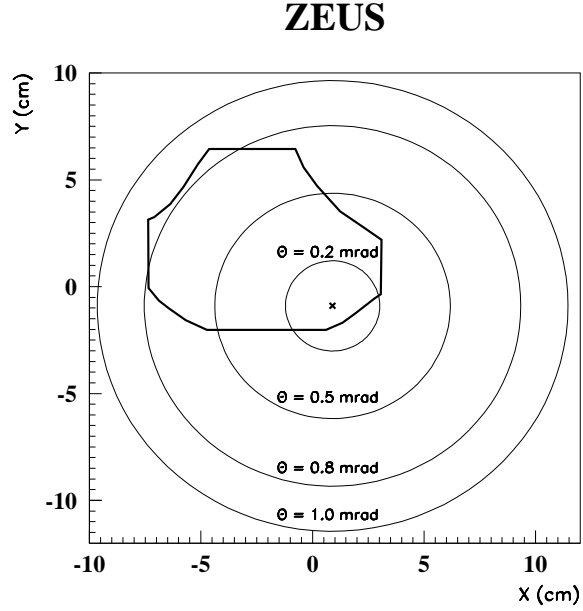}}
\caption{
Contours of constant neutron production angle compared to the
window of the geometric acceptance defined by the apertures of the
HERA beam-line elements (thick irregular curve).
The zero-degree point is marked by an $\bf{\times}$.
}
\label{fig:beamspot_isoth}
\end{figure}

\begin{figure}[hbp!]
\epsfysize=8cm
\centerline{\epsffile{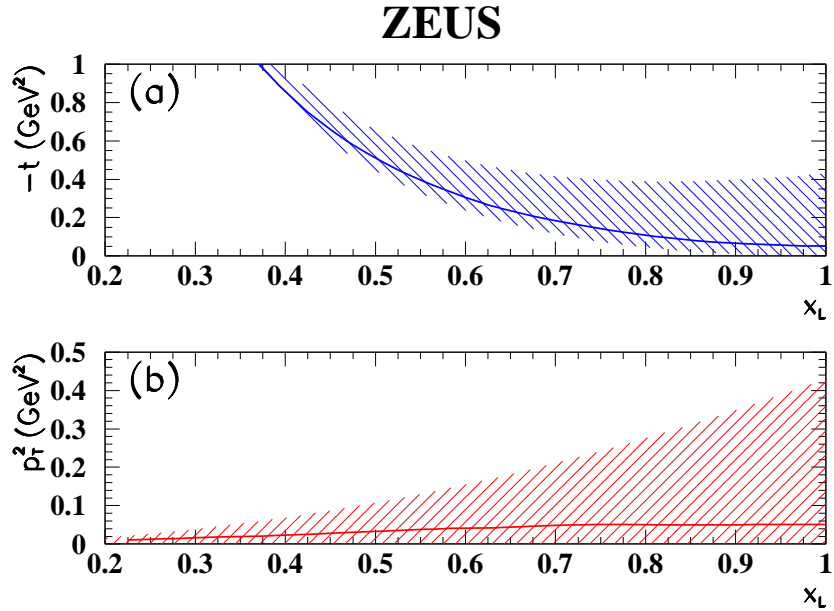}}
\caption{
The kinematic regions in (a) $t$ and (b) $\ptsq$ covered by the angular
acceptance of the FNC ($\theta <$0.8 mrad) are shown as shaded bands. The 
solid lines indicate the average $t$ and $\ptsq$ values, uncorrected for
acceptance, as a function of $\xl$.
}
\label{fig:kine_range}
\end{figure}

The HERA magnet apertures limit the FNC acceptance to neutrons with 
production angles less than $\theta_n^{\rm max} =0.8$ mrad,
that is to transverse
momenta $\ptmax = E_n\theta_n^{\rm max} =0.656\,\xl $~GeV. 
Only about one quarter of the
azimuth extends to $\theta_n^{\rm max}$, as can be seen
from the outline of the aperture shown in Fig.~\ref{fig:beamspot_isoth}.
The overall acceptance of the FNC, taking account of
beam-line geometry, inactive material,
beam tilt and angular spread, as well as
the angular distribution of the neutrons, is $\sim$20\% at low $\xl$, where the
$\pt$ range covered is small, but
increases monotonically, exceeding 30\% at high $\xl$, as seen in 
Table~\ref{etabxlerr}. The dominant influence on these values
is the lack of complete azimuthal coverage illustrated in 
Fig.~\ref{fig:beamspot_isoth}.
The kinematic region in $t$ and $\ptsq$
covered by the FNC as a function of $\xl$
is shown in Fig.~\ref{fig:kine_range}(a) and (b), respectively.
Although the acceptance extends to $\ptsq\simeq 0.4$ GeV$^2$,
the mean value of $\ptsq$ for the accepted data
is less than 0.05 GeV$^2$\cite{zeuslndijet}.

Charged particles produced at small 
angles and energies $\lap$ 450 GeV are swept away vertically by the
HERA bending magnets.
Higher-energy positively charged particles in the
energy range from about 450 GeV to about 600 GeV
are bent into the top towers above the beam pipe.
Neutrons are easily distinguished from protons, since the latter
deposit most of their energy in the top four towers of the FNC.
Protons with energies $\gap$ 600 GeV do not exit the beam pipe
in the vicinity of the FNC.

Photons, which populate only the lower $\xl$ range of the data,
are separated from neutrons using the energy-weighted
root-mean-square vertical width of the showers. After this
separation, the correction for photon contamination in the final
neutron sample is
negligible.

Timing information from the FNC was also used to eliminate backgrounds.
One source of background is due to the random overlap of energetic neutrons 
from proton beam-gas interactions with genuine DIS events.
The rate of FNC energy deposits above 400 GeV was monitored and, from
comparisons with the bunch-crossing rate, the overlap was estimated 
to be less than a few percent~\cite{fnc2} and was ignored.
This background was also checked using randomly
triggered events and events with an energy deposit in the FNC greater than 1000
GeV.

The calibration and monitoring\cite{calor97} of the FNC
followed the methods developed for the test
calorimeters\cite{fnc1,fnc2}. The gain of each PMT was
obtained by scanning the FNC with a $^{60}$Co 
radioactive source. Changes in gain during data taking were monitored 
using energy deposits from interactions of the HERA proton beam with
residual gas in the beam pipe.
The overall energy scale was set from the kinematic end-point
of 820 GeV by fitting data from proton beam-gas interactions 
with energy greater than 600 GeV
to the shape of the spectrum 
expected from one-pion exchange\cite{bishari,holtmann}. 
The accuracy ($\pm$ 2\%) of the energy-scale determination was limited 
by the uncertainty of the shape of the
beam-gas spectrum at the kinematic endpoint.

\subsection{Experimental conditions}

The data were taken in three ranges
of $Q^2$.
The first, consisting of photoproduced events where the 
scattered positron was detected
in the LUMI calorimeter, is characterised by
photon virtuality $Q^2 < 0.02$ GeV$^2$;
the second, consisting of intermediate-$Q^2$ events 
where the scattered positron was
detected in the BPC detector, is characterised by photon 
virtuality $0.1 < Q^2 < 0.74$ GeV$^2$; and
the third, originating from deep inelastic scattering,
where the scattered positron
was detected in the CAL, is characterised
by photon virtuality $Q^2 > 4$ GeV$^2$. The three samples are
denoted, respectively, as PHP, intermediate-$Q^2$ and DIS.

The data for the PHP sample come from
four short dedicated runs in 1996, taken with a minimum-bias
trigger, and correspond to an
integrated luminosity of 49 nb$^{-1}$ covering the full $\xl$ range.
For $\xl \gap 0.5$, 2 pb$^{-1}$ of data were also
taken in 1995 with the FNC included in the trigger.  
The other data samples
were taken in
1995-97, and correspond to
integrated luminosities of approximately 14 pb$^{-1}$ for the 
intermediate-$Q^2$ and
41 pb$^{-1}$ for the DIS samples.

\subsection{Trigger conditions}

A three-level trigger was used to select the events.
The trigger decision relied primarily on the energies deposited in 
the CAL, the BPC or the LUMI and was based on electromagnetic
energy, total transverse energy and missing transverse momentum.
CAL timing cuts were used to reject beam-gas interactions and cosmic rays.
No FNC requirements were placed on the trigger, except for part of 
the PHP sample in which events  were triggered by energy 
deposits in the FNC, LUMI and RCAL.
This data sample is useful 
only for $\xl > 0.52$, where the FNC triggering efficiency 
is better than 50\%.  The efficiency is 100\% for $\xl \gap 0.7$. 

\subsection{Reconstruction of kinematic variables}

The reconstruction of the lepton variables depends 
on the device in which the
scattered positron was detected. 
The LUMI measures only the energy and
the horizontal deflection of the scattered positron: such
particles have scattering angles
$\theta_e<6$ mrad, so that the
virtuality of the exchanged photon is limited to $Q^2<0.02$ GeV$^2$,
with a mean value $\approx 2\cdot 10^{-3}$ GeV$^2$.
In the BPC, both the energy and the position of the scattered 
positron  were measured, and the kinematic variables
estimated from these measurements\cite{zeusf2bpc}.
When the positron was detected in the CAL, the double angle (DA) method\cite{DAREF} was used to reconstruct the 
DIS variables from the scattering 
angles of the positron and the hadronic system.
The latter was determined 
from the hadronic energy flow measured in the CAL.

The impact position of the neutron on the FNC 
was not measured with sufficient accuracy to be used 
event-by-event. All distributions were obtained by integrating over
the angle of the produced neutrons up to the maximum
observable angle, $\theta_n^{\rm max}$= 0.8 mrad.

\section{Data selection}

The PHP, intermediate-$Q^2$
and DIS samples were first selected without any FNC requirement.
The PHP events had to satisfy criteria on the CAL energy deposits or 
CTD tracks~\cite{zeussigmatot} and required 
a scattered positron in the LUMI detector with $10<E_e'<18$ GeV, where $E_e'$
is the energy of the scattered positron.
The intermediate-$Q^2$ and DIS events had to satisfy  
$35 \le \delta \le 65$ GeV, where $\delta = \sum (E-P_Z) = 
\sum E(1-\cos \theta)$, with the
sum running over all calorimeter cells; $E$ and $\theta$ are the energy and
polar angle, respectively, for a CAL cell. For the intermediate-$Q^2$
 data, $\delta$ also
includes $2E'_e$.
The events also had to
have a reconstructed vertex with $-40<Z<100$ cm for the intermediate-$Q^2$
sample and 
$|Z|<40$ cm for the DIS events.
The intermediate-$Q^2$ events had to have
a scattered positron of $E'_e>7$ GeV within the 
fiducial volume of the BPC\cite{zeusf2bpc}, while the
DIS events had to have
a scattered positron of $E_e'>10$ GeV within the 
fiducial volume of the CAL\cite{zeusf294,zeusf293}.

The final event selection required, in addition to a good LUMI, 
BPC or CAL positron as defined above, a good FNC neutron candidate satisfying
the following conditions:
\begin{itemize}
\item between 164 GeV and 820 GeV ($0.2<\xl <1.0$) 
      of deposited energy in the FNC;
\item energy less than that corresponding to one minimum-ionising particle
      deposited in the counter
      farthest\footnote{
      The farthest counter, 199 cm from the FNC, 
      was chosen to minimise backsplash (albedo) corrections.}
      from the front of the FNC, 
      to reject neutrons
      that shower in front of the calorimeter;
\item no signal from the veto counter consistent with a shower from a
      previous bunch to reject pile-up energy deposits; 
\item the maximum energy deposit should be 
      in towers 6-9 
      in the bottom front section of the FNC,
      to reject protons that are 
      bent into the top towers by the vertical bending-magnets in
      the proton beam line;
\item a shower with both energy and vertical width consistent with 
      that of a hadron,
      to reject electromagnetic energy 
deposits\footnote{This cut is possible since the two-dimensional distribution of energy and shower width shows a clear separation between the wide, high-energy hadronic showers and narrower lower-energy deposits arising from photon-induced showers.}.
\end{itemize}

The final neutron-tagged sample consisted of 912 PHP, 8168 intermediate-$Q^2$ 
and 60102 DIS events selected from 
56960, 100599 and 642015 inclusive events, respectively. The
photoproduction background was 1.7\% for the DIS sample and 3\% for
the intermediate-$Q^2$ sample; since, as discussed below, ratios
of neutron-tagged and inclusive cross sections are used in this
analysis, this contamination has a negligible effect.

The data were divided into $\xl $ bins in the range $0.2< \xl < 1.0$,
as shown in  Table~\ref{etabxlerr}. In each bin, the acceptance 
extends from 0 to 0.8 mrad in neutron production angle,
representing a $\pt $ range from 
0 to $0.656\, \xl $ GeV.

Different binnings were used for the three data sets.
For the PHP data, a single bin was chosen with 
$Q^2<0.02$ GeV$^2$ and $0.345 < y < 0.636$,
corresponding to a $W$ range of $176 < W < 240$ GeV with a
mean $<W>$ = 209 GeV.
For the intermediate-$Q^2$ data ($0.1 < Q^2 < 0.74$ GeV$^2$), the same 
34 bins were used as for
the ZEUS inclusive $\gamma ^* p$ cross-section measurements\cite{zeusf2bpc}.
For the DIS data ($Q^2>4$ GeV$^2$), 
25 bins were used, as shown in Appendix I.

\section{The ratio method}
\label{sec:f2ln3}

To minimise systematic uncertainties, it is advantageous to use the ratio of cross sections rather than the absolute cross sections themselves.
The primary measurement, the fraction of events with a leading neutron,
is experimentally robust: a luminosity measurement is not required;
uncertainties due both to the finite acceptance of the central calorimeter
and to triggering inefficiencies are minimised; accurate Monte
Carlo modeling of the inclusive hadronic final state
is not needed; and the effects of radiative corrections are reduced.

The measured events were binned in
$Q^2$.  Depending on the $Q^2$ region, the events were further
binned in $x$ or $y$.
The ratio $r(x,Q^2,\xl)$ is defined as  
\begin{equation}
r(x,Q^2,\xl)=
\frac{d^3\sigma^{ep\rightarrow e^{\prime}Xn}/dx \,dQ^2 \,d\xl }
{d^2\sigma ^{ep \rightarrow e^{\prime}X}/dx \,dQ^2}\Delta \xl =
\frac{n_{\rm obs}(x,Q^2,\xl)}{A(\xl)N_{\rm obs}(x,Q^2)},  \label{eq:ratio}
\end{equation}
where $N_{\rm obs}(x,Q^2)$ is the number of events observed in a bin
of ($x$, $Q^2$)
and $n_{\rm obs}(x,Q^2,\xl)$ is that subset of events
with a neutron in an $\xl$ bin of width $\Delta \xl $; 
$A(\xl)$ is the acceptance and 
reconstruction efficiency for leading neutrons and is
discussed in the next section.

Inspection of Eqs.~(\ref{eq:inc}), 
(\ref{eq:ln3}) and (\ref{eq:ratio}) shows that, after appropriate
bin centering,
\begin{equation}
{F}_2^{\mbox{{\rm\tiny LN(3)}}}(x,Q^2,\xl)
=\frac{r(x,Q^2,\xl)}{\Delta \xl}F_2(x,Q^2) (1+\Delta), \label{eq:fn}
\end{equation}
where
\[
\Delta \simeq \delta - \delta_{\rm \tiny LN} -\delta_{\rm det}.
\]
The additional correction term, $\delta_{\rm det}$, is the difference between
the acceptance of the main detector
for all events and for the subset of events with a 
leading neutron. The overall correction,
$\Delta$, is small and was neglected (see Section~\ref{sec:corrections}).

Equation (\ref{eq:fn}) can be expressed in terms of cross sections by using
Eqs. (\ref{eq:sigtot}) and (\ref{eq:sigln}):
\begin{equation}
\frac{d\sigma^{\gamma^\ast p \rightarrow Xn}(W,Q^2)}{d\xl}=
\frac{r(W,Q^2,\xl)}{\Delta \xl}\sigma^{\gamma^\ast p}_{\rm tot}(W,Q^2).
\label{eq:dsigln}
\end{equation}

This equation is also  valid for 
photoproduction ($Q^2$=0).  It is evident from
Eq. (\ref{eq:ratio}) 
that, within an ($x$, $Q^2$) bin,
$r$ is the ratio of the cross section
for producing a leading neutron with $\theta_n<\theta_n^{\rm max}$
in a given $\xl$ bin to the total cross section.

\section{Corrections and systematic uncertainties}
\label{sec:systematics}

\subsection{Acceptance and corrections}
\label{sec:acceptance}

For the most part, trigger inefficiencies and acceptance uncertainties
in the main ZEUS detector 
are unimportant for the present analysis since ratios of
FNC-tagged events to inclusive events are measured. 
The observed distributions integrated over the ($x,Q^2$)
plane are not a faithful reproduction the true distributions 
because of varying
trigger configurations and finite acceptances. 
Therefore, when integrated distributions were
studied, the events were weighted, bin-by-bin,
by the inclusive cross section, 
to produce the correct $Q^2$ and $x$ (or $y$) distributions.
Since the ratio of the  neutron-tagged sample to the
inclusive sample is not strongly dependent on $x$ or $Q^2$ (see below), 
this weighting procedure does not 
significantly change the shape of any of the distributions of
the ratio. However, the statistical uncertainty is increased since
low-$Q^2$ events are under-represented in comparison to high-$Q^2$
events. 

The intermediate-$Q^2$
 and DIS data, triggered by the scattered positron, have a trigger
efficiency that is independent of the details of the final state. However,
the photoproduction data must also be corrected for the
CAL trigger efficiency, which is dependent
on the final state.
For example,
exclusive vector meson production, $ep \rightarrow eVp$, has a low
trigger efficiency and no neutrons in the final state.
As a result, the measured fraction of events tagged with a leading
neutron for photoproduction must be corrected downwards.
A study of each exclusive final state and its corresponding
trigger efficiency showed that the
measured fraction of tagged events in photoproduction should be
reduced by ($5\pm 2.5$)\%. This reduction was applied 
to the photoproduction data presented in this paper. 

The acceptance and reconstruction efficiency, $A(\xl)$, 
of the FNC was obtained using
the full simulation of the ZEUS detector based on
GEANT 3.13\cite{geant}.
The simulation includes the effect of inactive material
along the neutron path. The tilt ($ \lap 70\,\mu$rad in $X$
and $Y$, but varying from year to year) and divergence 
($\approx $ 70 (100) $\mu$rad in the $X~(Y)$ direction)
of the proton beam were also simulated.

The relative uncertainty on the absolute energy scale of the FNC is 
$\pm2\%$~\cite{calor97}. This introduces the normalisation uncertainties 
shown in Table~\ref{etabxlerr}.

\begin{figure}[htbp!]
\epsfysize=14cm
\centerline{\epsffile{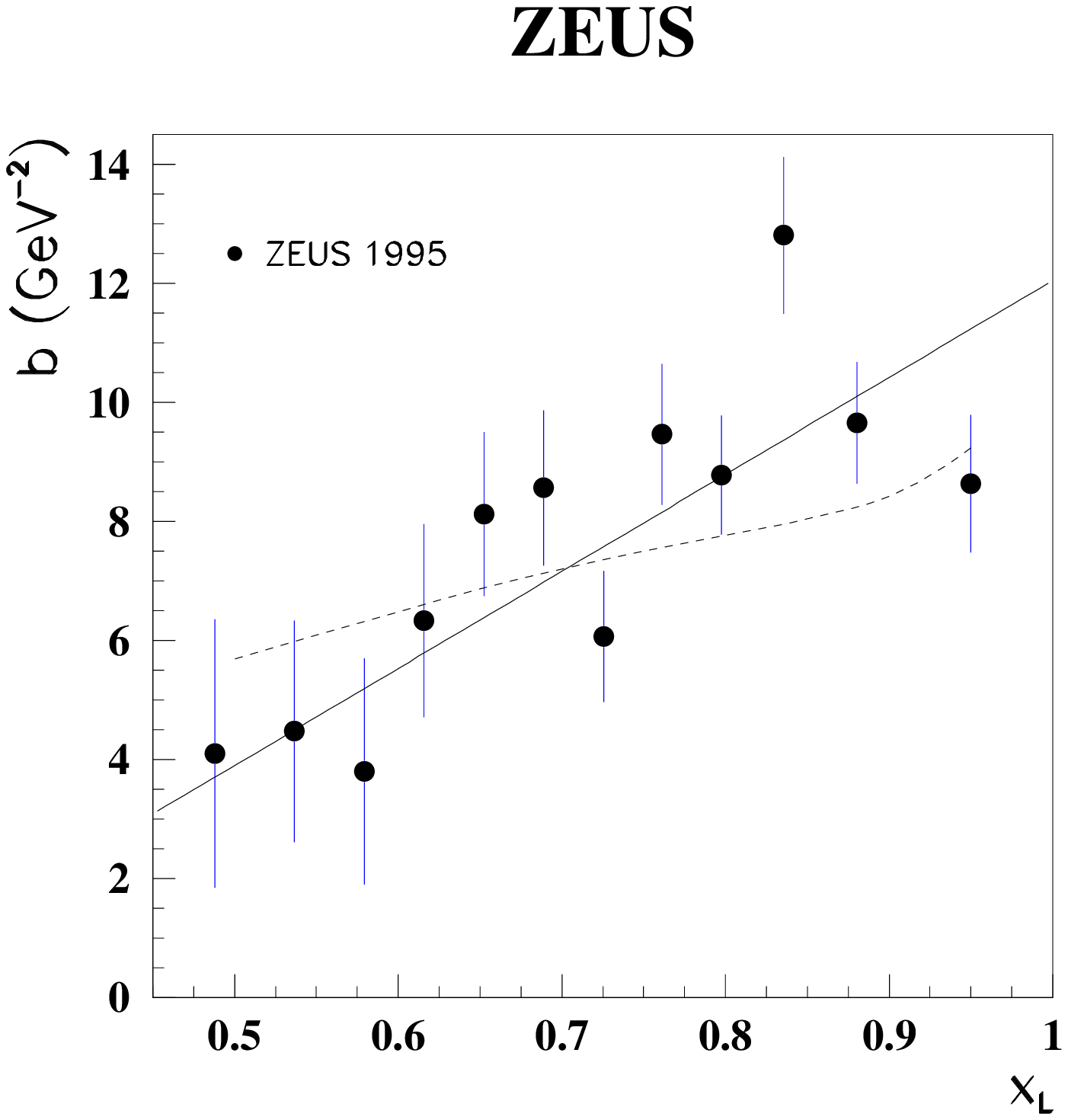}}
\caption{
The slope, $b$, of the exponential $\ptsq$ distribution as a function of $\xl$
for the 1995 DIS data set.
The solid line is given by $b(\xl) = (16.3\xl -4.25)$ GeV$^{-2}$. The dashed
line is the result obtained using the effective flux of 
Eq. (\ref{eq:eff_flux}). The vertical error bars display the statistical 
uncertainties only.
}
\label{fig:bvsxl}
\end{figure}

Because of the poor position resolution of the FNC,
the angular distribution of the neutrons
could not be measured directly.
Instead, it was constrained by the distribution of measured
$X$ positions of the neutrons.
The $X$ position of energy deposits in the FNC was determined from the ratio of pulse heights in the two readout channels. For the 1995 data taking, the corresponding rms resolution was 3.2 cm. The resolution degraded in subsequent years because of radiation damage to the scintillator. 
For fixed $\xl $, the $\ptsq$ distribution was expressed as
\begin{equation}
 \frac{dN}{d\ptsq}\propto e^{-b(\xl )\ptsq}.
\label{eq:bslope}
\end{equation}
The slope $b$ that best represents the data was determined by binning
the $X$-position data in $\xl $ and comparing it to
the Monte Carlo expectation for different $b$ slopes.
 The results for the 1995 DIS data set, given in Table~\ref{btab} and
 shown in Fig.~\ref{fig:bvsxl},
indicate that the exponential $\ptsq$ slope for neutron  production rises
with increasing $\xl$ and may be parameterised by
$b(\xl )$ = $(16.3~\xl -4.25)$ GeV$^{-2}$, as indicated by the solid line
(for $\xl < 0.26$, $b$ = 0 was used for the the acceptance corrections).
Also shown in Fig.~\ref{fig:bvsxl}, as
the dashed curve, is the slope parameter resulting from the effective OPE 
flux of Eq. (\ref{eq:eff_flux}). The predicted slopes are in reasonable
accord with the data.

The polar-angle acceptance of the FNC extends from 0 to 0.8 mrad
with an azimuthal acceptance shown in Fig.~\ref{fig:beamspot_isoth}.
All of the results presented in this paper have been corrected for
this azimuthal acceptance.
Since $\ptmax = 0.656\,\xl $ GeV, the sensitivity of the
acceptance to the $b$ slope decreases rapidly as $\xl $ decreases. 
As a systematic check, the above value of $b(\xl )$
was varied by $\pm2$ GeV$^{-2}$, which is the approximate uncertainty
in $b$. The uncertainty on the acceptance 
reaches 6\% at the highest $\xl $ values. 
The acceptances and their uncertainties, which are highly correlated 
between all of the bins, are summarised in Table~\ref{etabxlerr}.

\subsection{Systematic uncertainties}

The corrections applied to the acceptance lead to normalisation uncertainties.
The largest corrections and their typical values and  uncertainties  are:
\begin{itemize}
\item the increased acceptance of the FNC  when the LPS is not inserted 
      into the beam, since less material is present ($-5\pm 1$\%);
\item the veto cut,
      which avoids pile-up effects but removes events 
      that should have been counted ($+7\pm 2$\%);
\item shower albedo from the face of the FNC, 
      which triggers the veto counters
      and eliminates good events ($+1.5\pm 1$\%);
\item noise in the veto counters associated with the beam ($+3\pm 1.5$\%);
\item veto-counter inefficiency ($-1.5\pm 1.5$\%);
\item the random overlap of beam-gas events with $ep$ interactions ($-2.4\pm 0.2$\%).
\end{itemize}
The total normalisation uncertainty for the intermediate-$Q^2$
and DIS data is $\pm$4\%.
The corresponding uncertainty for the PHP
data is $\pm$5\%, which is highly correlated with the DIS uncertainty.

The photoproduction data taken with the FNC trigger
were corrected for the trigger inefficiency and 
were used only for $\xl >0.52$, the region where the trigger efficiency
is above 50\%. In this region,
 they were normalised to the data collected without the FNC-trigger
requirement. This leads to an additional normalisation
uncertainty for these data of $\pm$4\%.

\subsection{Relative corrections}
\label{sec:corrections}

The relative-correction term, $\Delta$, of Eq. (\ref{eq:fn})  
accounts for differences
in the corrections for the neutron-tagged and inclusive 
samples.
Only the difference between the radiative correction for the neutron-tagged
sample and that for the inclusive sample,
($\delta - \delta_{\mbox{\rm\tiny LN}}$, 
is relevant in this analysis\footnote{Possible effects due to the 
longitudinal structure
function and $Z^0$ exchange are also assumed to be negligible.}.
The program HECTOR\cite{hector}, which
provides a leading-logarithmic estimate of the radiative correction to
the Born term in $ep$ scattering, was
used to calculate the relative-correction term by modeling
inclusive neutron production at large $\xl$ via one-pion exchange.
The difference in the neutron-tagged and inclusive
correction terms increases with $x$ and $\xl$,
but remains small throughout most of the kinematic region, reaching 
(3-4)\% at $x = 0.06$ for $\xl \gap 0.9$ and 
at $x = 0.1$ for $\xl \gap 0.8$.

The final term in $\Delta$ is due to 
the correction factor, $\delta_{\rm det}$, which accounts for the
differences in the central-detector acceptance, migrations and bin-centering
for all events and for those events with a tagged neutron. 

Since $\Delta$ is smaller
than the overall systematic uncertainties of the FNC, it is
neglected in the present analysis.  

\section{General characteristics of the data} 
\label{sec:charact}

\subsection{Neutron energy spectra}
\label{sec:spectrum}

The neutron yields in bins
of $\xl$ and $Q^2$ are given in Table~\ref{etabxlall} for the three data
samples.
The data are corrected for the azimuthal acceptance and are
integrated over production angles $0< \theta_n<
\theta^{\rm max}_n$, where $\theta^{\rm max}_n$ =  0.8 mrad.  
As discussed earlier, this fixed angular 
range corresponds to a $\pt$ range which is $\xl $ dependent.
The spectra are normalised to the measured number of events 
in the same ($x,Q^2$) bin but without the leading-neutron requirement.
Modifying Eq. (\ref{eq:ratio}) to take corrections into account yields
\[
r(x,Q^2,x_L)\equiv \frac
{1}{A(\xl)}
\frac{\sum_{x}\sum_{Q^2}n_{\rm corr}(x,Q^2,\xl)}
{\sum_{x}\sum_{Q^2}N_{\rm corr}(x,Q^2)},
\]
where the subscript indicates that the observed number of events
has been corrected bin by bin
for the effect of the trigger configurations
discussed in Section~\ref{sec:systematics}.
The sum is over the chosen range of $Q^2$ and all measured bins of
$x$ in that range. 

\begin{figure}[thb]
\hspace{-1cm}
\epsfysize=9cm
\centerline{\epsffile{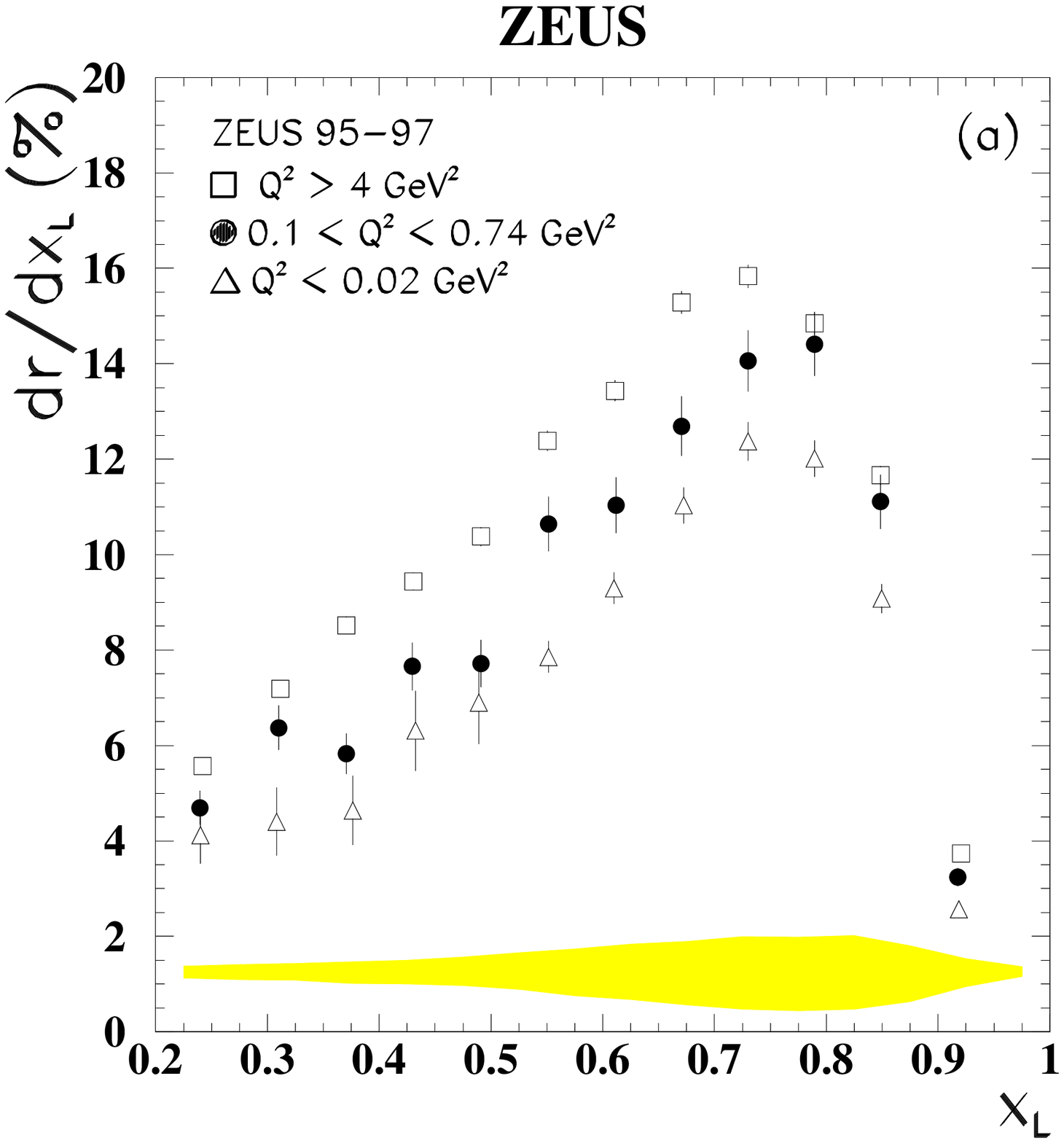}}
\epsfysize=9cm
\centerline{\epsffile{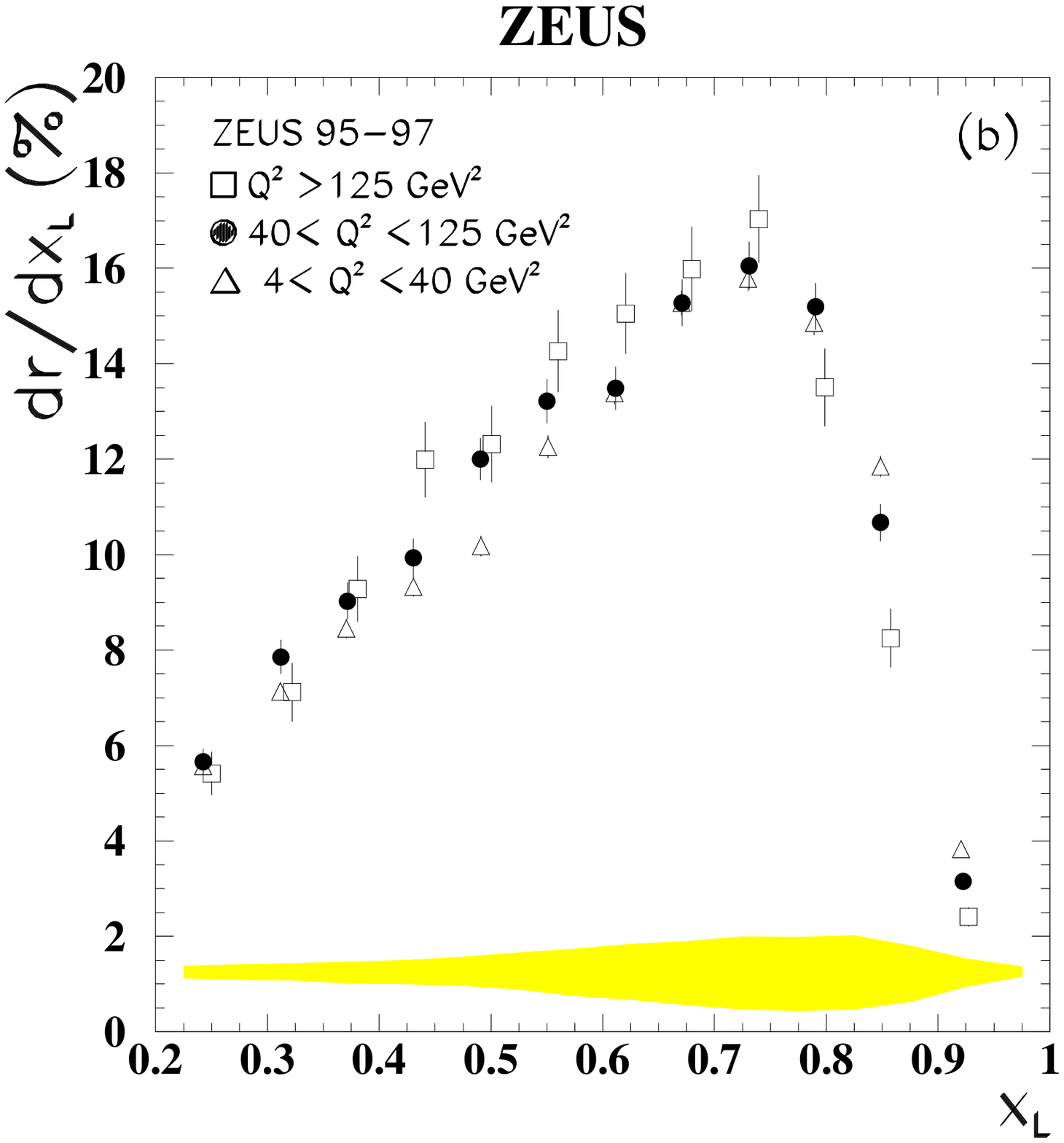}}
\caption{ (a) Energy spectra, $dr/d\xl$, of neutrons
with $\theta_n < 0.8$ mr 
in PHP, in the intermediate-$Q^2$ region and in DIS. 
The PHP data have a trigger-correction uncertainty
of 2.5\% which can move them relative to the intermediate-$Q^2$
and DIS data and, for $\xl >0.52$, have
an additional normalisation uncertainty of $\pm$4\%.
(b)
Energy spectra, $dr/d\xl$, of neutrons with $\theta_n < 0.8$ mr in DIS
for three $Q^2$ selections.
The points 
in (b) are offset slightly in $\xl$ for improved visibility. 
The shaded bands show
the systematic uncertainty due to the acceptance of the FNC. 
There is also a normalisation uncertainty of $\pm$5 ($\pm$4)\% for 
the PHP (intermediate-$Q^2$ and DIS) data.
The energy-scale uncertainty of $\pm$2\%  is equivalent to
a stretch or compression along the abscissa. Not shown are the correlated
normalisation uncertainties due to the energy-scale uncertainty
given in Table~\ref{etabxlerr}.
}
\label{fig:energy_spectrum}
\end{figure}

Figure~\ref{fig:energy_spectrum}(a) shows the $\xl$ spectra,
as $dr/d\xl$, for the
PHP, intermediate-$Q^2$ and DIS regions.
Between 5\% and 10\% of the inclusive events have a leading neutron in the
measured kinematic range.
The ratios are similar in shape
for these three kinematic regions,  
although the fractional yield of neutrons 
slowly increases with $Q^2$. The neutron spectra increase slowly with $\xl$
(mostly due to the increased $\pt$ acceptance),
reach a broad peak near $\xl =0.75$, and then
rapidly decrease to zero at $\xl  = 1$. 
The neutron rate increases by about 20\% between the PHP and DIS data. 

Figure~\ref{fig:energy_spectrum}(b) shows the DIS neutron spectra 
in three bins of $Q^2$.
Although, for $\xl  < 0.8$, the high-$Q^2$ data tend to
lie above the low-$Q^2$ data, the increase in neutron yield with
$Q^2$ is much less pronounced than that seen in
Fig.~\ref{fig:energy_spectrum}(a).  In the
region $0.64<\xl<0.82$, the neutron yield in DIS is approximately
independent of $Q^2$. For the highest-$\xl $ points,
the high-$Q^2$ data lie below the low-$Q^2$ data. 

\subsection{Determination of ${\sigma}^{\gamma p \rightarrow Xn}$}

\begin{figure}[htbp!]
\epsfysize=9.3cm
\centerline{\epsffile{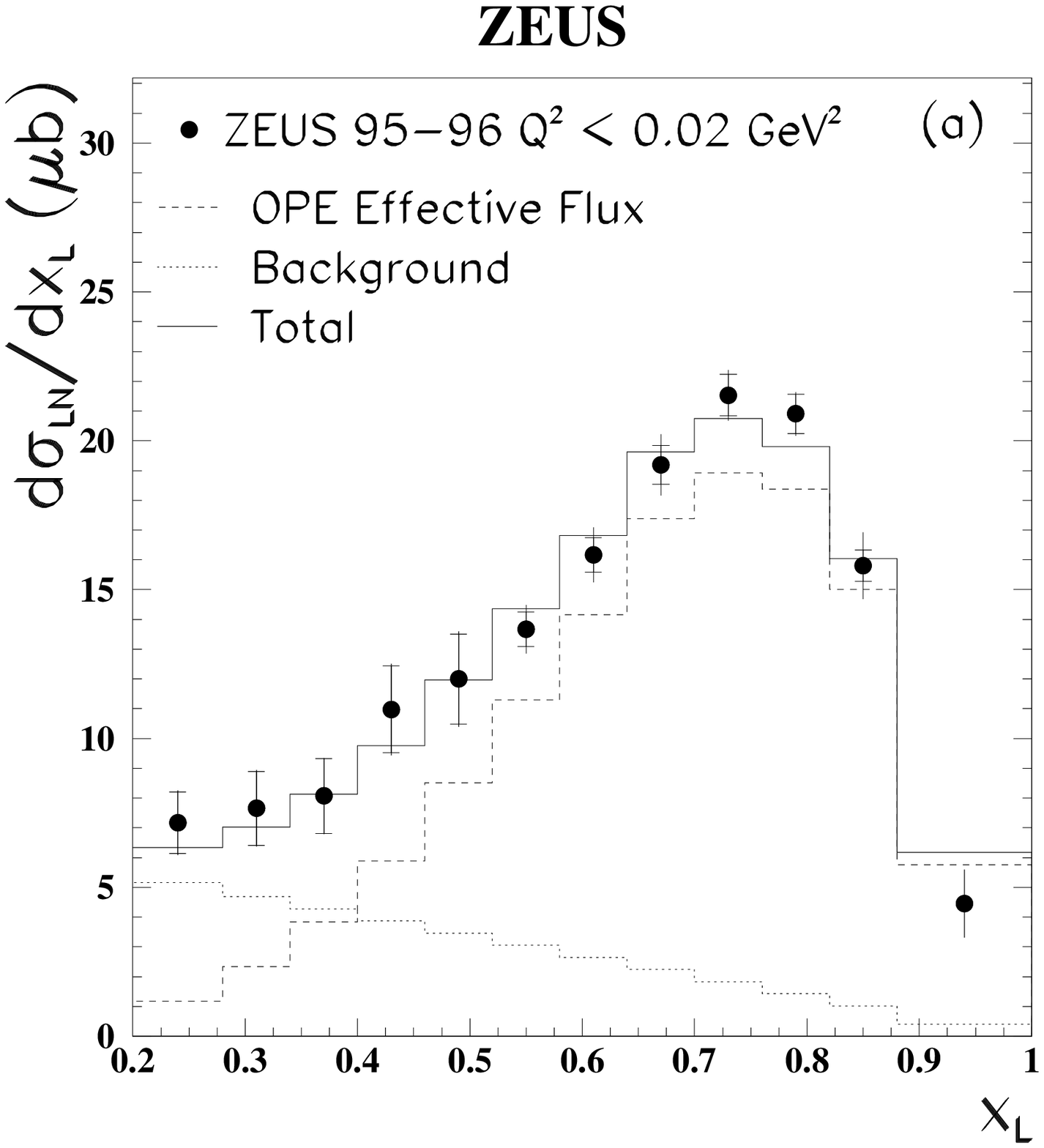}}
\epsfysize=9cm
\centerline{\epsffile{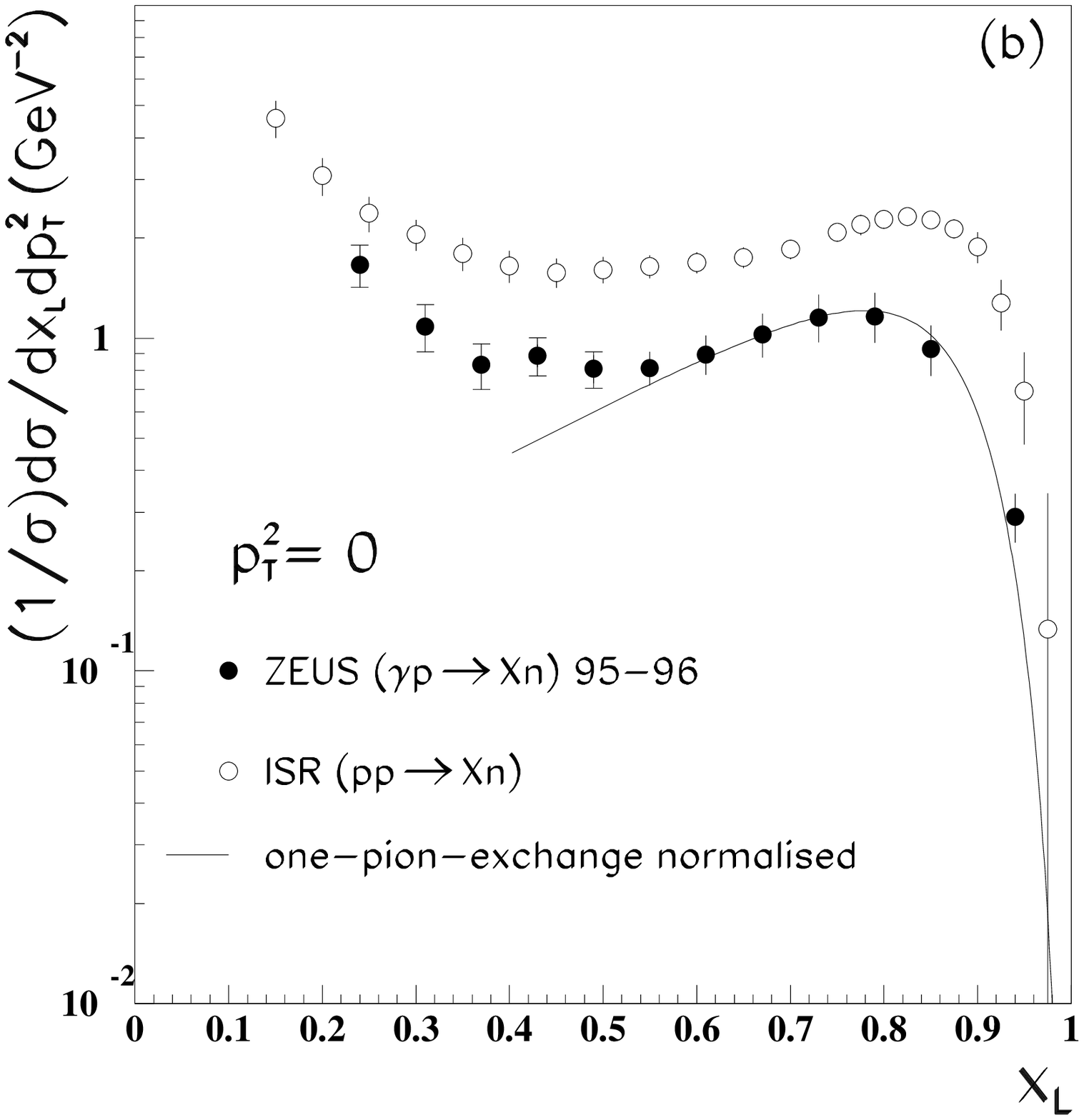}}
\caption{(a) The differential cross section for photoproduction
of leading neutrons with $\theta_n <$ 0.8 mrad, 
${d{\sigma}^{\rm LN}}/{d\xl}$, at $\langle W \rangle
=207$ GeV. The solid histogram shows
the result of a fit to the data using the OPE effective flux 
factor, $f_{\rm eff}$ of Eq.~(\ref{eq:eff_flux}), shown as the dashed histogram, plus a 
background term $\propto (1-\xl $), shown as the dotted histogram.
(b) The normalised cross section for $\gamma p \rightarrow Xn$ at $\ptsq$ = 0
(solid points). Also shown (open circles)
are data from the ISR for $pp \rightarrow Xn$~\cite{flauger}. 
The curve is
the expectation of the one-pion-exchange model for hadron-hadron data
scaled by 0.41. In both (a) and (b), the statistical uncertainties are shown by 
the inner error bars and the statistical and systematic uncertainties
added in quadrature by the outer bars. Points for $\xl <0.5$ ($\xl >0.5$)
are dominated by statistical (systematic) uncertainties (see Table~\ref{etabxlerr}).}
\label{fig:energy-spectrum-fits}
\end{figure}

Figure \ref{fig:energy-spectrum-fits}(a) shows the differential cross section,
$d{\sigma}^{\rm LN}/d\xl $, 
for the photoproduction reaction $\gamma p \rightarrow Xn$
at a mean $\gamma p$ cms energy $\langle W \rangle$ = 207 GeV. This result 
is obtained using Eq.
 (\ref{eq:dsigln}) with $\sigma^{\gamma p}_{\rm tot}$ = $174\pm 1 \pm 13$
 $\mu$b at $\langle W \rangle$ = 209 GeV~\cite{sigtot}.
Integrating $r$ over the range $0.2<\xl <1.0$ gives
\[
r = 5.72\pm 0.12(stat.)\pm 0.41(syst.)~\%,
\]
corresponding to an observed cross section in the above kinematic range of
\[
{\sigma}^{\gamma p \rightarrow Xn} = 10.0 \pm 0.2 (stat.)
\pm 1.0 (syst.)~\mu{\rm b}.
\]
The curves in Fig. \ref{fig:energy-spectrum-fits}(a)
are based on the effective flux of Eq.~(\ref{eq:eff_flux}). 
For photoproduction, the $\xl$ dependence of the 
leading-neutron cross section in this model can be expressed as the 
integral over $t$ (or $\ptsq$) of Eq. (\ref{eq:opef}):
\[
\frac{1}{\sigma} \frac{d\sigma^{\rm \tiny LN}}{d\xl} \propto
    \: (1-\xl )^{\alpha_{\pom}(0)-1}
   \int \feff (\xl ,t)dt,
\]
where it has been 
assumed that the $\gamma\pi$ total cross section has the same 
slow energy dependence as observed for other hadronic cross sections, 
characterised by the Pomeron intercept, 
$\alpha_\pom (0)) \sim 
1.08$~\cite{donnachie}. 

The solid histogram in Fig.~\ref{fig:energy-spectrum-fits}(a)
shows the result of a fit using the shape 
given by Eq.~(\ref{eq:eff_flux}) plus a background term proportional to $(1-\xl)$. 
The fit\footnote{The highest-$\xl$ bin was excluded from the fit
because of the large systematic uncertainty arising from the energy
scale.} 
uses only statistical uncertainties and gives a good description of the data.
Background terms of higher order in $(1-\xl)$ are not  
necessary.

The H1 collaboration also finds agreement between their leading-neutron 
data\cite{h1f2lb} in the DIS regime and
an OPE model. Simple leading-order OPE models (POMPYT and RAPGAP) also
give a good description of the ZEUS data on dijets
with a leading neutron in photoproduction\cite{zeuslndijet}.
A recent NLO calculation~\cite{klasen}
 based on OPE is also in good agreement with
the neutron-tagged dijet data.

\subsection{Vertex factorisation}
\label{sec:vfact}

      Leading baryon production in hadronic reactions has often been
analysed in the context the Triple Regge expansion~\cite{field,tripregge}. 
If the pion
trajectory dominates leading neutron production, as the data 
reported in this paper suggest, and if the $s^'\prime$ values are sufficiently high, then the
$\pi$-$\pi$-Pomeron triple-Regge vertex will dominate.
This is a reasonable approximation since $s^\prime$ is
typically 80 GeV and
factorisation should hold. The cross-section ratios are therefore given by
\[
\frac{d^2 \sigma^{ap \rightarrow X n}(\xl,\ptsq)}{d\xl \,d\ptsq}\left /
\frac{d^2 \sigma^{bp \rightarrow X n}(\xl,\ptsq)}{d\xl \,d\ptsq}\right .
=\frac{\sigma_{\rm tot}^{ap} }{\sigma_{\rm tot}^{bp} },
\]
or, with $a=\gamma$ and $b=p$,
\begin{equation}
\frac{1}{\sigma_{\rm tot}^{\gamma p} }\frac{d^2 \sigma^{\gamma p \rightarrow X n}
(\xl,\ptsq)}{d\xl\, d\ptsq}
=\frac{1}{\sigma_{\rm tot}^{pp} } \frac {d^2 \sigma^{pp \rightarrow X n}
(\xl,\ptsq)}{d\xl \,d\ptsq}.
\label{eq:vfact}
\end{equation}
To test this relation, 
the PHP data were integrated over $\ptsq$, assuming that the functional
form of Eq.~(\ref{eq:bslope}) also applies to photoproduction. 
The results for the PHP sample,
 expressed as the normalised cross section at $\ptsq$ = 0, are shown 
in Fig.~\ref{fig:energy-spectrum-fits}(b).
These data are compared with ISR data~\cite{flauger} from the reaction
$pp \rightarrow Xn$. The shapes of the two distributions are similar but
the PHP data are about a factor of two below
the ISR $pp$ results for $\xl >0.3$. These distributions,
when integrated over all $\ptsq$ and 
$\xl$, represent the average number of neutrons per
event, $<n>$. Using Eq. (\ref{eq:bslope}) to integrate over all 
$\ptsq$, the ZEUS PHP data yield $<n> \approx  0.18$ for the 
region $\xl >$ 0.28. 
The value 0.375$\pm$0.075 for $\theta_n <$ 150 mrad
per hemisphere per inelastic $pp$ collision was obtained 
at the ISR~\cite{engler} and is a
factor of two larger than the value obtained in the current analysis.
 
The curve in Fig.~\ref{fig:energy-spectrum-fits}(b),
based on the OPE model with the effective flux of Eq. (\ref{eq:eff_flux}),
gives a good representation of the shape of the ZEUS PHP data for $\xl > 0.6$.
In summary, while the shapes of the $\gamma p$ and $pp$ distributions
are similar,
the cross-section scaling expressed in Eq.~(\ref{eq:vfact}) is broken by about 
a factor of two.
This breaking  suggests that other Regge trajectories, with their associated
interferences, are contributing  differently to the photo- and
hadroproduction reactions.

\subsection{Absorptive effects}
\label{sec:abs}

In $\gamma^* p$ scattering,
the transverse size of the virtual photon decreases with increasing
$Q^2$, reducing the likelihood that
the produced neutron rescatters on the hadronic component of the photon. 

\begin{figure}[htbp!]
\epsfysize=19cm
\centerline{\epsffile{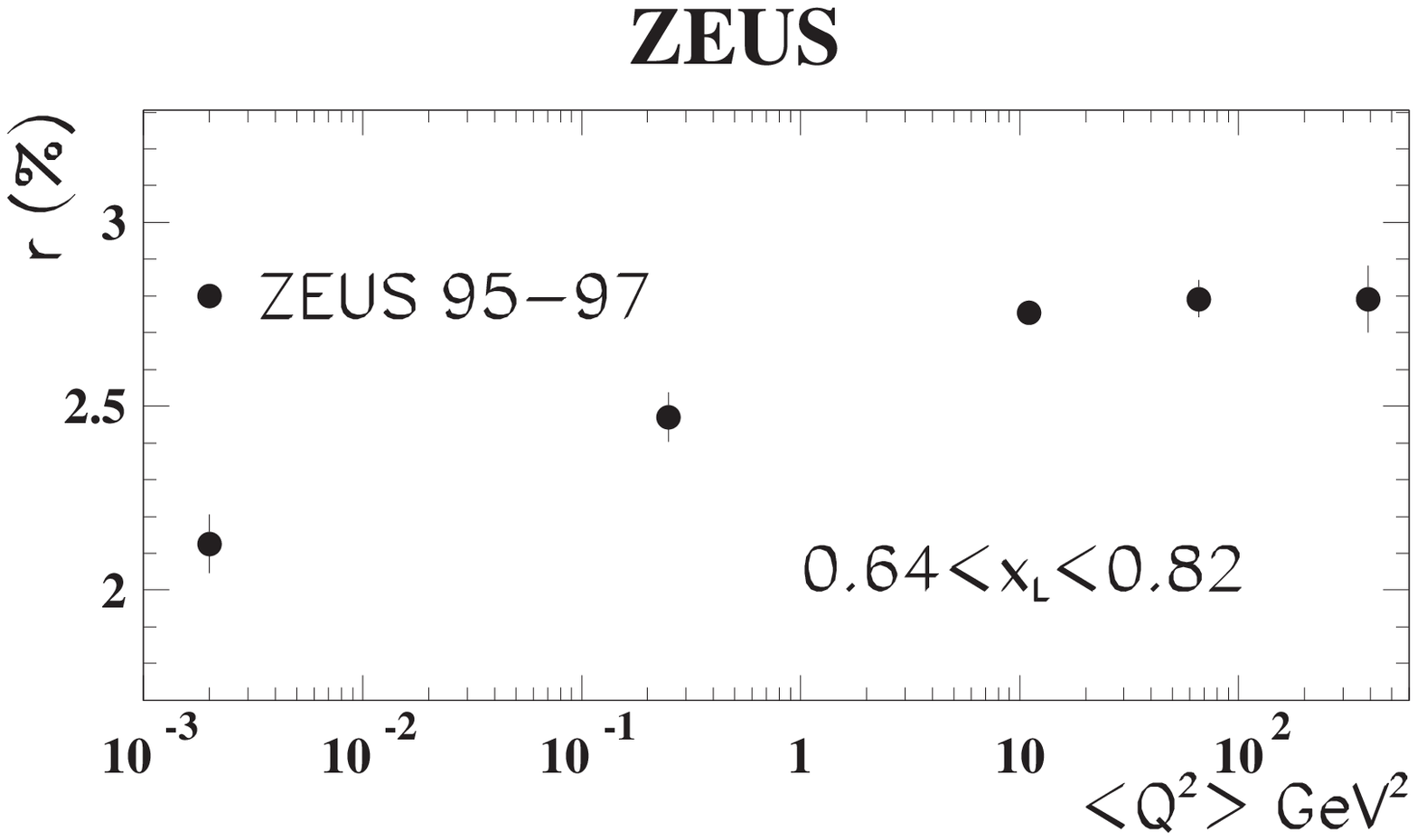}}
\vspace{-8cm}
\caption{The ratio $r$ for $0.64<\xl <0.82$ as a function of $Q^2$. The ratio
$r$ is plotted at the average value of $Q^2$ for the bin.}
\label{fig:absorption}
\end{figure}

At very high $Q^2$, the virtual photon acts as a point-like probe.
The HERA data thus offer the opportunity to study scattering
with a target of fixed size (the proton) and a projectile of variable size
(the photon). 
Figure~\ref{fig:energy_spectrum}(a) shows a clear
reduction in the relative yield of neutrons in PHP and in the 
intermediate-$Q^2$ region compared to that in DIS at higher $Q^2$.
This reduction, which is displayed as a function of $Q^2$ in 
Fig.~\ref{fig:absorption} for $0.64 < \xl < 0.82$, may be attributed to
absorptive effects.
Recently, a similar rise in the leading-proton yield as 
$Q^2$ is increased from the PHP region to $Q^2 >$ 2.5 GeV$^2$ has been
observed by the H1 collaboration~\cite{01-062}.

\begin{figure}[htbp!]
\epsfysize=19cm
\centerline{\epsffile{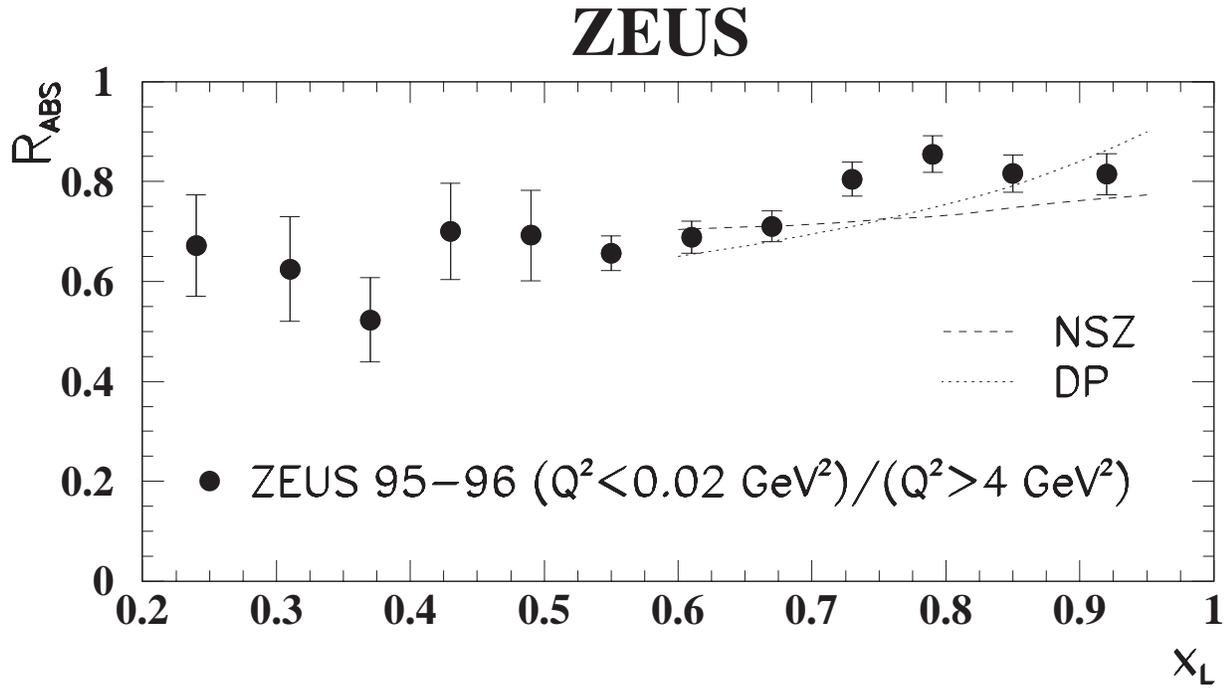}}
\vspace{-8cm}
\caption{The ratio of the neutron energy spectrum in PHP
to that in DIS. The curves labelled NSZ\cite{nszak} and DP\cite{alesio}
are theoretical predictions of the effects of absorption, 
($Q^2 < 0.02$ GeV$^2$)/($Q^2 > 10$ GeV$^2$). The vertical
bars display the statistical uncertainties.
}
\label{fig:absorptionxl}
\end{figure}

The $\xl $
dependence of this effect is shown in 
Fig.~\ref{fig:absorptionxl}, where the ratio $\Rabs (\xl )$ is defined as
\[
  \Rabs (\xl )\equiv \frac{r(Q^2<0.02\;{\rm GeV}^2,\xl )}
{r(Q^2>4\;{\rm GeV}^2,\xl )}. \nonumber
\]
In order to minimise sensitivity to drifts in the relative energy
scales,
which particularly affects the higher values of $\xl$, only PHP
and DIS data taken simultaneously were used to evaluate the
seven $\Rabs$ points highest in $\xl$. Figure~\ref{fig:absorptionxl}
shows that $\Rabs$
is approximately constant, although some variation is 
observed for the higher $\xl$ values.

The curves in Fig.~\ref{fig:absorptionxl} 
are theoretical predictions based
on OPE models\cite{nszak,alesio} for the difference 
in absorption between $Q^2 >$ 10 GeV$^2$ and photoproduction.
The $t$ dependence of the absorptive correction
has been evaluated over a kinematic region similar to
that of the present analysis 
to give the $t$-averaged correction.
The calculations~\cite{nszak,alesio} predict that absorption at the 
level of $\sim$10\% may still be present for $Q^2 >$ 10 GeV$^2$.

In conclusion,
the characteristics of the production of leading neutrons in the range
$0.64 < \xl < 0.82$, 
i.e. the rapid rise in rate with $Q^2$ and the saturation for the DIS
data, are in broad agreement with the expectation from absorptive effects.

\subsection{Tests of limiting fragmentation and factorisation in $x$ and $Q^2$}
\label{sec:sigmaln}

The hypothesis of limiting fragmentation\cite{limiting_frag_a} 
states that, in the  high-energy limit, the 
cross section  for the inclusive production of particle $c$ 
in the fragmentation region of $b$ in the reaction
$a+b \rightarrow c+X$ will approach a constant value.
This suggestion, which was
based on rather general geometrical arguments, was supported by
measurements made at the ISR~\cite{bellettini}. The expectation at HERA is 
that neutron production in the proton-fragmentation region of $ep$ 
collisions  will be independent of $Q^2$ and $W$~\cite{limiting_frag_b}.

A more differential form of factorisation than that discussed in
Section~\ref{sec:vfact} states that the dependence
of the cross section on the
lepton variables ($x$ and $Q^2$) should be independent of 
the baryon variables ($\xl $ and $\ptsq$).
In terms of the structure function $F^{\mbox{\rm\tiny LN(4)}}_2$,
this can be written as
\[
F_2^{\mbox{\rm\tiny LN(4)}}(x,Q^2,\xl,\pt)= f(\xl ,\pt)\cdot F(x,Q^2),
\]
where $f$ and $F$ are arbitrary functions.
The quantity ${F}^{\mbox{\rm\tiny LN(3)}}_2$ is then given by
\begin{equation}
{F}_2^{\mbox{\rm\tiny LN(3)}}(x,Q^2,\xl)= \tilde{f}(\xl)\cdot F(x,Q^2),
\label{eq:fact}
\end{equation}
where
\[
 \tilde{f}
(\xl) = \int_{0}^{\ptmax (\xl )} f(\xl ,\pt) d\pt.
\]

\begin{figure}[htbp!]
\epsfysize=15cm
\centerline{\epsffile{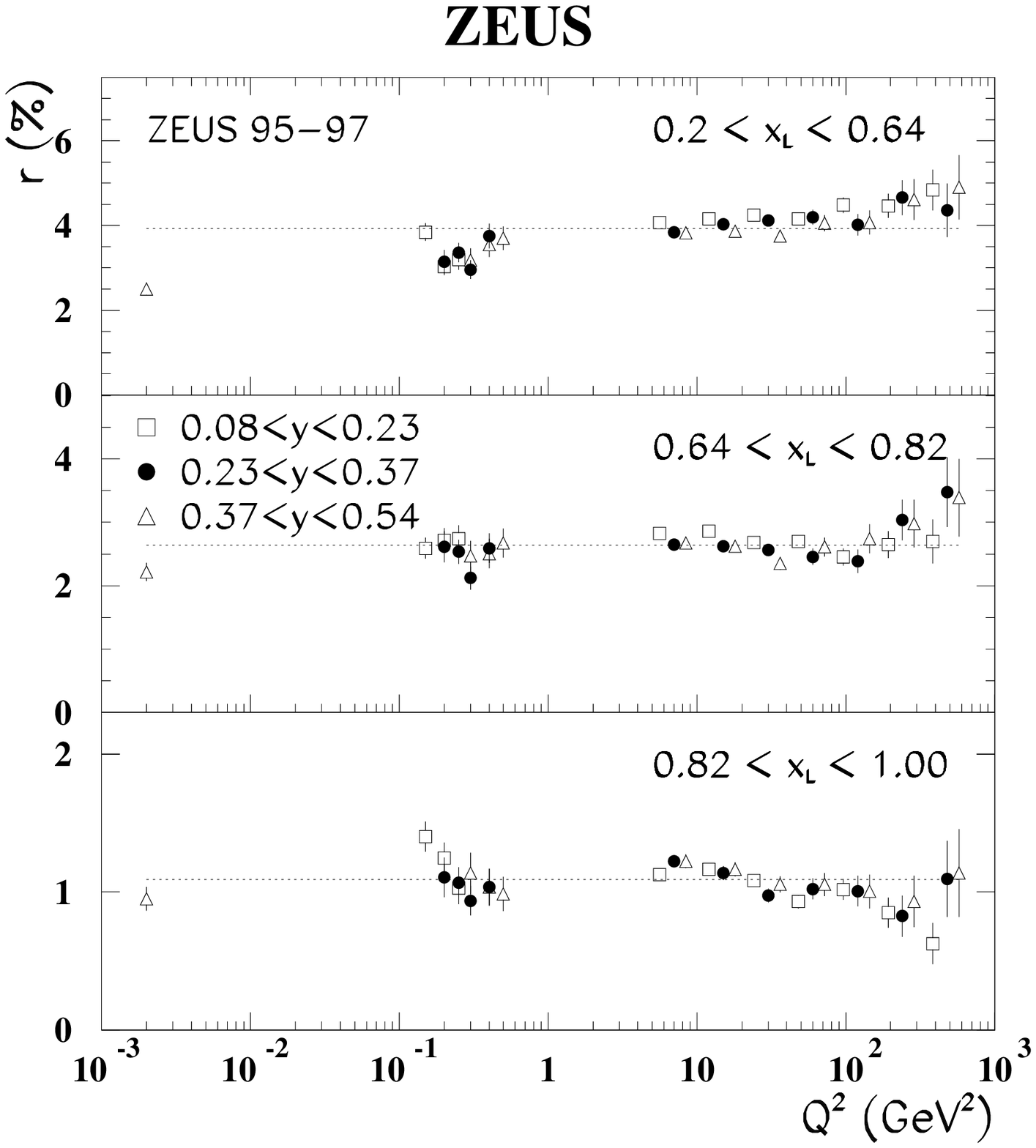}}

\caption{
Neutron production ($\theta_n \le 0.8$ mrad) 
as a fraction of the inclusive cross section
in the PHP ($Q^2<0.02$ GeV$^2$), 
intermediate-$Q^2$ ($0.1<Q^2<0.74$ GeV$^2$)
and DIS ($Q^2>4$ GeV$^2$) kinematic regions
as a function of $Q^2$ in bins of $y$ and $\xl $.
The PHP point is plotted at the 
average value of $\langle Q^2 \rangle=2\cdot 10^{-3}$ GeV$^2$ for the sample. 
The DIS points are offset in $Q^2$ for visibility. 
The horizontal dotted lines are plotted at the mean $r$ of
the plot to guide the eye. Only statistical
uncertainties are plotted for the DIS and intermediate-$Q^2$
regions. For the PHP point, the uncertainty plotted
is the statistical uncertainty added in quadrature with the
PHP relative normalisation uncertainty.
}
\label{fig:ratio-ybins-all}
\end{figure}

To investigate the $W$ dependence, the data may be studied as a function 
of $y$, or $x$, for fixed $Q^2$.
Figure~\ref{fig:ratio-ybins-all} compares
the cross-section ratio, $r$, as a function of $Q^2$ 
at fixed $y$ by choosing
three bins of $y$ 
for a low ($0.20<\xl <0.64$), a medium ($0.64<\xl <0.82$) and a high 
($0.82<\xl <1.0$) $\xl $ range. 
At low $\xl $ and fixed $y$, $r$ rises with $Q^2$,
approximately doubling over the kinematic range
of five orders of magnitude in $Q^2$.
As $\xl$ increases, the slope of $r$ as a function of $Q^2$ decreases
and becomes negative in the high-$\xl$ range.

\begin{figure}[htb]
\centerline{\epsffile{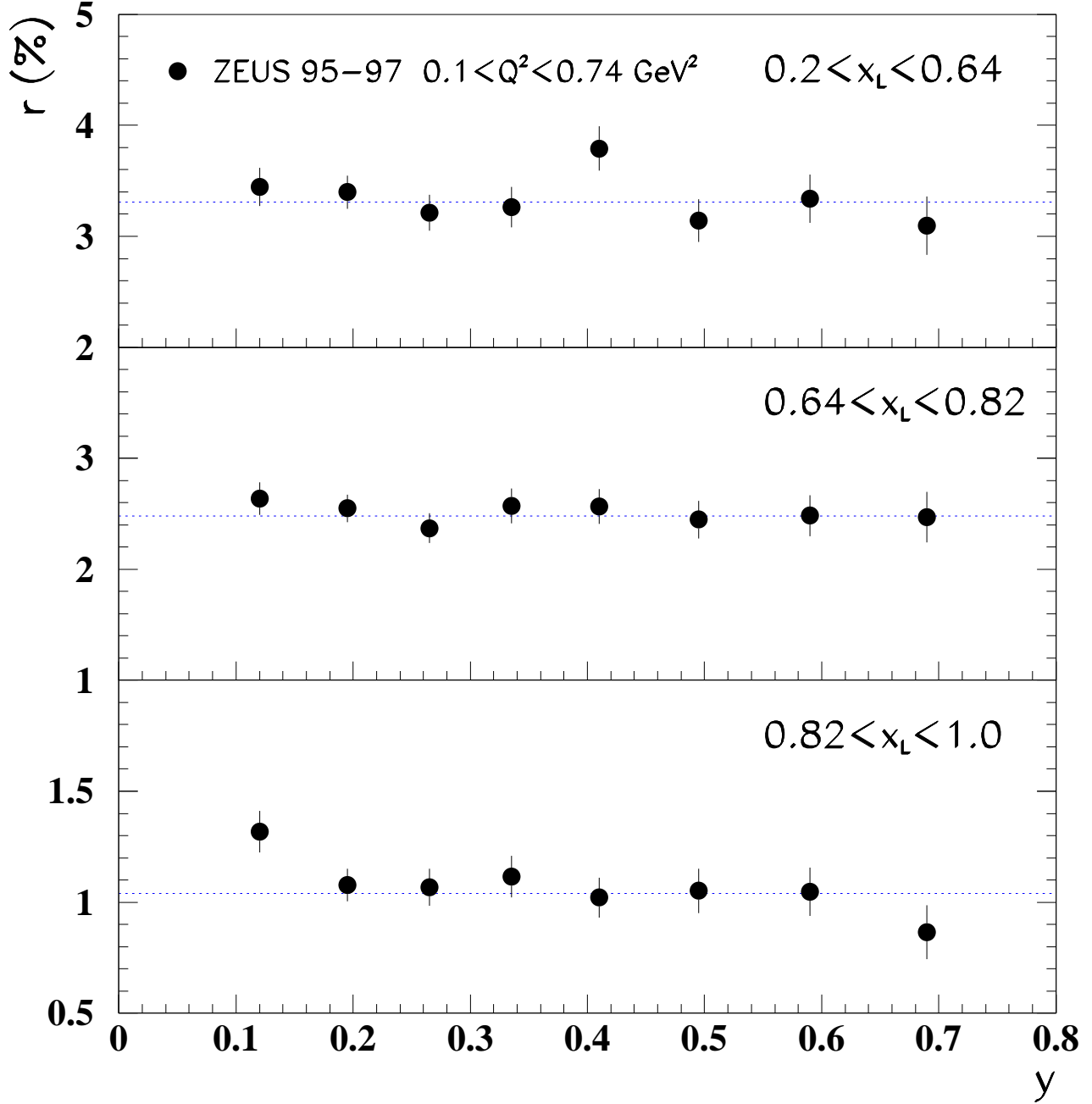}}
\caption{
Neutron production ($\theta_n \le 0.8$ mrad) 
for the intermediate-$Q^2$ region, $0.1<Q^2<0.74$ GeV$^2$,
as a fraction of the inclusive cross section
and as a function of $y$ for the low ($0.2<\xl <0.64$), medium
($0.64<\xl <0.82$), and high ($0.82<\xl <1$) $\xl $ 
ranges.
The dotted lines show the mean values of the ratio for each $\xl $ range.
Not shown are the correlated
systematic uncertainties given in Table~\ref{etabxlerr}.
}
\label{fig:bpc-all-3}

\end{figure}

Figure~\ref{fig:bpc-all-3} shows the values of $r$ for the 
intermediate-$Q^2$ data, separately
for the low, medium and high ranges of $\xl $. 
There is little variation of the ratio, $r$,
as a function of $y$. The dotted horizontal lines in the figures,
which show the average values of the points in each $\xl $ bin, are a 
satisfactory representation of the data.

\begin{figure}[htbp!]
\centerline{\epsffile{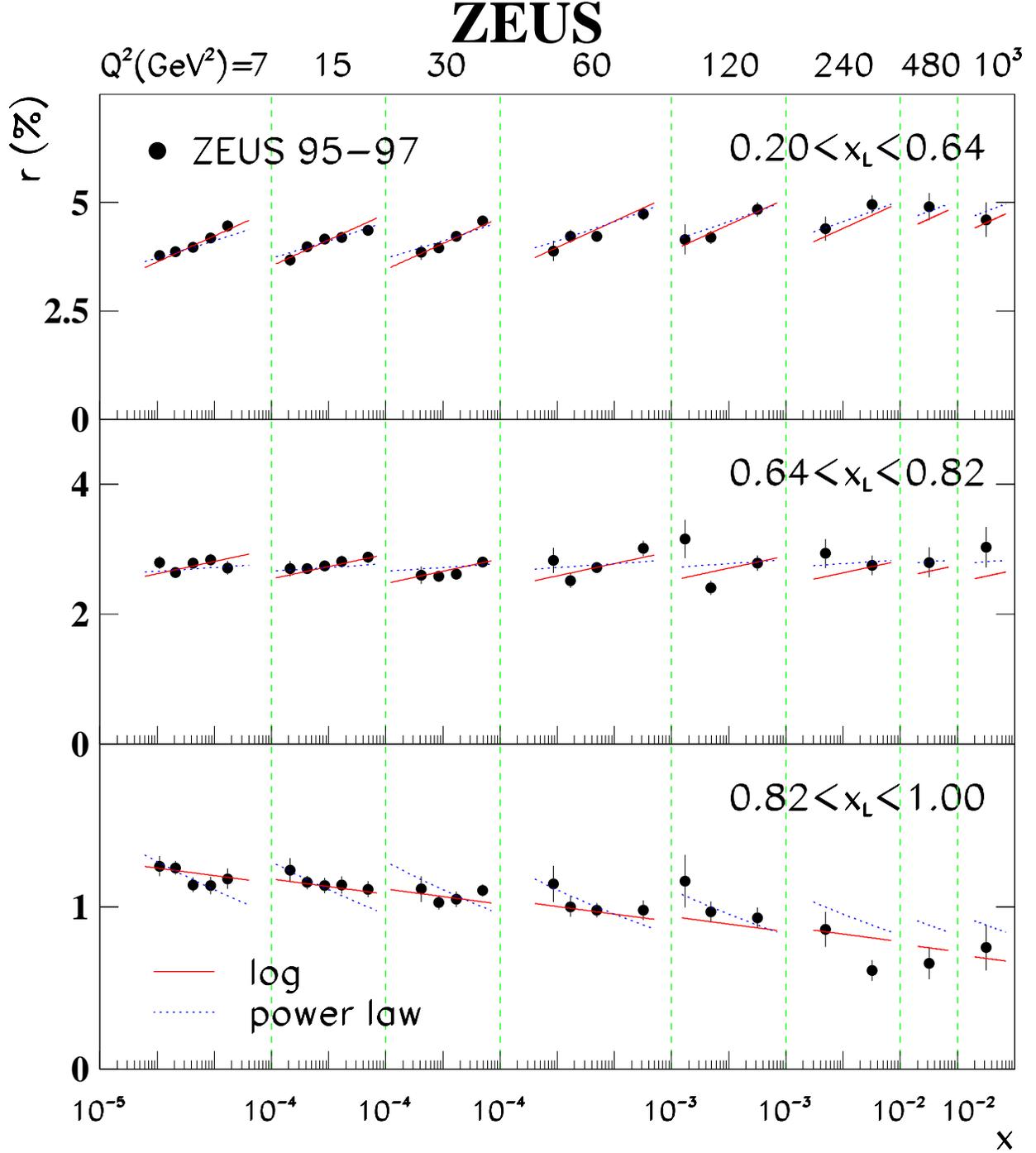}}
\caption{Neutron production ($\theta_n \le 0.8$ mrad) for the DIS region,
$Q^2>4$ GeV$^2$,
as a fraction of the inclusive cross section
and as a function of $x$ for the low ($0.2<\xl <0.64$), medium
($0.64<\xl <0.82$), and high ($0.82<\xl <1$) $\xl $ 
ranges, in the indicated bins of $Q^2$.
The dotted lines show the result of fitting a power law in $x$ to the ratio.
The solid lines show the result of a fit to the ratio
linear in both $\ln x$ and $\ln Q^2$, as discussed in the text.
Not shown are the correlated
systematic uncertainties given in Table~\ref{etabxlerr}.
}
\label{fig:ratio-dis}
\end{figure}

Figure~\ref{fig:ratio-dis} shows $r$ for the 
DIS data as a function of $x$ in bins of $Q^2$ and $\xl $.
The same trends as in Fig.~\ref{fig:ratio-ybins-all} are observed.
At fixed $Q^2$, there is a small, but systematic, dependence of
$r$ on $x$. 
At low $\xl $, 
the ratio rises with $x$; for $0.64 < \xl < 0.82$, the ratio is
approximately independent of $x$;
at high $\xl $, it falls with increasing $x$.
The dotted lines show
the results of fits of the form 
$(x_1/x)^{\lambda}$, where $x_1$ and 
$\lambda$ are fit parameters.
Each $\xl$ range is fitted separately. 

These fits provide a good representation
of the data except at the highest $\xl$ at high
$Q^2$. 
The data can also be fit to the form 
$a+b\ln\left( x/x_0 \right)+c\ln \left( Q^2/Q^2_0 \right)$, 
where $a$, $b$ and $c$ are fit parameters and
$Q^2_0$ and $x_0$ are arbitrary scales.
This gives a good representation of the data in all
$\xl$ bins, as shown by the solid lines.

In summary, 
the observed weak, logarithmic, variation 
as the size of the virtual photon changes
represents a breakdown of limiting
fragmentation and the factorisation as expressed in Eq.~(\ref{eq:fact}).
However,
in the interval $0.64 < \xl <0.82$, $r$ for the DIS data is, to a good
approximation, independent of the leptonic variables  and is only a function 
of $\xl$. For this $\xl$ range, Eq. (\ref{eq:fn}) implies that
\begin{eqnarray}
F_2^{\rm \tiny LN(3)}(x,Q^2,\xl) \propto F_2(x,Q^2)\, .
\label{eq:pxxch}
\end{eqnarray}

\section{Determination of ${F}_2^{\rm\tiny LN(3)}$}
\label{sec:f2ln}

\begin{figure}[htbp!]
\centerline{\epsffile{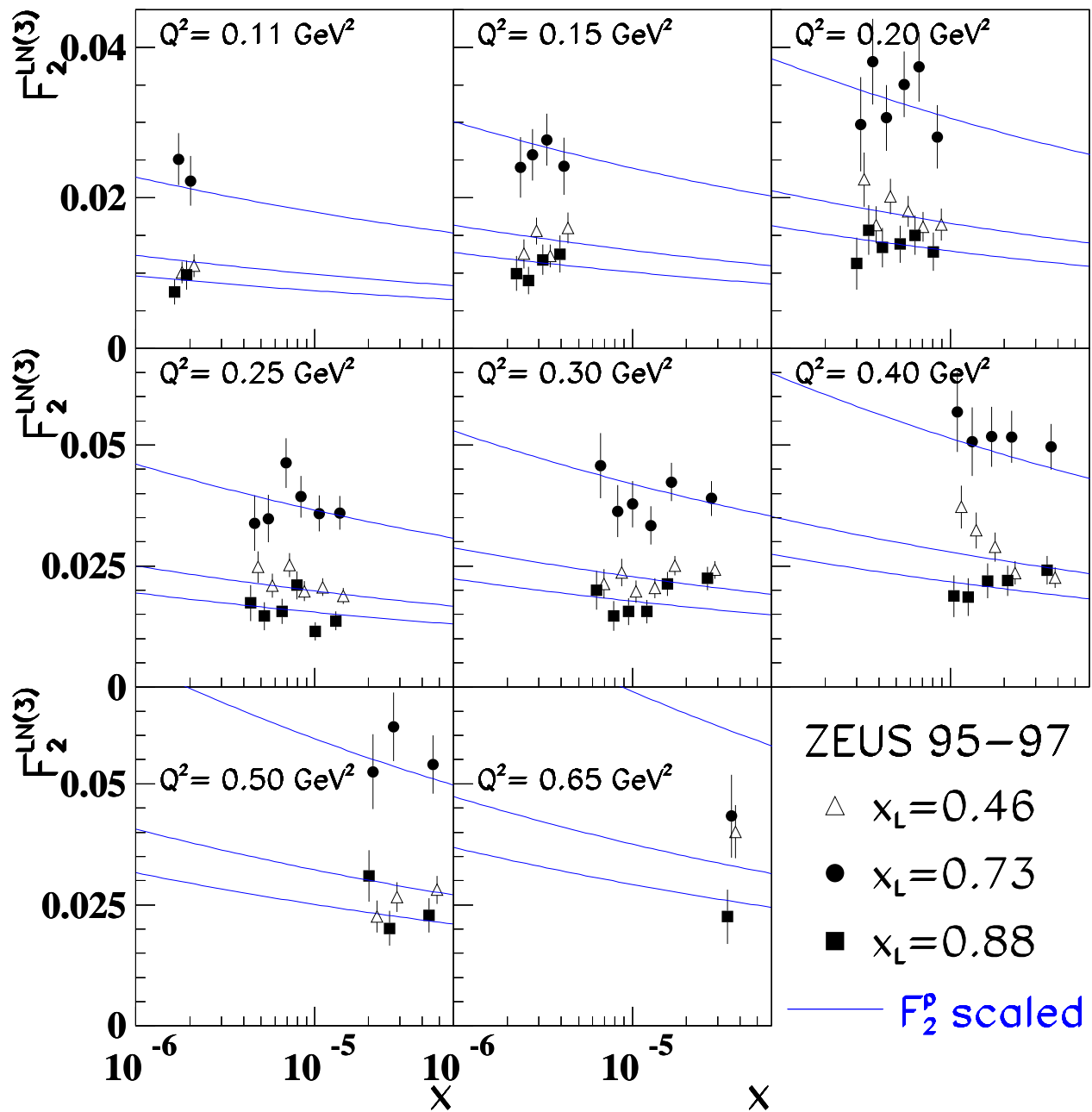}}
\epsfysize=14cm
\caption{${F}^{\mbox{\rm\tiny LN(3)}}_2$ for $\theta_n<0.8$ mrad for $0.20 < \xl <0.64$, $0.64 < \xl < 0.82$ and $0.82 < \xl < 1.0$ in the intermediate-$Q^2$ region, $0.1 < Q^2 <0.74\;{\rm GeV^2}$, as a function of $x$. The 
uncertainties shown are statistical added in quadrature with the uncertainties from the ZEUS Regge fit to $F_2$. The curves show $F_2(x,Q^2)$ for the proton scaled by the average value of $r$ divided by the bin width $\Delta \xl $. Not shown is an additional correlated systematic uncertainty of $\pm$8.8\% from the acceptance, the energy scale and the overall normalisation uncertainty.
}
\label{fig:f2ln-bpc}
\end{figure}

\begin{figure}[htbp!]
\centerline{\epsffile{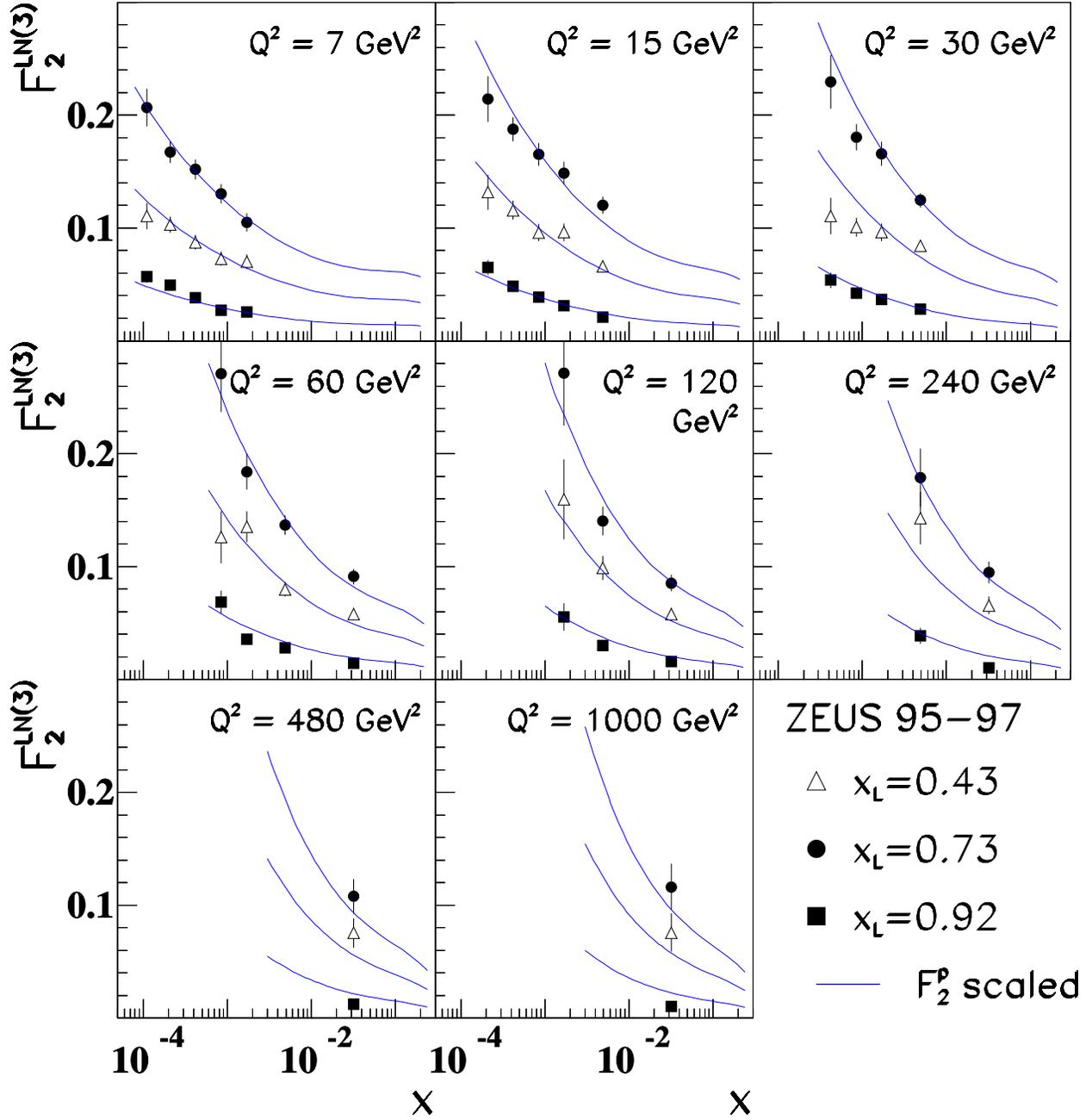}}
\epsfysize=14cm
\caption{
${F}^{\mbox{\rm\tiny LN(3)}}_2$ for $\theta_n<0.8$ mrad
for $0.40 < \xl <0.46$, $0.70 < \xl < 0.76$ and $0.88 < \xl < 1.0$
in the DIS region,
$Q^2 >4\;{\rm GeV^2}$, as a function of $x$. 
The uncertainties shown are statistical added in quadrature with the uncertainties from the ZEUS NLO QCD fit.
The curves show $F_2(x,Q^2)$ for the proton scaled by
the average value of $r$ divided by the
bin width $\Delta \xl $. Not shown is an additional correlated 
systematic uncertainty of $\pm$8.8\% from the acceptance, the 
energy-scale and the overall normalisation uncertainty. 
}
\label{fig:f2ln-2}
\end{figure}

The neutron-tagged structure function,
${F}_2^{\mbox{{\rm\tiny LN(3)}}}$, was determined using 
Eq. (\ref{eq:fn}). The $F_2(x,Q^2)$ values were taken from 
the ZEUS Regge fit\cite{zeusbpt} in the intermediate-$Q^2$ region, 
$Q^2 < 1$ GeV$^2$, and from
the ZEUS NLO QCD fit\cite{zeusphenlowq} for $Q^2 > 1$ GeV$^2$.
Figures~\ref{fig:f2ln-bpc} (intermediate-$Q^2$ data) 
and \ref{fig:f2ln-2} (DIS data) show 
${ F}^{\mbox{\rm\tiny LN(3)}}_2$ for three representative ranges of
$\xl$ as a function of $x$ at fixed values of $Q^2$. The normalisation of the
three sets of fixed-$\xl$ data varies because the range of $\ptsq$
changes with $\xl$ when integrating over a fixed $\theta_n$ range. 
To guide the eye, the line
shows the ZEUS Regge or NLO fit appropriately scaled
by $r$ in each $\xl$ bin
divided by $\Delta \xl$.  The $x$ and $Q^2$ dependence
of ${F}^{\mbox{\rm\tiny LN(3)}}_2$ is similar to that of $F_2$,
as expected since, as shown in Figs.~\ref{fig:ratio-ybins-all}
and~\ref{fig:ratio-dis}, the ratio $r$ has only a weak dependence on $x$ and 
$Q^2$ at fixed $\xl $.

\begin{figure}[htb]
\includegraphics[width=\linewidth]{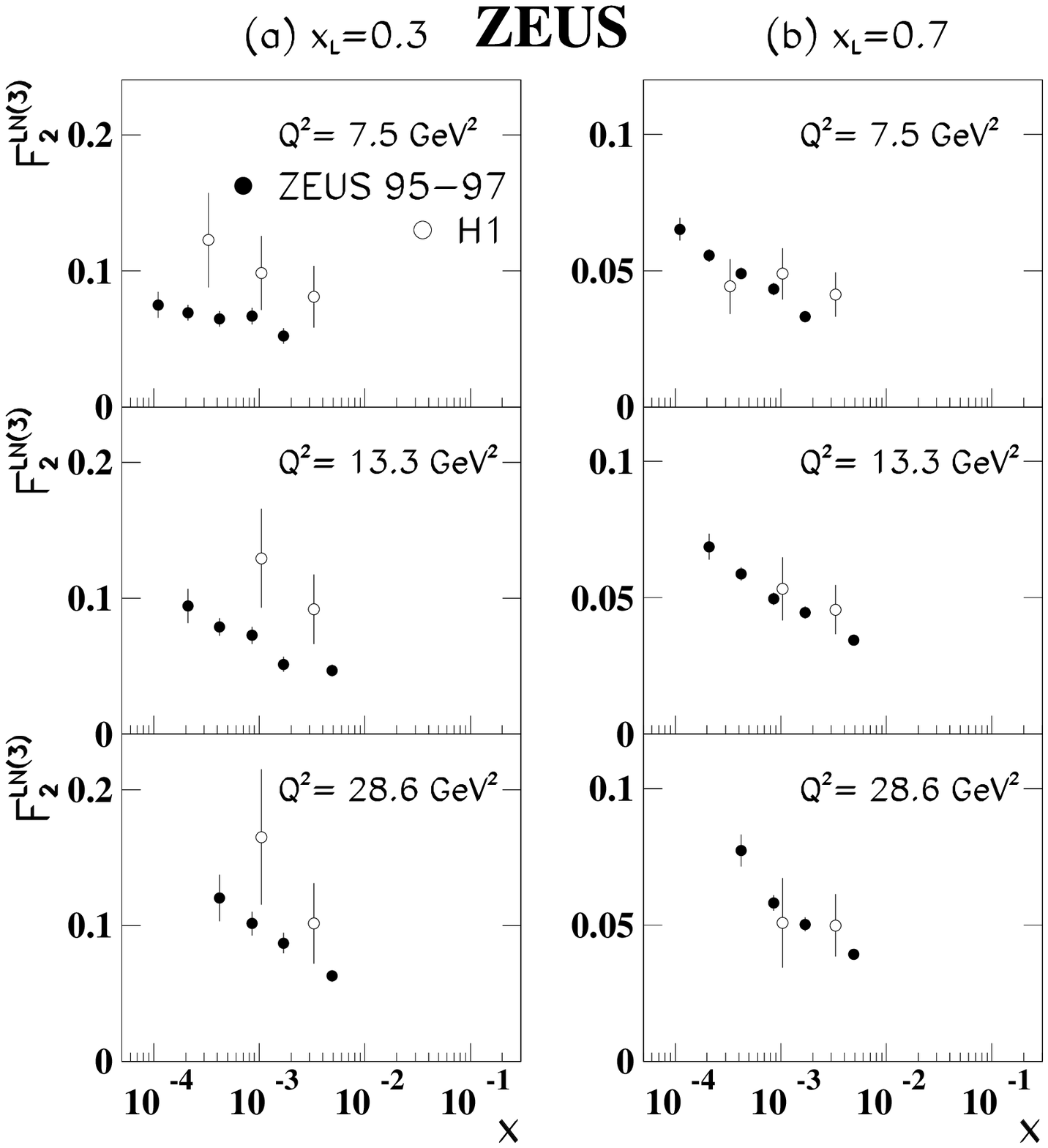}
\caption{
A comparison of the ZEUS and H1 values of
${F}^{\mbox{\rm\tiny LN(3)}}_2$ at 
(a) $\xl =0.3$, where the $\pt $ ranges covered by the two
measurements coincide, and (b) $\xl =0.7$, 
where the ZEUS data have been adjusted to 
the transverse-momentum range covered by H1, $\pt  < 0.2$ GeV, using the
form of Eq.~(\ref{eq:bslope}).
For ZEUS, the statistical uncertainties
are added in quadrature with that
from the ZEUS NLO QCD fit. A 
normalisation uncertainty of 6.7\% (7\%) remains at
$\xl = 0.3$ (0.7) due to the
acceptance, energy scale and overall normalisation uncertainties. 
For H1, the statistical and systematic
uncertainties were added in quadrature. The uncertainties on the H1 points are dominated by the $\xl$-dependent uncertainties and so are strongly correlated. There is an additional
normalisation uncertainty of 5.7\% (not shown). 
}
\label{fig:h1-compare}
\end{figure}

The H1 Collaboration has published values
of ${F}_2^{\mbox{\rm\tiny LN}(3)}$
in the kinematic region $2\le Q^2 \le 50$ GeV$^2$ and 
$6\cdot 10^{-5} \le x \le 6\cdot 10^{-3}$\cite{h1f2lb}. 
This region overlaps that
covered by the ZEUS data for $Q^2>4$ GeV$^2$.
Although the H1 data cover the ZEUS range in $\xl $, the 
distribution of the neutrons is integrated only up
to $\ptmax $ = 0.2 GeV, so
the ZEUS and H1 data can only be directly compared at
$\xl =0.3$, corresponding to $\ptmax $ = 0.197 GeV.
For higher $\xl $, the ZEUS
values must be reduced to account for 
the smaller $\pt $ range measured by H1. 
Figure~\ref{fig:h1-compare}(a) compares
the ZEUS and H1 values for ${F}_2^{\mbox{\rm\tiny LN}(3)}$
at $\xl =0.3$ for the three bins of $Q^2$ where the data overlap. 
Figure~\ref{fig:h1-compare}(b) shows
the same comparison at $\xl =0.7$, where
the ZEUS data have been adjusted to 
the transverse-momentum range $\ptmax $  = 
0.2 GeV using the form of Eq. (\ref{eq:bslope}).
The shapes of the distributions are in good agreement, as is
the normalisation for the higher-$\xl$ selection. At low $\xl$,
although the H1 values lie systematically above the corresponding ZEUS measurements, the H1 uncertainties are highly correlated. In addition, contributions from photon showers, which populate the low $\xl$ region, have not been subtracted from the H1 data.

\section{OPE and the pion structure function}

In a one-pion-exchange model, the structure function
$F_2^{\rm\tiny LN(4)}$ can be written as
\begin{eqnarray}
F_2^{\rm \tiny LN(4)}(x,Q^2,\xl,t)=f_{\pi/p}(\xl,t)F_2^{\pi}(x/(1-\xl),Q^2,t)
(1-\Delta_{\rm abs}(Q^2,\xl,t)),
\label{eq:pxch}
\end{eqnarray}
where $f_{\pi/p}$ is the flux of pions in the proton, $F_2^{\pi}$
is the structure function of the pion at a virtuality $t$, and
$\Delta_{\rm abs}$ is
the absorptive correction before a $t$-integration is performed.  

There is now no simple
factorisation of $F_2^{\rm \tiny LN(4)}$, although if $F_2^{\pi}$ 
can be expressed as a power law in $x_{\pi}$, 
where the Bjorken variable of the pion 
is defined as $x_{\pi}=x/(1-\xl)$, as is the case for $F_2$
of the proton with respect to $x$~\cite{zeusphenlowq}, the
approximate factorisation of Eq. (\ref{eq:fact})
will be restored.  

Comparison of  Eq. (\ref{eq:fact}) with Eq. (\ref{eq:fn}) 
(neglecting $\Delta$) shows that 
if $F$ is identified with $F_2$, then $r/\Delta \xl$
may be identified with $\tilde{f}$.
Given that $r$ for the DIS data with $0.64 < \xl < 0.82$ (where OPE dominates,
see Section~\ref{sec:comp}) is only a function of $\xl$, as
described in Section~\ref{sec:sigmaln}, Eqs. (\ref{eq:pxxch}) and 
(\ref{eq:pxch}) imply that 
\begin{eqnarray}
F_2^{\pi}(x_{\pi},Q^2) \propto F_2 (x,Q^2)\, .
\nonumber
\end{eqnarray}

\subsection{Competing processes to OPE}
\label{sec:comp}

\begin{figure}[htb]
\epsfysize=12cm
\centerline{\epsffile{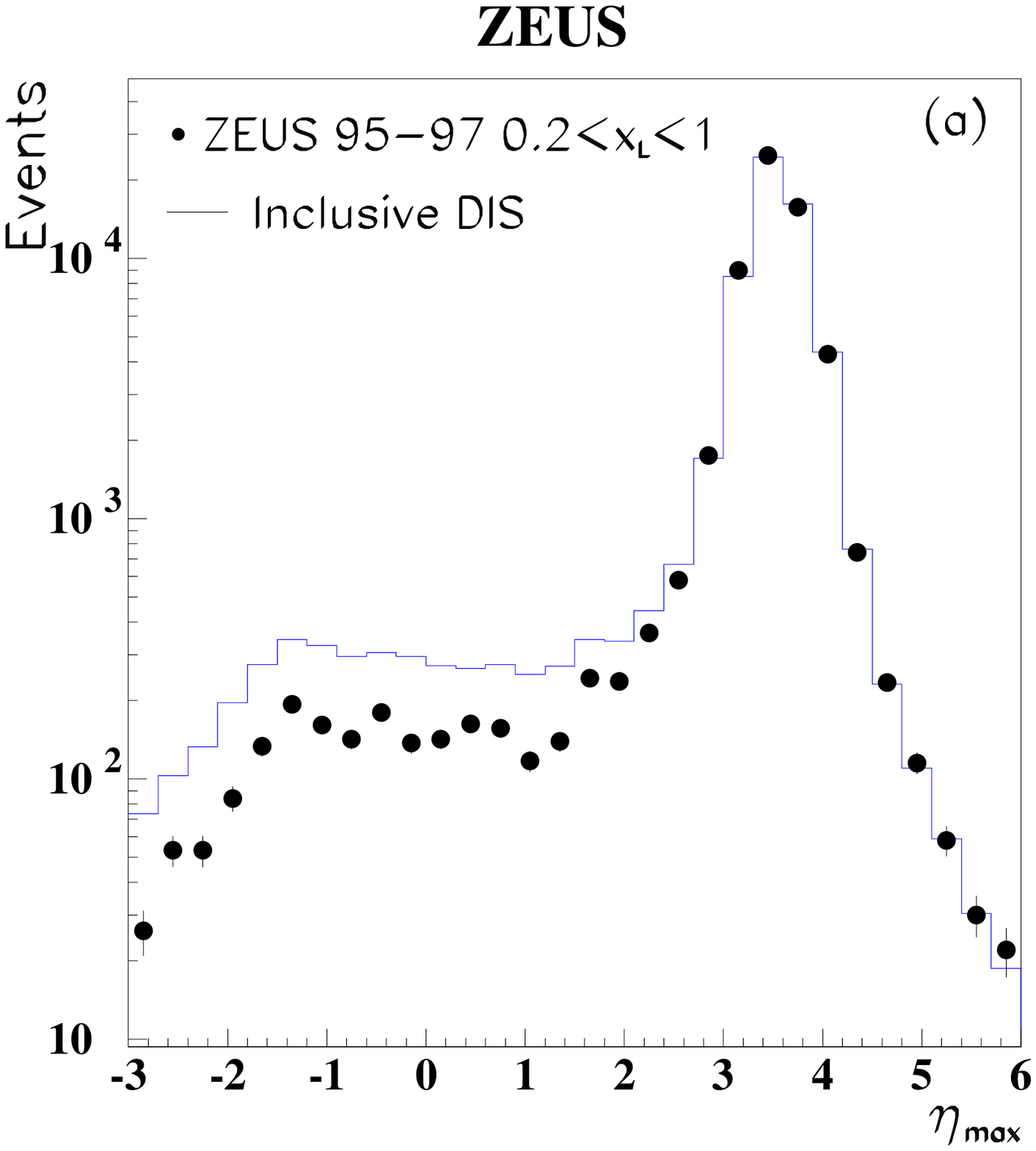}}
\epsfysize=6cm
\centerline{\epsffile{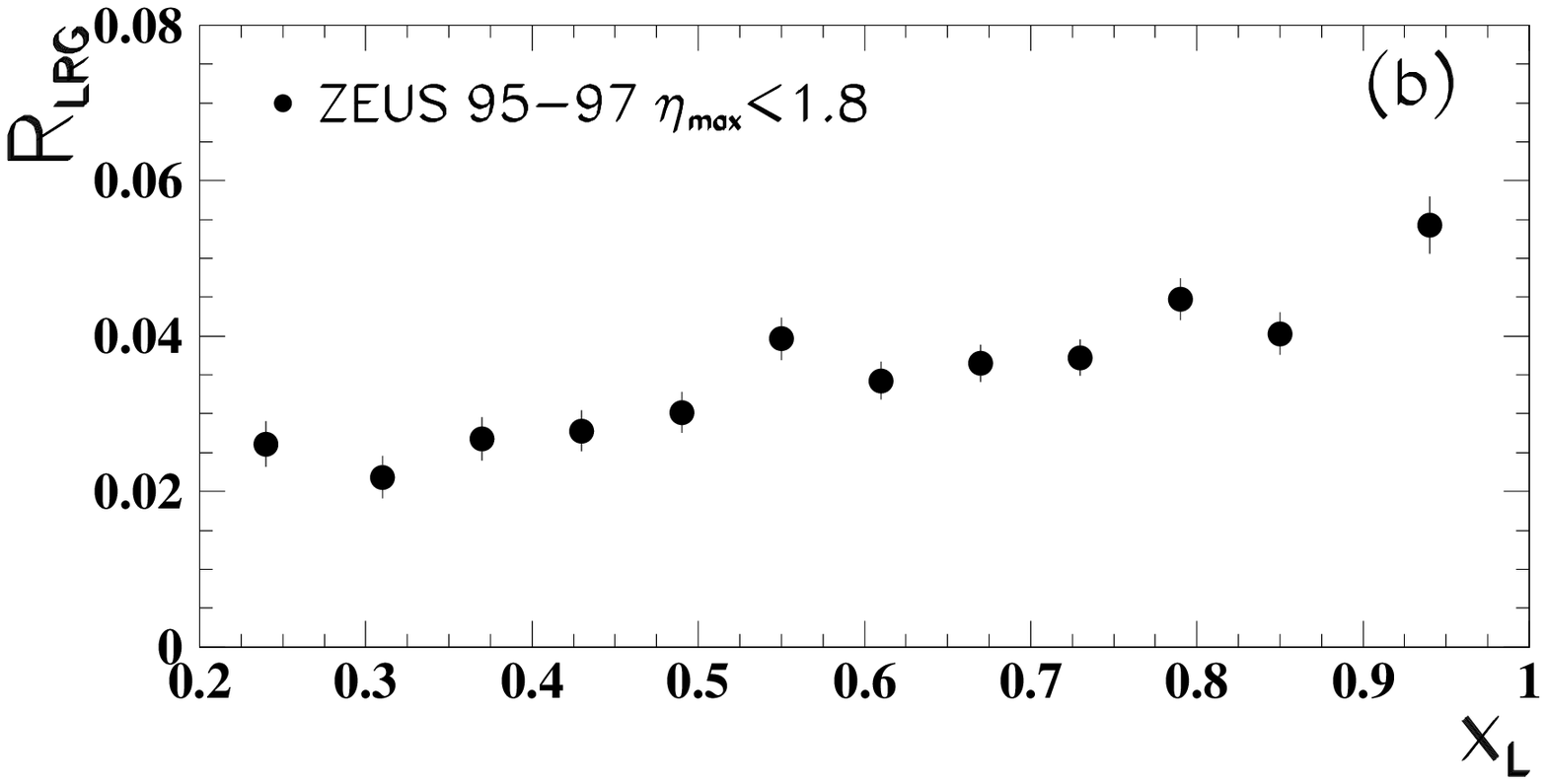}}
\caption{(a) The $\etamax$ distribution in DIS for 
neutron-tagged events (points).
The histogram shows the inclusive $\etamax$ distribution,
normalised to the neutron-tagged data for $\etamax >1.8$;
(b) the fraction, $\rm R_{LRG}$,  of leading neutron events with $Q^2>4$ 
GeV$^2$ passing the LRG criterion $\etamax <1.8$, as a function of $\xl $. 
}
\label{fig:etamax_dis}
\end{figure}

Several processes which compete with pion exchange as the
mechanism for leading neutron production were ignored in Eq. (\ref{eq:pxch}),
namely:
\begin{itemize}
\item diffractive dissociation in which the dissociated system decays to a 
state including a neutron

Diffractively produced events can be selected
by requiring the presence of a large rapidity gap in the
hadronic final state.  For such events, the
mass of the dissociated proton system is restricted to
low values, \mbox{$M_{\mbox{\rm\tiny N}}\lap 4$ GeV}. 

An event is said to have a large rapidity gap (LRG) in the ZEUS 
detector if the pseudorapidity
of the most-forward energy deposit with energy greater 
than 400 MeV ($\etamax$) is less than 1.8\cite{zeusdiff95}.
Figure~\ref{fig:etamax_dis}(a) shows the $\etamax $ distributions
for both the neutron-tagged and inclusive DIS
samples, where the latter has been normalised to the neutron-tagged sample
 for $\etamax > 1.8$.
For both $\etamax <1.8$ and $\etamax >1.8$, the shape of the neutron-tagged
distribution is similar to that of the inclusive distribution; 
however, there are relatively fewer LRG 
events in the neutron-tagged sample. 
The LRG events represent only 4\% of the total
number of DIS events with neutrons in the measured kinematic
region, but are 7\% of the total number of DIS events. 
A reduction in the fraction
of LRG events with a final-state neutron is expected since
only proton diffractive dissociation or diffractive meson exchange 
(the Deck effect\cite{deck}) 
can contribute. 

To investigate a possible $\xl$-dependence of the contribution of
diffractive events, Fig.~\ref{fig:etamax_dis}(b)
shows the ratio, $R_{\rm LRG}$,  of the
neutron-tagged DIS events, selected by the LRG criterion,
to all neutron-tagged DIS events, as a function of
$\xl $. The rise by a factor of three over the $\xl$ range shows that
the LRG neutron-tagged events 
have a harder neutron energy spectrum than that
of the inclusive neutron-tagged sample. 
It is clear that diffractive events
are not a major source of leading neutrons at any value of $\xl$.
For the region $0.64 < \xl < 0.82$, $R_{\rm LRG}$ is $0.039 \pm 0.001~(stat.)$;

\item $\rho$ and $a_2$ exchange

Theoretical studies of neutron production 
in $ep$ collisions~\cite{holtmann,kopeliovich} 
suggest that isovector exchanges other than the pion
contribute less than 10\% to neutron 
production at $\xl = 0.73$ and for the $\pt $ range of the present data.
This is quite different than for leading
proton production, where isoscalar Regge
exchange provides the dominant contribution\cite{sns,h1f2lb};

\item isovector exchange leading to $\Delta$ production

The $p\rightarrow \Delta(1236)$ transition, formed
by $\pi$, $\rho$ and $a_2$ exchange, can also contribute
to neutron production\cite{pumplin,holtmann,kopeliovich,nsss,thomas}.
In this case, the neutron, which comes
from the decay $\Delta^0\rightarrow n\pi^0$ 
or $\Delta^+\rightarrow n\pi^+$,
no longer has an energy determined by 
the energy of the exchanged meson.
The neutron energy spectrum peaks near $\xl \approx 0.5$
and extends only to $\xl\approx 0.7$\cite{sns}. It thus gives a small
contribution in the $0.64 < \xl < 0.82$ bin.
A comparison of the data on $p\rightarrow n$
and $p\rightarrow\Delta^{++}$\cite{erwin_delta,higgins,barish,dao}
in charge-exchange reactions at Fermilab
indicates that only about 6\% of the forward neutrons come from
the $\Delta$ channel. This observation agrees with
theoretical estimates of the $\Delta\pi$ contribution to the 
Fock state of the proton, which is approximately half that of
$n\pi$\cite{holtmann,thomas}.
A calculation~\cite{nsss} shows that the contribution of
$\rho/a_2$ exchange, plus the $\Delta$
contribution, to the hadronic charge-exchange reaction $pp \rightarrow Xn$
could be as high as 30\%. Since no analogous calculation exists for DIS,
this only provides an indication of a possible background to the neutron
production discussed in this paper;

\item models other than one-particle exchange

Monte Carlo studies, using standard DIS generators, show~\cite{fnc2}
that these processes have a rate of neutron production a factor of 
three lower than the data and produce a
neutron energy spectrum with the wrong shape, 
peaking at values of $\xl$ below 0.3.
\end{itemize}

In summary, the expectation for the processes listed above  
at $\langle \xl \rangle$ = 0.73 and $\langle \pt ^2 \rangle$ = 0.08 
GeV$^2$ is that they contribute
of the order of 20\% of the leading neutron 
production. This estimate can be checked using the measured 
neutron-energy spectrum. 
The OPE fit to the differential cross-section
$d\sigma/d\xl $ shown in Fig.~\ref{fig:energy-spectrum-fits}(a)
suggests that, at $\xl =0.73$, the residual background to OPE is
$\lap 10$\%, in reasonable 
accord with the studies discussed above.

\subsection{The pion structure function and the photon-pion total cross
section}
\label{sec:efflux}

Having established that the leading neutron data are dominated by pion
exchange in the range $0.64 < \xl < 0.82$, the OPE model can be used to
determine the structure function of the pion, $F_2^{\pi}$.
The quantity ${F}_2^{\rm\tiny LN(3)}$, defined in Section~\ref{sec:kinematics},
was obtained by integrating $F_2^{\rm\tiny LN(4)}$ over $t$ (or $\theta_n$).  
The $t$ dependence of $F_2^\pi$ 
is absorbed into the flux factor that describes the hadronic
charge-exchange data~\cite{erwin,pickup,engler,
robinson,flauger,hanlon,hartner,eisenberg,blobel,abramowicz};
the structure function of the real pion is then given by
$F_2^{\pi}(x_{\pi},Q^2)\equiv F_2^{\pi}(x_{\pi},Q^2,t=m_\pi^2)$.

Integrating Eq. (\ref{eq:pxch}) over $t$ and rearranging 
leads to
\begin{equation}
F_2^{\pi}(x_{\pi},Q^2)=\Gamma(Q^2,\xl) {F}_2^{\rm \tiny LN(3)}(x,Q^2,\xl) ,
\label{eq:f2pi1}
\end{equation}
where
$\Gamma(Q^2,\xl)$ is the inverse of the pion flux factor integrated 
over the measured $t$ region and
corrected for the $t$-averaged absorptive effect, $\delta_{\rm abs}$:
\[
\Gamma(Q^2,\xl)=\frac{1}{(1-\delta_{\rm abs}(\xl,Q^2))
\int_{t_{\rm min}}^{t_{\rm max}} f_{\pi/p}(\xl,t) dt }.
\]
As discussed in Section~\ref{sec:abs}, the theoretical 
expectation \cite{alesio} is that  
$\delta_{\rm abs}(\xl,Q^2)$ is less than 10\% for $Q^2 > $10 GeV$^2$.  

Thus, Eq. (\ref{eq:f2pi1}) shows that, for the DIS region,
$F_2^{\pi}(x_\pi,Q^2)$  is approximately proportional to 
${F}_2^{\rm \tiny LN(3)}(x,Q^2,\xl)$
for a fixed $\xl$.
In analogy to Eq.~(\ref{eq:f2pi1}), 
the photon-pion total cross section can be written as
\begin{equation}
\sigma^{\gamma \pi}_{\rm tot}(W\sqrt{1-\xl})=
{\Gamma(Q^2=0,\xl)} \cdot
\frac{d{\sigma}^{\gamma p \rightarrow Xn}(W,Q^2=0)}{d\xl}.
\label{eq:gampi}
\end{equation}

\subsection{The pion-flux normalisation}

To determine the normalisation of the pion flux, and hence $F_2^{\pi}$
and $\sigma_{tot}^{\gamma \pi}$, a theoretical
model is necessary for the form factor $F(\xl,t)$ in the effective
flux factor of Eq. (\ref{eq:flux}).
  Unfortunately, there is no consensus as to which of
the competing models~\cite{vogt} is most appropriate.  As examples of 
the extremes
from the available range of the models, two possibilities are considered
here. The evaluation of the normalisation is made at $\xl = 0.73$, where the
ratio $r$ is approximately independent of $x$ and $Q^2$ and OPE is
expected to dominate.

\begin{enumerate}

\item{ One possibility is to use the same flux as is employed in
hadronic reactions.
The data on the hadron charge-exchange reactions $pn\rightarrow Xp$, 
$pp \rightarrow Xn$  and $\pi n\rightarrow Xp$~\cite{erwin,pickup,engler,
robinson,flauger,hanlon,hartner,
eisenberg,blobel,abramowicz} are in good agreement with 
OPE and the Bishari
flux factor~\cite{bishari} given by Eq.~(\ref{eq:eff_flux}).
Using Eq. (\ref{eq:opef}), the cross section may be written as
\[
 \sigma^{pn\rightarrow Xp}=\int \int f_{\rm eff} (\xl,t)\sigma^{\pi p}_{\rm tot} dt \,d\xl ,
\]
where
$\sigma ^{\pi p}_{\rm tot}$ is the $\pi p$ total cross section and the integration
is over the kinematic range of the $pn\rightarrow Xp$ measurement.
The flux factor, $f_{\rm eff}$, can again
be interpreted as an $\it{effective}$ pion flux which already
incorporates the effects of absorption, 
processes other than OPE and off-mass-shell effects.  

Assuming that all of these effects are the same for the photoproduction
reaction $\gamma p \rightarrow Xn$ as for the hadronic processes 
$pn \rightarrow Xp$ and $pp \rightarrow Xn$ then
\[
 {\Gamma(Q^2=0,\langle \xl \rangle=0.73)}=\frac{\Delta \xl}{\int \int_{t_{\rm min}}^{t_{\rm max}}
f_{\rm eff}(\langle \xl \rangle= 0.73,t)dt \, d\xl} = \frac{0.18}{0.0775} =2.32, 
\]
for the range $0.64 < \xl < 0.82$. 

Using Eq.~(\ref{eq:gampi}) with $d\sigma^{\gamma p\rightarrow Xn}/d\xl$ 
evaluated from
Fig.~\ref{fig:energy-spectrum-fits}(a)
at $\xl$ = 0.73, 
$\sigma^{\gamma \pi}_{\rm tot}(W=107 ~{\rm GeV})$
is $50\pm 2$ $\mu$b. 
Assuming a Regge dependence such that $\sigma \propto (W^2)^{\alpha_{\pom}(0)-1}$ with
$\alpha_\pom(0)= 1.08$~\cite{donnachie} yields $\sigma^{\gamma 
\pi}_{\rm tot}(W=207 ~{\rm GeV})$ =$56\pm 2$ $\mu$b.
Given that $\sigma^{\gamma p}_{\rm tot}(W=209~ {\rm GeV}) = 
174\pm 13$ $\mu$b~\cite{sigtot}, 
$\sigma^{\gamma \pi}_{\rm tot}/\sigma^{\gamma p}_{\rm tot}$ is 
then $0.32\pm 0.03$.

Using Fig.~\ref{fig:absorption}
to estimate the difference
in the absorption for DIS and photoproduction as 22\% leads, for $Q^2 > $ 4 GeV$^2$,
 to
\[
F_2^{\pi}(x_{\pi},Q^2)=1.81\,{F}_2^{\rm\tiny {LN(3)}}(x,Q^2,\langle \xl \rangle=0.73),
\]
or, combining Eqs. (\ref{eq:fn}) and (\ref{eq:f2pi1}), 
\begin{equation}
F_2^{\pi}(x_{\pi},Q^2)=10.05\cdot r(x,Q^2,\langle \xl \rangle=0.73) 
\cdot {F}_2(x,Q^2).
\label{eq:f2pi2}
\end{equation}
}
\item {As a second possibility, the additive quark model~\cite{aqm} can be 
used, which predicts that, at $\xl$ = 0.73,
\begin{equation}
 \frac{\sigma^{\gamma \pi}_{\rm tot}(W\sqrt{1-\xl})}
{\sigma^{\gamma p}_{\rm tot}(W)}=\frac{2}{3}(1-\xl)^{\alpha_\pom(0)-1}=0.60,
\label{eq:aqm}
\end{equation}
provided that 
$\sigma^{\gamma \pi}_{\rm tot}$ and $\sigma^{\gamma p}_{\rm tot}$
have the same energy dependence at high energies, governed
by the Pomeron intercept, $\alpha_\pom(0)= 1.08$.
Combining Eq. (\ref{eq:dsigln}) with Eqs. (\ref{eq:gampi}) and (\ref{eq:aqm}) 
and using Table~\ref{etabxlall} gives
\[
  \Gamma(Q^2=0,\langle \xl \rangle=0.73)=\frac{2}{3} (1-0.73)^{\alpha_\pom(0)-1}
\frac{\Delta \xl}{r(Q^2=0,\langle\xl \rangle=0.73)} = \frac{0.60}{0.1251}=4.8.
\]
Using $\Rabs $ to adjust for the difference in absorptive
effects between
photoproduction and DIS, thereby reducing $\Gamma$, yields, for $Q^2>$ 4 GeV$^2$:
\[
F_2^{\pi}(x_{\pi},Q^2)=3.74{F}_2^{\rm \tiny LN(3)}(x,Q^2,\langle \xl \rangle
=0.73)
\]
\begin{equation}
=20.78\cdot r(x,Q^2,\langle \xl \rangle=0.73) \cdot {F}_2(x,Q^2).
\label{eq:f2pi3}
\end{equation}
}
\end{enumerate}
In summary, methods 1 and 2 differ in normalisation by about a factor of two.
This is approximately
the same factor that was observed in Fig.~\ref{fig:energy-spectrum-fits}(b)
when comparing the neutron rates presented here
and the $pp$ data from the ISR.

Other methods give factors between those described above.
For example, models of the pion 
flux factor based on $\bar{d}/\bar{u}$ data 
give form factors, $F(\xl ,t)< 1$~\cite{pollock,vogt} 
and, therefore, a value of $\Gamma$ that is higher than that given by method 
1. Normalisation of $F_2^\pi$ using existing measurements at high $x_\pi$
\cite{na24,wa70,na3,na10,e573,e615,e609}, as parameterised, for example, by
Gl$\ddot{\rm u}$ck, Reya and Vogt~\cite{grv_pi},
gives a $\Gamma$ value that is lower than that from method 2.


\subsection{Results and comparison with models}

\begin{figure}[htbp!]
\epsfysize=16cm
\centerline{\epsffile{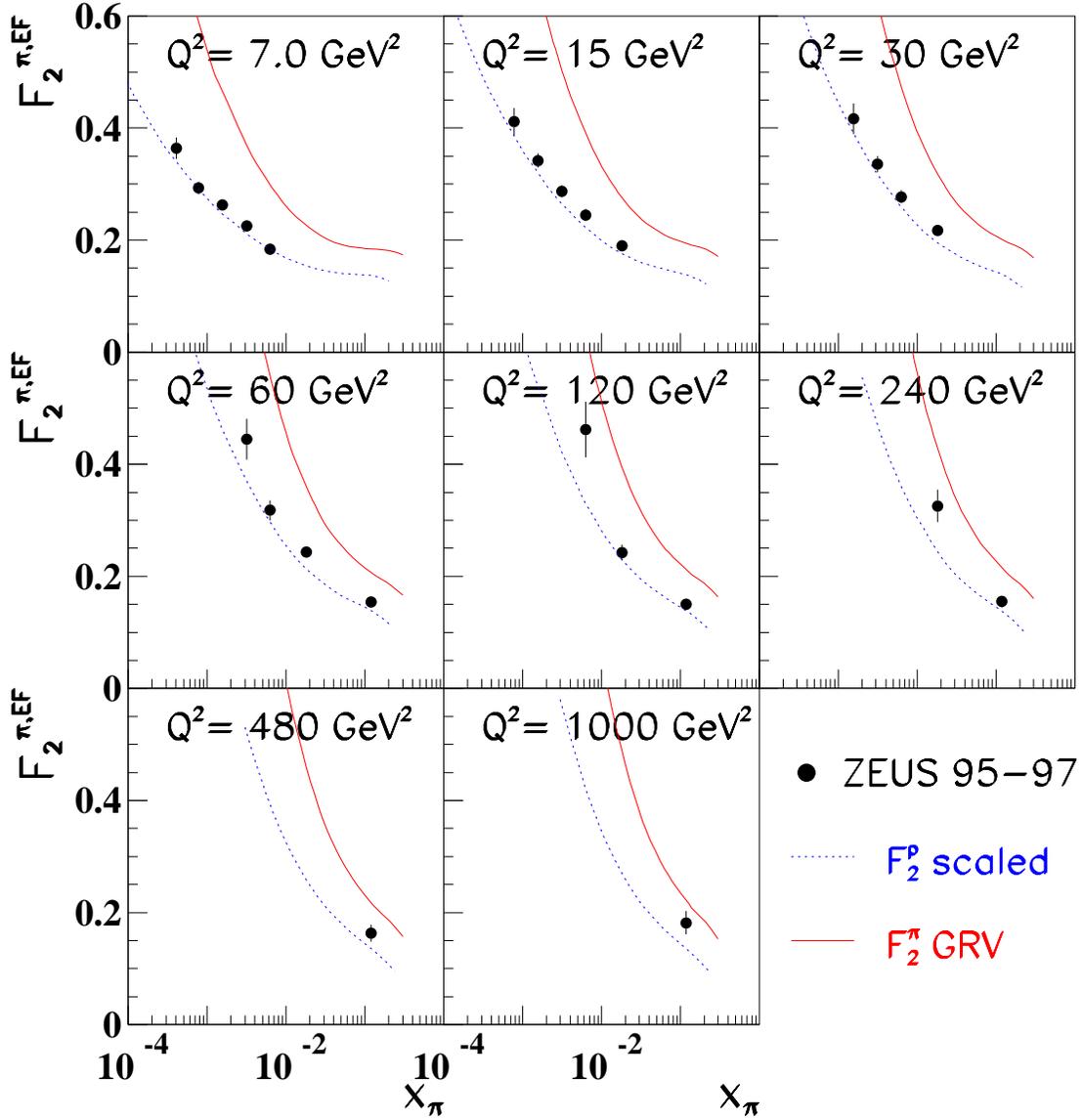}}

\caption{$F^{\pi}_2$ as a function of $x_\pi$ for the pion in
bins of $Q^2$ determined for $0.64 < \xl  < 0.82 $, using Eq. (\ref{eq:f2pi2}).
The pion flux used to 
determine $F^{\pi}_2$ is the effective OPE flux (EF) used in hadron-hadron
charge-exchange reactions, Eq.~(\ref{eq:eff_flux}).
The uncertainty shown on $F^{\pi,\rm{EF}}_2$ arises from
the statistical uncertainty due to the leading neutron
added in quadrature with the uncertainty on $F_2$. Not shown are the correlated
systematic uncertainties given in Table~\ref{etabxlerr}. The dotted lines are
$F_2(x,Q^2)$ for the proton, scaled by 0.361. The solid curves are $F^{\pi}_2$ 
from the GRV parameterisation~\cite{grv_pi}.
}
\label{fig:f2pi-1}
\end{figure}

Figure \ref{fig:f2pi-1} shows $F_2^{\pi, EF}(x_\pi,Q^2)$, which is 
$F_2^{\pi}(x_{\pi},Q^2)$ as a function of $x_\pi$ evaluated from Eq. 
(\ref{eq:f2pi2}).
This
evaluation has been made for the range $0.64 < \xl < 0.82$ where OPE
is expected to dominate and where the fraction of events with a leading
neutron is observed to be approximately independent of $x$ and $Q^2$.
Evaluations for nearby ranges of $\xl $ give consistent results. 
Although the normalisation of $F_2^\pi$ is uncertain to a factor of two, 
the $x$ and $Q^2$
dependence is well measured.  The dotted lines show the shape of 
$F_2$ of the proton, scaled by 0.361. It is striking that $F_2^\pi$ has
approximately the same $x$ and $Q^2$ dependence as $F_2$ of the proton.

\begin{figure}[htbp!]
\centerline{\epsffile{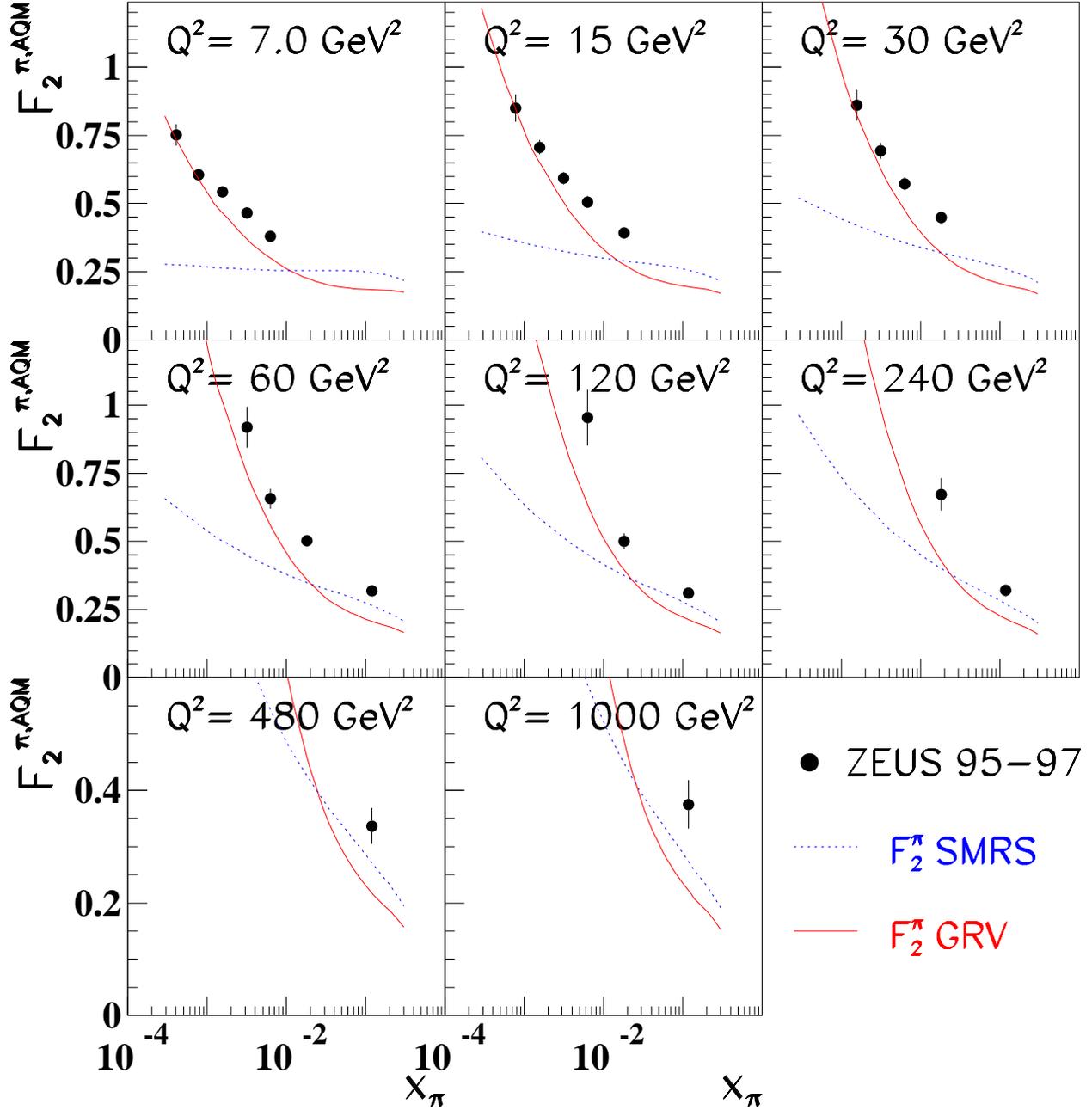}}
\epsfysize=14cm
\caption{$F^{\pi}_2$ as a function of $x_\pi$ for the pion in bins of $Q^2$
determined for $0.64 < \xl  < 0.82 $. The pion flux used to 
determine $F^{\pi}_2$ is the flux obtained using the additive quark model (AQM)
of Eq.~(\ref{eq:f2pi3}).
The uncertainty shown on $F^{\pi,\rm{AQM}}_2$ arises from
the statistical uncertainty due to the leading neutron
added in quadrature with the uncertainty on $F_2$. Not shown are the correlated
systematic uncertainties given in Table~\ref{etabxlerr}. 
The solid curves are $F^{\pi}_2$ from the GRV parameterisation~\cite{grv_pi}
while the dotted curves are from the Sutton {\it et al.} parameterisation~\cite{sutton}.
}
\label{fig:f2pi-2}
\end{figure}

Figure \ref{fig:f2pi-2} shows $F_2^{\pi, AQM}$ using the additive quark model
to normalise the cross section.
The comparison of these results with the solid curves, which represent the 
absolute predictions of Gl$\ddot{\rm u}$ck, Reya and Vogt (GRV)~\cite{grv_pi},
shows that the normalisation is such that the curves are closer to the data
than when using the effective flux shown in Fig.~\ref{fig:f2pi-1}, although the data are still higher than the predictions, particularly at the lower
values of $Q^2$. 

There have been several recent attempts to understand the
quark structure of the pion on the basis of simple
theoretical ideas:
\begin{enumerate}
\item In the model of Gl$\ddot{\rm u}$ck, Reya and Vogt\cite{grv_pi}, the pion
sea is generated dynamically by the QCD evolution equations
from the valence-quark and valence-gluon distributions of
the pion at a low scale of 0.3 GeV$^2$. The valence
gluon is, by assumption, proportional to the valence quark.
The valence quarks are taken from fits to experimental data 
\cite{na24,wa70,na3,na10,e573,e615,e609} at high $x_\pi$. 
This model qualitatively 
describes the shape of the measurement of $F^{\pi}_2$ as seen in
Figs~\ref{fig:f2pi-1} and~\ref{fig:f2pi-2};
\item The prediction for $F_2^{\pi}$ from Sutton 
et al. (SMRS)~\cite{sutton} is also shown in Figure \ref{fig:f2pi-2}. 
This analysis is an NLO QCD fit
to Drell-Yan and prompt-photon data from fixed-target experiments.

Since the data corresponds to $x_\pi > 0.2$, the predictions
for the sea region are uncertain. However, the SMRS fits, when used in
an OPE Monte Carlo simulation, gave a good description of the ZEUS data on
leading neutrons plus dijets in photoproduction~\cite{zeuslndijet}.
The latter correspond to the region
$x_{\pi}> 10^{-2}$, whereas the present data extend to much lower $x_{\pi}$ and
show a very different shape than the $F_2^{\pi}$ from the 
SMRS parameterisation;

\item In the constituent model of 
Altarelli et al\cite{altarelli1},
the structure function of any hadron is determined by that of its
constituent quarks, and, in particular, the pion structure function
can be predicted from the known nucleon structure function\cite{altarelli2}. 
The structure function of a hadron $h$, $F^h_2$,  at low $x$, where
the contribution of the constituent valence quarks is negligible, 
is given by
\[
   F^h_2(x,Q^2)=\int^1_x f_h(z)F^q_2(x/z,Q^2) dz,
\]
where $f_h(z)$ is the density of quarks
carrying momentum fraction $z$ of the hadron $h$, 
and $F^q_2$ is the structure function of the valence quarks.
It is assumed that, at low $x$, the structure function
$F^q_2$ is independent of the flavour.
If, as $x\rightarrow 0$, $F^q_2(x)\sim x^{-\lambda(Q^2)}$,
where $\lambda(Q^2)$ is independent of $x$\cite{zeusphenlowq},
all of the $F^h_2$ structure functions are 
proportional to the same function at low $x$.
This prediction is in agreement with the present data since 
$F^{\pi}_2\propto F_2$.
This model gives a normalisation for
$F^{\pi}_2$ close to $F_2^{\pi,AQM}$;
\item Nikolaev, Speth and Zoller have explicitly studied
the small-$x$ behavior of $F^{\pi}_2$ using a colour-dipole
BFKL-Regge expansion\cite{nszol}. They find that a good
approximation is:
\[
    F^{\pi}_2(x,Q^2)\simeq \frac{2}{3}F_2^p\left( \frac{2}{3}x,Q^2 \right).
\]
Since $F_2$ has a power-law behaviour in $x$ for the ZEUS
kinematic region, the predicted shape is in reasonable agreement with
the data.
\end{enumerate}
Thus, the data presented in this paper 
provide additional constraints on the shape of the pion
structure function for values of $x_{\pi} < 10^{-2}$.

\section{Conclusions}

Leading neutron production for $\xl >0.2$ and $\theta_n <0.8$
mrad has been studied in neutral current $ep$ collisions at
HERA from photoproduction to $Q^2 \sim 10^4$ GeV$^2$. 
The cross section for the 
production of leading
neutrons has been determined as a ratio $r$ relative to the inclusive 
neutral current cross section, thereby reducing 
the systematic uncertainty considerably. 

For the above angular range, which defines a transverse-momentum acceptance
of $\pt < 0.656~\xl$ GeV, the neutron energy spectrum 
exhibits a broad peak at $\xl \approx 0.75$.
It is approximately one third of its maximum at $\xl =0.2$, and
falls to zero approaching the kinematic limit at $\xl =1$.
The rate of neutron production in the
measured phase space in photoproduction is about half the rate
observed in hadroproduction and thus the simplest form of 
vertex factorisation is broken.
A comparison of the neutron yield in the three kinematic regions 
of photoproduction, intermediate $Q^2$ and DIS shows an increase of
about 20\% between photoproduction and DIS, 
saturating for $Q^2>$ 4 GeV$^2$. This can be attributed to the
decrease in absorptive effects as the transverse size of the incident
photon decreases. 

Generally, there is no strong dependence of the ratio $r$ on $x$ and $Q^2$;
however, at low $\xl $ and fixed $y$, the ratio $r$ rises with $Q^2$,
approximately doubling over the kinematic range
of five orders of magnitude in $Q^2$.
For  the high-$\xl $ range, there is a tendency for $r$ to 
decrease at fixed $y$. Thus, limiting fragmentation and factorisation
do not hold. 

The structure-function ${F}_2^{\rm \tiny LN(3)}$ has been measured
over three orders of magnitude
in $x$ and $Q^2$ and for the range $0.2 < \xl < 0.8$ with $\theta_n < 0.8$ 
mrad.
The structure function of the pion, $F^{\pi}_2$,
has been extracted, up to uncertainties in the overall normalisation, in the framework of a one-pion-exchange
model in the range $0.64 < \xl <0.82$, where one-pion-exchange dominates. 
It has approximately the same $x$ and $Q^2$ dependence as
$F_2$ of the proton. The data provide new constraints on the shape of the pion structure function for $x_{\pi} < 10^{-2}$.

\section*{Acknowledgements}

We are especially grateful to the
DESY Directorate whose encouragement and financial support made
possible the construction and installation of the FNC.
We are also happy to acknowledge the efforts and forbearance of
the DESY accelerator group, whose machine components are crowded by the FNC,
and the support of the DESY computing staff.

We thank G.\ Altarelli, L.\ Frankfurt, N.\ Nikolaev, 
M.\ Strikman and A.\ Szczurek for their help with
many of the theoretical issues involved in this study.

This study was only made possible by the physics insight and hard work of G. Levman, to whom we owe a great debt of gratitude. 

\section*{Appendix I: Tables of relative cross sections}

This appendix presents tables of the relative cross sections
for leading neutron production in neutral current interactions,
$ep\rightarrow e'Xn$. The cross sections are measured relative 
to the total inclusive neutral current cross section,
$ep\rightarrow e'X$.

The $\xl $ bins, the acceptance and the acceptance and energy-scale 
uncertainties for
each bin are given in Table~\ref{etabxlerr}. Only for
$0.88<\xl <1$ is the centroid of the bin, 0.92,
significantly different from the bin centre, 0.94. 

The values of $b$, the exponential slope of the $\ptsq$
distribution, are given in Table~\ref{btab}.

Table~\ref{etabxlall} gives the relative differential cross section 
in $\xl $, $(1/\sigma)(d\sigma^{\rm LN} /d\xl )$, for
leading neutron production integrated over $x$ and $Q^2$ for
the three kinematic regimes: a) photoproduction ($Q^2<0.02$ GeV$^2$),
(b) the intermediate-$Q^2$ region ($0.1<Q^2<0.74$ GeV$^2$),
and (c) DIS ($Q^2>4$ GeV$^2$).

A change in the energy scale represents a stretch or
compression (dilation) of the energy distribution. 
Although the dilation is only $\pm 2$\%, 
it becomes the dominant normalisation
systematic uncertainty
at high $\xl $, where the energy spectrum is
falling quickly (Table~\ref{etabxlall}).

The next twelve tables give the fraction of events with a leading neutron in DIS ($Q^2>4$ GeV$^2$)
in bins of $Q^2$ and $x$, while the final three tables give the fraction of events with a leading neutron in the
intermediate-$Q^2$ region ($0.1 < Q^2 < 0.74$ GeV$^2$) in bins of $Q^2$ and $y$.

\newpage

\begin{table} 
 \begin{center} 
 \begin{tabular}{|c|c|c|c|c|} \hline 
 $\xl$ & $<\xl>$ & Acceptance & Acceptance & Energy-scale \\ 
 range & & & uncertainty & uncertainty contribution \\ 
 & &(\%) & (\%) & (\%) \\ 
\hline
   0.20 -  0.28 &  0.24 & 21.5 & +2/-2 & -5/+4 \\ 
\hline
   0.28 -  0.34 &  0.31 & 21.2 & +2/-2 & -5/+4 \\ 
\hline
   0.34 -  0.40 &  0.37 & 21.7 & +2/-3 & -5/+4 \\ 
\hline
   0.40 -  0.46 &  0.43 & 22.2 & +3/-3 & -5/+4 \\ 
\hline
   0.46 -  0.52 &  0.49 & 22.5 & +3/-3 & -4/+5 \\ 
\hline
   0.52 -  0.58 &  0.55 & 21.9 & +4/-3 & -4/+5 \\ 
\hline
   0.58 -  0.64 &  0.61 & 23.0 & +4/-4 & -4/+4 \\ 
\hline
   0.64 -  0.70 &  0.67 & 22.9 & +5/-5 & -4/+4 \\ 
\hline
   0.70 -  0.76 &  0.73 & 23.3 & +5/-5 & -2/+2 \\ 
\hline
   0.76 -  0.82 &  0.79 & 23.7 & +5/-6 & +2/-1 \\ 
\hline
   0.82 -  0.88 &  0.85 & 25.4 & +6/-6 & +5/-8 \\ 
\hline
   0.88 -  1.00 &  0.92 & 32.1 & +6/-6 & +25/-26 \\ 
\hline 
 \end{tabular} 
 \end{center} 
 \caption{\emcap The $\xl$ bins, their acceptance, and the acceptance 
uncertainty. The right hand column shows the contribution from the
energy-scale uncertainty. Note that the energy-scale 
uncertainty is completely correlated between bins. 
\label{etabxlerr} } 
 \end{table} 

\begin{table} 
 \begin{center} 
 \begin{tabular}{|c|c|c|} \hline 
 $\langle \xl \rangle$ & $b$ & $ \pm (stat.)$ \\ 
 & GeV$^{-2}$ & GeV$^{-2}$ \\ 
\hline
 0.49 &  4.10 &  1.10 \\ 
\hline
 0.54 &  4.47 &  1.00 \\ 
\hline
 0.58 &  3.80 &  1.10 \\ 
\hline
 0.62 &  6.33 &  1.00 \\ 
\hline
 0.65 &  8.12 &  0.90 \\ 
\hline
 0.69 &  8.56 &  0.90 \\ 
\hline
 0.73 &  6.06 &  0.80 \\ 
\hline
 0.76 &  9.46 &  0.90 \\ 
\hline
 0.80 &  8.78 &  0.80 \\ 
\hline
 0.84 & 12.81 &  1.10 \\ 
\hline
 0.88 &  9.65 &  0.90 \\ 
\hline
 0.95 &  8.63 &  1.10 \\ 
\hline 
 \end{tabular} 
 \end{center} 
 \caption{\emcap The values of the $b$ slopes and their statistical errors 
in bins of $\xl$ for the 1995 DIS data.
\label{btab} } 
 \end{table}

\begin{table} 
 \begin{center} 
 \begin{tabular}{|c|c|c|c|c|} \hline 
 $\xl$ & $<\xl>$ & $Q^2<0.02$ GeV$^2$ & $0.1<Q^2<0.74$ GeV$^2$ & $Q^2>4$ GeV$^2$ \\ 
 range & & $meas. \pm stat.$ & $meas. \pm stat.$ & $meas. \pm stat.$ \\ 
 & & & (\%) & (\%) \\ 
\hline
  0.20 -  0.28 &  0.24 &  4.12 $\pm$  0.60 &  4.70 $\pm$  0.36 &  5.57 $\pm$  0.13 \\ 
\hline
  0.28 -  0.34 &  0.31 &  4.40 $\pm$  0.71 &  6.37 $\pm$  0.46 &  7.19 $\pm$  0.17 \\ 
\hline
  0.34 -  0.40 &  0.37 &  4.64 $\pm$  0.72 &  5.83 $\pm$  0.42 &  8.51 $\pm$  0.18 \\ 
\hline
  0.40 -  0.46 &  0.43 &  6.31 $\pm$  0.84 &  7.66 $\pm$  0.50 &  9.44 $\pm$  0.19 \\ 
\hline
  0.46 -  0.52 &  0.49 &  6.90 $\pm$  0.87 &  7.72 $\pm$  0.49 & 10.38 $\pm$  0.20 \\ 
\hline
  0.52 -  0.58 &  0.55 &  7.94 $\pm$  0.34 & 10.64 $\pm$  0.57 & 12.38 $\pm$  0.22 \\ 
\hline
  0.58 -  0.64 &  0.61 &  9.40 $\pm$  0.34 & 11.03 $\pm$  0.59 & 13.43 $\pm$  0.22 \\ 
\hline
  0.64 -  0.70 &  0.67 & 11.15 $\pm$  0.38 & 12.69 $\pm$  0.62 & 15.28 $\pm$  0.24 \\ 
\hline
  0.70 -  0.76 &  0.73 & 12.51 $\pm$  0.41 & 14.06 $\pm$  0.64 & 15.83 $\pm$  0.24 \\ 
\hline
  0.76 -  0.82 &  0.79 & 12.15 $\pm$  0.39 & 14.41 $\pm$  0.67 & 14.85 $\pm$  0.23 \\ 
\hline
  0.82 -  0.88 &  0.85 &  9.18 $\pm$  0.31 & 11.11 $\pm$  0.56 & 11.66 $\pm$  0.20 \\ 
\hline
  0.88 -  1.00 &  0.92 &  2.59 $\pm$  0.09 &  3.25 $\pm$  0.19 &  3.74 $\pm$  0.07 \\ 
 \hline 
 \end{tabular} 
 \end{center} 
 \caption{\emcap Values in percent of $(1/\sigma)(d\sigma^{LN}/d\xl) = 
dr/d\xl$ for 
leading neutrons with $\theta_n < 0.8$ mrad. For the PHP data,
$Q^2<0.02$ GeV$^2$, the normalisation uncertainty is $\pm$5\%, and for 
$\xl>0.52$ 
there is a further normalisation uncertainty of $\pm$4\%. The 
statistical error 
for the PHP data with $\xl>0.52$ includes a 
contribution from the uncertainty in the FNC-trigger 
efficiency. For the intermediate-$Q^2$
and DIS data, the normalisation uncertainty is $\pm$4\%.
\label{etabxlall} } 
 \end{table}

\newpage
\begin{table}
\centerline{\Large $ 0.2 < x_L < 0.28 $ }
\vspace{0.25cm}
\begin{center} 
 \begin{tabular}{|c|c|c|c|c|} \hline 
 $Q^2$ & $Q^2$ & $\xbj$ & $\xbj$ & $ratio$ (\%) \\ 
 GeV$^2$ & range & & range & $meas. \pm stat$ \\ 
\hline
7 & 4 - 10 & 1.1 $\cdot 10^{-4}$ & 8.0 - 15.0 $\cdot 10^{-5}$ &  0.48 $\pm$  0.04 \\ 
 &  & 2.1 $\cdot 10^{-4}$ & 1.5 - 3.0 $\cdot 10^{-4}$ &  0.40 $\pm$  0.03 \\ 
 &  & 4.2 $\cdot 10^{-4}$ & 3.0 - 6.0 $\cdot 10^{-4}$ &  0.43 $\pm$  0.03 \\ 
 &  & 8.5 $\cdot 10^{-4}$ & 6.0 - 12.0 $\cdot 10^{-4}$ &  0.46 $\pm$  0.04 \\ 
 &  & 17.0 $\cdot 10^{-4}$ & 1.2 - 2.4 $\cdot 10^{-3}$ &  0.50 $\pm$  0.05 \\ 
\hline
15 & 10 - 20 & 2.1 $\cdot 10^{-4}$ & 1.5 - 3.0 $\cdot 10^{-4}$ &  0.40 $\pm$  0.05 \\ 
 &  & 4.2 $\cdot 10^{-4}$ & 3.0 - 6.0 $\cdot 10^{-4}$ &  0.39 $\pm$  0.03 \\ 
 &  & 8.5 $\cdot 10^{-4}$ & 6.0 - 12.0 $\cdot 10^{-4}$ &  0.42 $\pm$  0.03 \\ 
 &  & 17.0 $\cdot 10^{-4}$ & 1.2 - 2.4 $\cdot 10^{-3}$ &  0.43 $\pm$  0.04 \\ 
 &  & 49.0 $\cdot 10^{-4}$ & 2.4 - 10.0 $\cdot 10^{-3}$ &  0.42 $\pm$  0.03 \\ 
\hline
30 & 20 - 40 & 4.2 $\cdot 10^{-4}$ & 3.0 - 6.0 $\cdot 10^{-4}$ &  0.55 $\pm$  0.06 \\ 
 &  & 8.5 $\cdot 10^{-4}$ & 6.0 - 12.0 $\cdot 10^{-4}$ &  0.50 $\pm$  0.04 \\ 
 &  & 17.0 $\cdot 10^{-4}$ & 1.2 - 2.4 $\cdot 10^{-3}$ &  0.43 $\pm$  0.03 \\ 
 &  & 49.0 $\cdot 10^{-4}$ & 2.4 - 10.0 $\cdot 10^{-3}$ &  0.49 $\pm$  0.03 \\ 
\hline
60 & 40 - 80 & 8.5 $\cdot 10^{-4}$ & 6.0 - 12.0 $\cdot 10^{-4}$ &  0.49 $\pm$  0.08 \\ 
 &  & 17.0 $\cdot 10^{-4}$ & 1.2 - 2.4 $\cdot 10^{-3}$ &  0.46 $\pm$  0.05 \\ 
 &  & 49.0 $\cdot 10^{-4}$ & 2.4 - 10.0 $\cdot 10^{-3}$ &  0.39 $\pm$  0.03 \\ 
 &  & 32.0 $\cdot 10^{-3}$ & 1.0 - 10.0 $\cdot 10^{-2}$ &  0.42 $\pm$  0.04 \\ 
\hline
120 & 80 - 160 & 17.0 $\cdot 10^{-4}$ & 1.2 - 2.4 $\cdot 10^{-3}$ &  0.53 $\pm$  0.12 \\ 
 &  & 49.0 $\cdot 10^{-4}$ & 2.4 - 10.0 $\cdot 10^{-3}$ &  0.42 $\pm$  0.05 \\ 
 &  & 32.0 $\cdot 10^{-3}$ & 1.0 - 10.0 $\cdot 10^{-2}$ &  0.53 $\pm$  0.05 \\ 
\hline
240 & 160 - 320 & 49.0 $\cdot 10^{-4}$ & 2.4 - 10.0 $\cdot 10^{-3}$ &  0.42 $\pm$  0.09 \\ 
 &  & 32.0 $\cdot 10^{-3}$ & 1.0 - 10.0 $\cdot 10^{-2}$ &  0.56 $\pm$  0.07 \\ 
\hline
480 & 320 - 640 & 32.0 $\cdot 10^{-3}$ & 1.0 - 10.0 $\cdot 10^{-2}$ &  0.44 $\pm$  0.09 \\ 
\hline
1000 & 640 - 10000 & 32.0 $\cdot 10^{-3}$ & 1.0 - 10.0 $\cdot 10^{-2}$ &  0.37 $\pm$  0.11 \\ 
\hline \hline 
 \multicolumn{5}{l}{Acceptance uncertainty: +2/-2 \%} \\\hline 
 \multicolumn{5}{l}{Energy scale uncertainty of $\pm$2 \%: -5/+4 \%} \\\hline 
 \multicolumn{5}{l}{Normalization error: 4 \%} \\\hline 
 \end{tabular} 
 \end{center} 
 \caption[The fraction of events with a leading neutron at $x_L=0.24$.]{\emcap The fraction of events with a leading neutron at $x_L=0.24$ in bins of $x$ and $Q^2$. \label{f2tabxl1} } 
 \end{table}

\newpage
\begin{table}
\centerline{\Large $ 0.28 < x_L < 0.34 $ }
\vspace{0.25cm}
\begin{center} 
 \begin{tabular}{|c|c|c|c|c|} \hline 
 $Q^2$ & $Q^2$ & $\xbj$ & $\xbj$ & $ratio$ (\%) \\ 
 GeV$^2$ & range & & range & $meas. \pm stat$ \\ 
\hline
7 & 4 - 10 & 1.1 $\cdot 10^{-4}$ & 8.0 - 15.0 $\cdot 10^{-5}$ &  0.34 $\pm$  0.04 \\ 
 &  & 2.1 $\cdot 10^{-4}$ & 1.5 - 3.0 $\cdot 10^{-4}$ &  0.37 $\pm$  0.03 \\ 
 &  & 4.2 $\cdot 10^{-4}$ & 3.0 - 6.0 $\cdot 10^{-4}$ &  0.41 $\pm$  0.03 \\ 
 &  & 8.5 $\cdot 10^{-4}$ & 6.0 - 12.0 $\cdot 10^{-4}$ &  0.50 $\pm$  0.04 \\ 
 &  & 17.0 $\cdot 10^{-4}$ & 1.2 - 2.4 $\cdot 10^{-3}$ &  0.46 $\pm$  0.04 \\ 
\hline
15 & 10 - 20 & 2.1 $\cdot 10^{-4}$ & 1.5 - 3.0 $\cdot 10^{-4}$ &  0.40 $\pm$  0.05 \\ 
 &  & 4.2 $\cdot 10^{-4}$ & 3.0 - 6.0 $\cdot 10^{-4}$ &  0.40 $\pm$  0.03 \\ 
 &  & 8.5 $\cdot 10^{-4}$ & 6.0 - 12.0 $\cdot 10^{-4}$ &  0.45 $\pm$  0.03 \\ 
 &  & 17.0 $\cdot 10^{-4}$ & 1.2 - 2.4 $\cdot 10^{-3}$ &  0.38 $\pm$  0.04 \\ 
 &  & 49.0 $\cdot 10^{-4}$ & 2.4 - 10.0 $\cdot 10^{-3}$ &  0.45 $\pm$  0.04 \\ 
\hline
30 & 20 - 40 & 4.2 $\cdot 10^{-4}$ & 3.0 - 6.0 $\cdot 10^{-4}$ &  0.47 $\pm$  0.06 \\ 
 &  & 8.5 $\cdot 10^{-4}$ & 6.0 - 12.0 $\cdot 10^{-4}$ &  0.49 $\pm$  0.04 \\ 
 &  & 17.0 $\cdot 10^{-4}$ & 1.2 - 2.4 $\cdot 10^{-3}$ &  0.51 $\pm$  0.04 \\ 
 &  & 49.0 $\cdot 10^{-4}$ & 2.4 - 10.0 $\cdot 10^{-3}$ &  0.50 $\pm$  0.03 \\ 
\hline
60 & 40 - 80 & 8.5 $\cdot 10^{-4}$ & 6.0 - 12.0 $\cdot 10^{-4}$ &  0.37 $\pm$  0.07 \\ 
 &  & 17.0 $\cdot 10^{-4}$ & 1.2 - 2.4 $\cdot 10^{-3}$ &  0.38 $\pm$  0.04 \\ 
 &  & 49.0 $\cdot 10^{-4}$ & 2.4 - 10.0 $\cdot 10^{-3}$ &  0.52 $\pm$  0.04 \\ 
 &  & 32.0 $\cdot 10^{-3}$ & 1.0 - 10.0 $\cdot 10^{-2}$ &  0.51 $\pm$  0.05 \\ 
\hline
120 & 80 - 160 & 17.0 $\cdot 10^{-4}$ & 1.2 - 2.4 $\cdot 10^{-3}$ &  0.24 $\pm$  0.08 \\ 
 &  & 49.0 $\cdot 10^{-4}$ & 2.4 - 10.0 $\cdot 10^{-3}$ &  0.43 $\pm$  0.05 \\ 
 &  & 32.0 $\cdot 10^{-3}$ & 1.0 - 10.0 $\cdot 10^{-2}$ &  0.56 $\pm$  0.05 \\ 
\hline
240 & 160 - 320 & 49.0 $\cdot 10^{-4}$ & 2.4 - 10.0 $\cdot 10^{-3}$ &  0.43 $\pm$  0.09 \\ 
 &  & 32.0 $\cdot 10^{-3}$ & 1.0 - 10.0 $\cdot 10^{-2}$ &  0.40 $\pm$  0.06 \\ 
\hline
480 & 320 - 640 & 32.0 $\cdot 10^{-3}$ & 1.0 - 10.0 $\cdot 10^{-2}$ &  0.55 $\pm$  0.11 \\ 
\hline
1000 & 640 - 10000 & 32.0 $\cdot 10^{-3}$ & 1.0 - 10.0 $\cdot 10^{-2}$ &  0.45 $\pm$  0.12 \\ 
\hline \hline 
 \multicolumn{5}{l}{Acceptance uncertainty: +2/-2 \%} \\\hline 
 \multicolumn{5}{l}{Energy scale uncertainty of $\pm$2 \%: -5/+4 \%} \\\hline 
 \multicolumn{5}{l}{Normalization error: 4 \%} \\\hline 
 \end{tabular} 
 \end{center} 
 \caption[The fraction of events with a leading neutron at $x_L=0.31$.]{\emcap The fraction of events with a leading neutron at $x_L=0.31$ in bins of $x$ and $Q^2$. \label{f2tabxl2} } 
 \end{table}

\newpage
\begin{table}
\centerline{\Large $ 0.34 < x_L < 0.4 $ }
\vspace{0.25cm}
\begin{center} 
 \begin{tabular}{|c|c|c|c|c|} \hline 
 $Q^2$ & $Q^2$ & $\xbj$ & $\xbj$ & $ratio$ (\%) \\ 
 GeV$^2$ & range & & range & $meas. \pm stat$ \\ 
\hline
7 & 4 - 10 & 1.1 $\cdot 10^{-4}$ & 8.0 - 15.0 $\cdot 10^{-5}$ &  0.45 $\pm$  0.04 \\ 
 &  & 2.1 $\cdot 10^{-4}$ & 1.5 - 3.0 $\cdot 10^{-4}$ &  0.46 $\pm$  0.03 \\ 
 &  & 4.2 $\cdot 10^{-4}$ & 3.0 - 6.0 $\cdot 10^{-4}$ &  0.52 $\pm$  0.03 \\ 
 &  & 8.5 $\cdot 10^{-4}$ & 6.0 - 12.0 $\cdot 10^{-4}$ &  0.50 $\pm$  0.04 \\ 
 &  & 17.0 $\cdot 10^{-4}$ & 1.2 - 2.4 $\cdot 10^{-3}$ &  0.55 $\pm$  0.05 \\ 
\hline
15 & 10 - 20 & 2.1 $\cdot 10^{-4}$ & 1.5 - 3.0 $\cdot 10^{-4}$ &  0.42 $\pm$  0.05 \\ 
 &  & 4.2 $\cdot 10^{-4}$ & 3.0 - 6.0 $\cdot 10^{-4}$ &  0.52 $\pm$  0.03 \\ 
 &  & 8.5 $\cdot 10^{-4}$ & 6.0 - 12.0 $\cdot 10^{-4}$ &  0.54 $\pm$  0.04 \\ 
 &  & 17.0 $\cdot 10^{-4}$ & 1.2 - 2.4 $\cdot 10^{-3}$ &  0.48 $\pm$  0.04 \\ 
 &  & 49.0 $\cdot 10^{-4}$ & 2.4 - 10.0 $\cdot 10^{-3}$ &  0.59 $\pm$  0.04 \\ 
\hline
30 & 20 - 40 & 4.2 $\cdot 10^{-4}$ & 3.0 - 6.0 $\cdot 10^{-4}$ &  0.45 $\pm$  0.06 \\ 
 &  & 8.5 $\cdot 10^{-4}$ & 6.0 - 12.0 $\cdot 10^{-4}$ &  0.49 $\pm$  0.04 \\ 
 &  & 17.0 $\cdot 10^{-4}$ & 1.2 - 2.4 $\cdot 10^{-3}$ &  0.55 $\pm$  0.04 \\ 
 &  & 49.0 $\cdot 10^{-4}$ & 2.4 - 10.0 $\cdot 10^{-3}$ &  0.58 $\pm$  0.03 \\ 
\hline
60 & 40 - 80 & 8.5 $\cdot 10^{-4}$ & 6.0 - 12.0 $\cdot 10^{-4}$ &  0.56 $\pm$  0.09 \\ 
 &  & 17.0 $\cdot 10^{-4}$ & 1.2 - 2.4 $\cdot 10^{-3}$ &  0.53 $\pm$  0.05 \\ 
 &  & 49.0 $\cdot 10^{-4}$ & 2.4 - 10.0 $\cdot 10^{-3}$ &  0.51 $\pm$  0.04 \\ 
 &  & 32.0 $\cdot 10^{-3}$ & 1.0 - 10.0 $\cdot 10^{-2}$ &  0.58 $\pm$  0.05 \\ 
\hline
120 & 80 - 160 & 17.0 $\cdot 10^{-4}$ & 1.2 - 2.4 $\cdot 10^{-3}$ &  0.52 $\pm$  0.12 \\ 
 &  & 49.0 $\cdot 10^{-4}$ & 2.4 - 10.0 $\cdot 10^{-3}$ &  0.51 $\pm$  0.05 \\ 
 &  & 32.0 $\cdot 10^{-3}$ & 1.0 - 10.0 $\cdot 10^{-2}$ &  0.54 $\pm$  0.05 \\ 
\hline
240 & 160 - 320 & 49.0 $\cdot 10^{-4}$ & 2.4 - 10.0 $\cdot 10^{-3}$ &  0.52 $\pm$  0.10 \\ 
 &  & 32.0 $\cdot 10^{-3}$ & 1.0 - 10.0 $\cdot 10^{-2}$ &  0.59 $\pm$  0.07 \\ 
\hline
480 & 320 - 640 & 32.0 $\cdot 10^{-3}$ & 1.0 - 10.0 $\cdot 10^{-2}$ &  0.63 $\pm$  0.11 \\ 
\hline
1000 & 640 - 10000 & 32.0 $\cdot 10^{-3}$ & 1.0 - 10.0 $\cdot 10^{-2}$ &  0.54 $\pm$  0.13 \\ 
\hline \hline 
 \multicolumn{5}{l}{Acceptance uncertainty: +2/-3 \%} \\\hline 
 \multicolumn{5}{l}{Energy scale uncertainty of $\pm$2 \%: -5/+4 \%} \\\hline 
 \multicolumn{5}{l}{Normalization error: 4 \%} \\\hline 
 \end{tabular} 
 \end{center} 
 \caption[The fraction of events with a leading neutron at $x_L=0.37$.]{\emcap The fraction of events with a leading neutron at $x_L=0.37$ in bins of $x$ and $Q^2$. \label{f2tabxl3} } 
 \end{table}

\newpage
\begin{table}
\centerline{\Large $ 0.4 < x_L < 0.46 $ }
\vspace{0.25cm}
\begin{center} 
 \begin{tabular}{|c|c|c|c|c|} \hline 
 $Q^2$ & $Q^2$ & $\xbj$ & $\xbj$ & $ratio$ (\%) \\ 
 GeV$^2$ & range & & range & $meas. \pm stat$ \\ 
\hline
7 & 4 - 10 & 1.1 $\cdot 10^{-4}$ & 8.0 - 15.0 $\cdot 10^{-5}$ &  0.51 $\pm$  0.04 \\ 
 &  & 2.1 $\cdot 10^{-4}$ & 1.5 - 3.0 $\cdot 10^{-4}$ &  0.55 $\pm$  0.03 \\ 
 &  & 4.2 $\cdot 10^{-4}$ & 3.0 - 6.0 $\cdot 10^{-4}$ &  0.55 $\pm$  0.04 \\ 
 &  & 8.5 $\cdot 10^{-4}$ & 6.0 - 12.0 $\cdot 10^{-4}$ &  0.55 $\pm$  0.04 \\ 
 &  & 17.0 $\cdot 10^{-4}$ & 1.2 - 2.4 $\cdot 10^{-3}$ &  0.62 $\pm$  0.05 \\ 
\hline
15 & 10 - 20 & 2.1 $\cdot 10^{-4}$ & 1.5 - 3.0 $\cdot 10^{-4}$ &  0.52 $\pm$  0.05 \\ 
 &  & 4.2 $\cdot 10^{-4}$ & 3.0 - 6.0 $\cdot 10^{-4}$ &  0.55 $\pm$  0.03 \\ 
 &  & 8.5 $\cdot 10^{-4}$ & 6.0 - 12.0 $\cdot 10^{-4}$ &  0.55 $\pm$  0.04 \\ 
 &  & 17.0 $\cdot 10^{-4}$ & 1.2 - 2.4 $\cdot 10^{-3}$ &  0.66 $\pm$  0.05 \\ 
 &  & 49.0 $\cdot 10^{-4}$ & 2.4 - 10.0 $\cdot 10^{-3}$ &  0.60 $\pm$  0.04 \\ 
\hline
30 & 20 - 40 & 4.2 $\cdot 10^{-4}$ & 3.0 - 6.0 $\cdot 10^{-4}$ &  0.41 $\pm$  0.05 \\ 
 &  & 8.5 $\cdot 10^{-4}$ & 6.0 - 12.0 $\cdot 10^{-4}$ &  0.46 $\pm$  0.03 \\ 
 &  & 17.0 $\cdot 10^{-4}$ & 1.2 - 2.4 $\cdot 10^{-3}$ &  0.54 $\pm$  0.04 \\ 
 &  & 49.0 $\cdot 10^{-4}$ & 2.4 - 10.0 $\cdot 10^{-3}$ &  0.65 $\pm$  0.03 \\ 
\hline
60 & 40 - 80 & 8.5 $\cdot 10^{-4}$ & 6.0 - 12.0 $\cdot 10^{-4}$ &  0.48 $\pm$  0.08 \\ 
 &  & 17.0 $\cdot 10^{-4}$ & 1.2 - 2.4 $\cdot 10^{-3}$ &  0.64 $\pm$  0.06 \\ 
 &  & 49.0 $\cdot 10^{-4}$ & 2.4 - 10.0 $\cdot 10^{-3}$ &  0.53 $\pm$  0.04 \\ 
 &  & 32.0 $\cdot 10^{-3}$ & 1.0 - 10.0 $\cdot 10^{-2}$ &  0.67 $\pm$  0.06 \\ 
\hline
120 & 80 - 160 & 17.0 $\cdot 10^{-4}$ & 1.2 - 2.4 $\cdot 10^{-3}$ &  0.65 $\pm$  0.14 \\ 
 &  & 49.0 $\cdot 10^{-4}$ & 2.4 - 10.0 $\cdot 10^{-3}$ &  0.59 $\pm$  0.06 \\ 
 &  & 32.0 $\cdot 10^{-3}$ & 1.0 - 10.0 $\cdot 10^{-2}$ &  0.64 $\pm$  0.06 \\ 
\hline
240 & 160 - 320 & 49.0 $\cdot 10^{-4}$ & 2.4 - 10.0 $\cdot 10^{-3}$ &  0.77 $\pm$  0.12 \\ 
 &  & 32.0 $\cdot 10^{-3}$ & 1.0 - 10.0 $\cdot 10^{-2}$ &  0.70 $\pm$  0.08 \\ 
\hline
480 & 320 - 640 & 32.0 $\cdot 10^{-3}$ & 1.0 - 10.0 $\cdot 10^{-2}$ &  0.77 $\pm$  0.12 \\ 
\hline
1000 & 640 - 10000 & 32.0 $\cdot 10^{-3}$ & 1.0 - 10.0 $\cdot 10^{-2}$ &  0.75 $\pm$  0.16 \\ 
\hline \hline 
 \multicolumn{5}{l}{Acceptance uncertainty: +3/-3 \%} \\\hline 
 \multicolumn{5}{l}{Energy scale uncertainty of $\pm$2 \%: -5/+4 \%} \\\hline 
 \multicolumn{5}{l}{Normalization error: 4 \%} \\\hline 
 \end{tabular} 
 \end{center} 
 \caption[The fraction of events with a leading neutron at $x_L=0.43$.]{\emcap The fraction of events with a leading neutron at $x_L=0.43$ in bins of $x$ and $Q^2$. \label{f2tabxl4} } 
 \end{table}

\newpage
\begin{table}
\centerline{\Large $ 0.46 < x_L < 0.52 $ }
\vspace{0.25cm}
\begin{center} 
 \begin{tabular}{|c|c|c|c|c|} \hline 
 $Q^2$ & $Q^2$ & $\xbj$ & $\xbj$ & $ratio$ (\%) \\ 
 GeV$^2$ & range & & range & $meas. \pm stat$ \\ 
\hline
7 & 4 - 10 & 1.1 $\cdot 10^{-4}$ & 8.0 - 15.0 $\cdot 10^{-5}$ &  0.51 $\pm$  0.04 \\ 
 &  & 2.1 $\cdot 10^{-4}$ & 1.5 - 3.0 $\cdot 10^{-4}$ &  0.59 $\pm$  0.03 \\ 
 &  & 4.2 $\cdot 10^{-4}$ & 3.0 - 6.0 $\cdot 10^{-4}$ &  0.57 $\pm$  0.04 \\ 
 &  & 8.5 $\cdot 10^{-4}$ & 6.0 - 12.0 $\cdot 10^{-4}$ &  0.67 $\pm$  0.05 \\ 
 &  & 17.0 $\cdot 10^{-4}$ & 1.2 - 2.4 $\cdot 10^{-3}$ &  0.69 $\pm$  0.05 \\ 
\hline
15 & 10 - 20 & 2.1 $\cdot 10^{-4}$ & 1.5 - 3.0 $\cdot 10^{-4}$ &  0.55 $\pm$  0.05 \\ 
 &  & 4.2 $\cdot 10^{-4}$ & 3.0 - 6.0 $\cdot 10^{-4}$ &  0.59 $\pm$  0.03 \\ 
 &  & 8.5 $\cdot 10^{-4}$ & 6.0 - 12.0 $\cdot 10^{-4}$ &  0.65 $\pm$  0.04 \\ 
 &  & 17.0 $\cdot 10^{-4}$ & 1.2 - 2.4 $\cdot 10^{-3}$ &  0.67 $\pm$  0.05 \\ 
 &  & 49.0 $\cdot 10^{-4}$ & 2.4 - 10.0 $\cdot 10^{-3}$ &  0.62 $\pm$  0.04 \\ 
\hline
30 & 20 - 40 & 4.2 $\cdot 10^{-4}$ & 3.0 - 6.0 $\cdot 10^{-4}$ &  0.50 $\pm$  0.06 \\ 
 &  & 8.5 $\cdot 10^{-4}$ & 6.0 - 12.0 $\cdot 10^{-4}$ &  0.51 $\pm$  0.04 \\ 
 &  & 17.0 $\cdot 10^{-4}$ & 1.2 - 2.4 $\cdot 10^{-3}$ &  0.66 $\pm$  0.04 \\ 
 &  & 49.0 $\cdot 10^{-4}$ & 2.4 - 10.0 $\cdot 10^{-3}$ &  0.71 $\pm$  0.03 \\ 
\hline
60 & 40 - 80 & 8.5 $\cdot 10^{-4}$ & 6.0 - 12.0 $\cdot 10^{-4}$ &  0.55 $\pm$  0.09 \\ 
 &  & 17.0 $\cdot 10^{-4}$ & 1.2 - 2.4 $\cdot 10^{-3}$ &  0.68 $\pm$  0.06 \\ 
 &  & 49.0 $\cdot 10^{-4}$ & 2.4 - 10.0 $\cdot 10^{-3}$ &  0.69 $\pm$  0.04 \\ 
 &  & 32.0 $\cdot 10^{-3}$ & 1.0 - 10.0 $\cdot 10^{-2}$ &  0.80 $\pm$  0.06 \\ 
\hline
120 & 80 - 160 & 17.0 $\cdot 10^{-4}$ & 1.2 - 2.4 $\cdot 10^{-3}$ &  0.92 $\pm$  0.16 \\ 
 &  & 49.0 $\cdot 10^{-4}$ & 2.4 - 10.0 $\cdot 10^{-3}$ &  0.64 $\pm$  0.06 \\ 
 &  & 32.0 $\cdot 10^{-3}$ & 1.0 - 10.0 $\cdot 10^{-2}$ &  0.86 $\pm$  0.07 \\ 
\hline
240 & 160 - 320 & 49.0 $\cdot 10^{-4}$ & 2.4 - 10.0 $\cdot 10^{-3}$ &  0.58 $\pm$  0.10 \\ 
 &  & 32.0 $\cdot 10^{-3}$ & 1.0 - 10.0 $\cdot 10^{-2}$ &  0.69 $\pm$  0.08 \\ 
\hline
480 & 320 - 640 & 32.0 $\cdot 10^{-3}$ & 1.0 - 10.0 $\cdot 10^{-2}$ &  0.74 $\pm$  0.12 \\ 
\hline
1000 & 640 - 10000 & 32.0 $\cdot 10^{-3}$ & 1.0 - 10.0 $\cdot 10^{-2}$ &  0.81 $\pm$  0.16 \\ 
\hline \hline 
 \multicolumn{5}{l}{Acceptance uncertainty: +3/-3 \%} \\\hline 
 \multicolumn{5}{l}{Energy scale uncertainty of $\pm$2 \%: -4/+5 \%} \\\hline 
 \multicolumn{5}{l}{Normalization error: 4 \%} \\\hline 
 \end{tabular} 
 \end{center} 
 \caption[The fraction of events with a leading neutron at $x_L=0.49$.]{\emcap The fraction of events with a leading neutron at $x_L=0.49$ in bins of $x$ and $Q^2$. \label{f2tabxl5} } 
 \end{table}

\newpage
\begin{table}
\centerline{\Large $ 0.52 < x_L < 0.58 $ }
\vspace{0.25cm}
\begin{center} 
 \begin{tabular}{|c|c|c|c|c|} \hline 
 $Q^2$ & $Q^2$ & $\xbj$ & $\xbj$ & $ratio$ (\%) \\ 
 GeV$^2$ & range & & range & $meas. \pm stat$ \\ 
\hline
7 & 4 - 10 & 1.1 $\cdot 10^{-4}$ & 8.0 - 15.0 $\cdot 10^{-5}$ &  0.71 $\pm$  0.05 \\ 
 &  & 2.1 $\cdot 10^{-4}$ & 1.5 - 3.0 $\cdot 10^{-4}$ &  0.70 $\pm$  0.03 \\ 
 &  & 4.2 $\cdot 10^{-4}$ & 3.0 - 6.0 $\cdot 10^{-4}$ &  0.66 $\pm$  0.04 \\ 
 &  & 8.5 $\cdot 10^{-4}$ & 6.0 - 12.0 $\cdot 10^{-4}$ &  0.73 $\pm$  0.05 \\ 
 &  & 17.0 $\cdot 10^{-4}$ & 1.2 - 2.4 $\cdot 10^{-3}$ &  0.76 $\pm$  0.06 \\ 
\hline
15 & 10 - 20 & 2.1 $\cdot 10^{-4}$ & 1.5 - 3.0 $\cdot 10^{-4}$ &  0.71 $\pm$  0.06 \\ 
 &  & 4.2 $\cdot 10^{-4}$ & 3.0 - 6.0 $\cdot 10^{-4}$ &  0.76 $\pm$  0.04 \\ 
 &  & 8.5 $\cdot 10^{-4}$ & 6.0 - 12.0 $\cdot 10^{-4}$ &  0.79 $\pm$  0.04 \\ 
 &  & 17.0 $\cdot 10^{-4}$ & 1.2 - 2.4 $\cdot 10^{-3}$ &  0.72 $\pm$  0.05 \\ 
 &  & 49.0 $\cdot 10^{-4}$ & 2.4 - 10.0 $\cdot 10^{-3}$ &  0.84 $\pm$  0.05 \\ 
\hline
30 & 20 - 40 & 4.2 $\cdot 10^{-4}$ & 3.0 - 6.0 $\cdot 10^{-4}$ &  0.72 $\pm$  0.07 \\ 
 &  & 8.5 $\cdot 10^{-4}$ & 6.0 - 12.0 $\cdot 10^{-4}$ &  0.72 $\pm$  0.04 \\ 
 &  & 17.0 $\cdot 10^{-4}$ & 1.2 - 2.4 $\cdot 10^{-3}$ &  0.79 $\pm$  0.05 \\ 
 &  & 49.0 $\cdot 10^{-4}$ & 2.4 - 10.0 $\cdot 10^{-3}$ &  0.81 $\pm$  0.04 \\ 
\hline
60 & 40 - 80 & 8.5 $\cdot 10^{-4}$ & 6.0 - 12.0 $\cdot 10^{-4}$ &  0.75 $\pm$  0.10 \\ 
 &  & 17.0 $\cdot 10^{-4}$ & 1.2 - 2.4 $\cdot 10^{-3}$ &  0.69 $\pm$  0.06 \\ 
 &  & 49.0 $\cdot 10^{-4}$ & 2.4 - 10.0 $\cdot 10^{-3}$ &  0.71 $\pm$  0.04 \\ 
 &  & 32.0 $\cdot 10^{-3}$ & 1.0 - 10.0 $\cdot 10^{-2}$ &  0.88 $\pm$  0.06 \\ 
\hline
120 & 80 - 160 & 17.0 $\cdot 10^{-4}$ & 1.2 - 2.4 $\cdot 10^{-3}$ &  0.77 $\pm$  0.15 \\ 
 &  & 49.0 $\cdot 10^{-4}$ & 2.4 - 10.0 $\cdot 10^{-3}$ &  0.80 $\pm$  0.07 \\ 
 &  & 32.0 $\cdot 10^{-3}$ & 1.0 - 10.0 $\cdot 10^{-2}$ &  0.88 $\pm$  0.07 \\ 
\hline
240 & 160 - 320 & 49.0 $\cdot 10^{-4}$ & 2.4 - 10.0 $\cdot 10^{-3}$ &  0.83 $\pm$  0.12 \\ 
 &  & 32.0 $\cdot 10^{-3}$ & 1.0 - 10.0 $\cdot 10^{-2}$ &  0.97 $\pm$  0.09 \\ 
\hline
480 & 320 - 640 & 32.0 $\cdot 10^{-3}$ & 1.0 - 10.0 $\cdot 10^{-2}$ &  0.88 $\pm$  0.13 \\ 
\hline
1000 & 640 - 10000 & 32.0 $\cdot 10^{-3}$ & 1.0 - 10.0 $\cdot 10^{-2}$ &  0.86 $\pm$  0.17 \\ 
\hline \hline 
 \multicolumn{5}{l}{Acceptance uncertainty: +4/-3 \%} \\\hline 
 \multicolumn{5}{l}{Energy scale uncertainty of $\pm$2 \%: -4/+5 \%} \\\hline 
 \multicolumn{5}{l}{Normalization error: 4 \%} \\\hline 
 \end{tabular} 
 \end{center} 
 \caption[The fraction of events with a leading neutron at $x_L=0.55$.]{\emcap The fraction of events with a leading neutron at $x_L=0.55$ in bins of $x$ and $Q^2$. \label{f2tabxl6} } 
 \end{table}

\newpage
\begin{table}
\centerline{\Large $ 0.58 < x_L < 0.64 $ }
\vspace{0.25cm}
\begin{center} 
 \begin{tabular}{|c|c|c|c|c|} \hline 
 $Q^2$ & $Q^2$ & $\xbj$ & $\xbj$ & $ratio$ (\%) \\ 
 GeV$^2$ & range & & range & $meas. \pm stat$ \\ 
\hline
7 & 4 - 10 & 1.1 $\cdot 10^{-4}$ & 8.0 - 15.0 $\cdot 10^{-5}$ &  0.78 $\pm$  0.05 \\ 
 &  & 2.1 $\cdot 10^{-4}$ & 1.5 - 3.0 $\cdot 10^{-4}$ &  0.79 $\pm$  0.04 \\ 
 &  & 4.2 $\cdot 10^{-4}$ & 3.0 - 6.0 $\cdot 10^{-4}$ &  0.81 $\pm$  0.04 \\ 
 &  & 8.5 $\cdot 10^{-4}$ & 6.0 - 12.0 $\cdot 10^{-4}$ &  0.78 $\pm$  0.05 \\ 
 &  & 17.0 $\cdot 10^{-4}$ & 1.2 - 2.4 $\cdot 10^{-3}$ &  0.88 $\pm$  0.06 \\ 
\hline
15 & 10 - 20 & 2.1 $\cdot 10^{-4}$ & 1.5 - 3.0 $\cdot 10^{-4}$ &  0.68 $\pm$  0.06 \\ 
 &  & 4.2 $\cdot 10^{-4}$ & 3.0 - 6.0 $\cdot 10^{-4}$ &  0.77 $\pm$  0.04 \\ 
 &  & 8.5 $\cdot 10^{-4}$ & 6.0 - 12.0 $\cdot 10^{-4}$ &  0.78 $\pm$  0.04 \\ 
 &  & 17.0 $\cdot 10^{-4}$ & 1.2 - 2.4 $\cdot 10^{-3}$ &  0.85 $\pm$  0.05 \\ 
 &  & 49.0 $\cdot 10^{-4}$ & 2.4 - 10.0 $\cdot 10^{-3}$ &  0.86 $\pm$  0.05 \\ 
\hline
30 & 20 - 40 & 4.2 $\cdot 10^{-4}$ & 3.0 - 6.0 $\cdot 10^{-4}$ &  0.76 $\pm$  0.07 \\ 
 &  & 8.5 $\cdot 10^{-4}$ & 6.0 - 12.0 $\cdot 10^{-4}$ &  0.80 $\pm$  0.04 \\ 
 &  & 17.0 $\cdot 10^{-4}$ & 1.2 - 2.4 $\cdot 10^{-3}$ &  0.75 $\pm$  0.04 \\ 
 &  & 49.0 $\cdot 10^{-4}$ & 2.4 - 10.0 $\cdot 10^{-3}$ &  0.84 $\pm$  0.04 \\ 
\hline
60 & 40 - 80 & 8.5 $\cdot 10^{-4}$ & 6.0 - 12.0 $\cdot 10^{-4}$ &  0.69 $\pm$  0.10 \\ 
 &  & 17.0 $\cdot 10^{-4}$ & 1.2 - 2.4 $\cdot 10^{-3}$ &  0.83 $\pm$  0.06 \\ 
 &  & 49.0 $\cdot 10^{-4}$ & 2.4 - 10.0 $\cdot 10^{-3}$ &  0.87 $\pm$  0.05 \\ 
 &  & 32.0 $\cdot 10^{-3}$ & 1.0 - 10.0 $\cdot 10^{-2}$ &  0.87 $\pm$  0.06 \\ 
\hline
120 & 80 - 160 & 17.0 $\cdot 10^{-4}$ & 1.2 - 2.4 $\cdot 10^{-3}$ &  0.52 $\pm$  0.12 \\ 
 &  & 49.0 $\cdot 10^{-4}$ & 2.4 - 10.0 $\cdot 10^{-3}$ &  0.80 $\pm$  0.06 \\ 
 &  & 32.0 $\cdot 10^{-3}$ & 1.0 - 10.0 $\cdot 10^{-2}$ &  0.83 $\pm$  0.06 \\ 
\hline
240 & 160 - 320 & 49.0 $\cdot 10^{-4}$ & 2.4 - 10.0 $\cdot 10^{-3}$ &  0.84 $\pm$  0.12 \\ 
 &  & 32.0 $\cdot 10^{-3}$ & 1.0 - 10.0 $\cdot 10^{-2}$ &  1.04 $\pm$  0.09 \\ 
\hline
480 & 320 - 640 & 32.0 $\cdot 10^{-3}$ & 1.0 - 10.0 $\cdot 10^{-2}$ &  0.89 $\pm$  0.13 \\ 
\hline
1000 & 640 - 10000 & 32.0 $\cdot 10^{-3}$ & 1.0 - 10.0 $\cdot 10^{-2}$ &  0.82 $\pm$  0.16 \\ 
\hline \hline 
 \multicolumn{5}{l}{Acceptance uncertainty: +4/-4 \%} \\\hline 
 \multicolumn{5}{l}{Energy scale uncertainty of $\pm$2 \%: -4/+4 \%} \\\hline 
 \multicolumn{5}{l}{Normalization error: 4 \%} \\\hline 
 \end{tabular} 
 \end{center} 
 \caption[The fraction of events with a leading neutron at $x_L=0.61$.]{\emcap The fraction of events with a leading neutron at $x_L=0.61$ in bins of $x$ and $Q^2$. \label{f2tabxl7} } 
 \end{table}

\newpage
\begin{table}
\centerline{\Large $ 0.64 < x_L < 0.7 $ }
\vspace{0.25cm}
\begin{center} 
 \begin{tabular}{|c|c|c|c|c|} \hline 
 $Q^2$ & $Q^2$ & $\xbj$ & $\xbj$ & $ratio$ (\%) \\ 
 GeV$^2$ & range & & range & $meas. \pm stat$ \\ 
\hline
7 & 4 - 10 & 1.1 $\cdot 10^{-4}$ & 8.0 - 15.0 $\cdot 10^{-5}$ &  0.86 $\pm$  0.06 \\ 
 &  & 2.1 $\cdot 10^{-4}$ & 1.5 - 3.0 $\cdot 10^{-4}$ &  0.91 $\pm$  0.04 \\ 
 &  & 4.2 $\cdot 10^{-4}$ & 3.0 - 6.0 $\cdot 10^{-4}$ &  0.92 $\pm$  0.05 \\ 
 &  & 8.5 $\cdot 10^{-4}$ & 6.0 - 12.0 $\cdot 10^{-4}$ &  1.00 $\pm$  0.06 \\ 
 &  & 17.0 $\cdot 10^{-4}$ & 1.2 - 2.4 $\cdot 10^{-3}$ &  0.85 $\pm$  0.06 \\ 
\hline
15 & 10 - 20 & 2.1 $\cdot 10^{-4}$ & 1.5 - 3.0 $\cdot 10^{-4}$ &  0.93 $\pm$  0.07 \\ 
 &  & 4.2 $\cdot 10^{-4}$ & 3.0 - 6.0 $\cdot 10^{-4}$ &  0.93 $\pm$  0.04 \\ 
 &  & 8.5 $\cdot 10^{-4}$ & 6.0 - 12.0 $\cdot 10^{-4}$ &  0.91 $\pm$  0.05 \\ 
 &  & 17.0 $\cdot 10^{-4}$ & 1.2 - 2.4 $\cdot 10^{-3}$ &  0.97 $\pm$  0.05 \\ 
 &  & 49.0 $\cdot 10^{-4}$ & 2.4 - 10.0 $\cdot 10^{-3}$ &  0.92 $\pm$  0.05 \\ 
\hline
30 & 20 - 40 & 4.2 $\cdot 10^{-4}$ & 3.0 - 6.0 $\cdot 10^{-4}$ &  0.97 $\pm$  0.08 \\ 
 &  & 8.5 $\cdot 10^{-4}$ & 6.0 - 12.0 $\cdot 10^{-4}$ &  0.86 $\pm$  0.05 \\ 
 &  & 17.0 $\cdot 10^{-4}$ & 1.2 - 2.4 $\cdot 10^{-3}$ &  0.86 $\pm$  0.05 \\ 
 &  & 49.0 $\cdot 10^{-4}$ & 2.4 - 10.0 $\cdot 10^{-3}$ &  0.95 $\pm$  0.04 \\ 
\hline
60 & 40 - 80 & 8.5 $\cdot 10^{-4}$ & 6.0 - 12.0 $\cdot 10^{-4}$ &  0.89 $\pm$  0.11 \\ 
 &  & 17.0 $\cdot 10^{-4}$ & 1.2 - 2.4 $\cdot 10^{-3}$ &  0.81 $\pm$  0.06 \\ 
 &  & 49.0 $\cdot 10^{-4}$ & 2.4 - 10.0 $\cdot 10^{-3}$ &  0.92 $\pm$  0.05 \\ 
 &  & 32.0 $\cdot 10^{-3}$ & 1.0 - 10.0 $\cdot 10^{-2}$ &  0.97 $\pm$  0.07 \\ 
\hline
120 & 80 - 160 & 17.0 $\cdot 10^{-4}$ & 1.2 - 2.4 $\cdot 10^{-3}$ &  0.82 $\pm$  0.15 \\ 
 &  & 49.0 $\cdot 10^{-4}$ & 2.4 - 10.0 $\cdot 10^{-3}$ &  0.91 $\pm$  0.07 \\ 
 &  & 32.0 $\cdot 10^{-3}$ & 1.0 - 10.0 $\cdot 10^{-2}$ &  0.95 $\pm$  0.07 \\ 
\hline
240 & 160 - 320 & 49.0 $\cdot 10^{-4}$ & 2.4 - 10.0 $\cdot 10^{-3}$ &  1.04 $\pm$  0.13 \\ 
 &  & 32.0 $\cdot 10^{-3}$ & 1.0 - 10.0 $\cdot 10^{-2}$ &  0.90 $\pm$  0.09 \\ 
\hline
480 & 320 - 640 & 32.0 $\cdot 10^{-3}$ & 1.0 - 10.0 $\cdot 10^{-2}$ &  1.10 $\pm$  0.14 \\ 
\hline
1000 & 640 - 10000 & 32.0 $\cdot 10^{-3}$ & 1.0 - 10.0 $\cdot 10^{-2}$ &  0.98 $\pm$  0.18 \\ 
\hline \hline 
 \multicolumn{5}{l}{Acceptance uncertainty: +5/-5 \%} \\\hline 
 \multicolumn{5}{l}{Energy scale uncertainty of $\pm$2 \%: -4/+4 \%} \\\hline 
 \multicolumn{5}{l}{Normalization error: 4 \%} \\\hline 
 \end{tabular} 
 \end{center} 
 \caption[The fraction of events with a leading neutron at $x_L=0.67$.]{\emcap The fraction of events with a leading neutron at $x_L=0.67$ in bins of $x$ and $Q^2$. \label{f2tabxl8} } 
 \end{table}

\newpage
\begin{table}
\centerline{\Large $ 0.7 < x_L < 0.76 $ }
\vspace{0.25cm}
\begin{center} 
 \begin{tabular}{|c|c|c|c|c|} \hline 
 $Q^2$ & $Q^2$ & $\xbj$ & $\xbj$ & $ratio$ (\%) \\ 
 GeV$^2$ & range & & range & $meas. \pm stat$ \\ 
\hline
7 & 4 - 10 & 1.1 $\cdot 10^{-4}$ & 8.0 - 15.0 $\cdot 10^{-5}$ &  0.95 $\pm$  0.06 \\ 
 &  & 2.1 $\cdot 10^{-4}$ & 1.5 - 3.0 $\cdot 10^{-4}$ &  0.90 $\pm$  0.04 \\ 
 &  & 4.2 $\cdot 10^{-4}$ & 3.0 - 6.0 $\cdot 10^{-4}$ &  0.96 $\pm$  0.05 \\ 
 &  & 8.5 $\cdot 10^{-4}$ & 6.0 - 12.0 $\cdot 10^{-4}$ &  0.98 $\pm$  0.05 \\ 
 &  & 17.0 $\cdot 10^{-4}$ & 1.2 - 2.4 $\cdot 10^{-3}$ &  0.92 $\pm$  0.06 \\ 
\hline
15 & 10 - 20 & 2.1 $\cdot 10^{-4}$ & 1.5 - 3.0 $\cdot 10^{-4}$ &  0.84 $\pm$  0.07 \\ 
 &  & 4.2 $\cdot 10^{-4}$ & 3.0 - 6.0 $\cdot 10^{-4}$ &  0.88 $\pm$  0.04 \\ 
 &  & 8.5 $\cdot 10^{-4}$ & 6.0 - 12.0 $\cdot 10^{-4}$ &  0.94 $\pm$  0.05 \\ 
 &  & 17.0 $\cdot 10^{-4}$ & 1.2 - 2.4 $\cdot 10^{-3}$ &  1.02 $\pm$  0.06 \\ 
 &  & 49.0 $\cdot 10^{-4}$ & 2.4 - 10.0 $\cdot 10^{-3}$ &  1.09 $\pm$  0.05 \\ 
\hline
30 & 20 - 40 & 4.2 $\cdot 10^{-4}$ & 3.0 - 6.0 $\cdot 10^{-4}$ &  0.85 $\pm$  0.08 \\ 
 &  & 8.5 $\cdot 10^{-4}$ & 6.0 - 12.0 $\cdot 10^{-4}$ &  0.83 $\pm$  0.04 \\ 
 &  & 17.0 $\cdot 10^{-4}$ & 1.2 - 2.4 $\cdot 10^{-3}$ &  0.93 $\pm$  0.05 \\ 
 &  & 49.0 $\cdot 10^{-4}$ & 2.4 - 10.0 $\cdot 10^{-3}$ &  0.96 $\pm$  0.04 \\ 
\hline
60 & 40 - 80 & 8.5 $\cdot 10^{-4}$ & 6.0 - 12.0 $\cdot 10^{-4}$ &  1.03 $\pm$  0.12 \\ 
 &  & 17.0 $\cdot 10^{-4}$ & 1.2 - 2.4 $\cdot 10^{-3}$ &  0.87 $\pm$  0.06 \\ 
 &  & 49.0 $\cdot 10^{-4}$ & 2.4 - 10.0 $\cdot 10^{-3}$ &  0.91 $\pm$  0.05 \\ 
 &  & 32.0 $\cdot 10^{-3}$ & 1.0 - 10.0 $\cdot 10^{-2}$ &  1.06 $\pm$  0.07 \\ 
\hline
120 & 80 - 160 & 17.0 $\cdot 10^{-4}$ & 1.2 - 2.4 $\cdot 10^{-3}$ &  1.11 $\pm$  0.17 \\ 
 &  & 49.0 $\cdot 10^{-4}$ & 2.4 - 10.0 $\cdot 10^{-3}$ &  0.83 $\pm$  0.06 \\ 
 &  & 32.0 $\cdot 10^{-3}$ & 1.0 - 10.0 $\cdot 10^{-2}$ &  0.94 $\pm$  0.07 \\ 
\hline
240 & 160 - 320 & 49.0 $\cdot 10^{-4}$ & 2.4 - 10.0 $\cdot 10^{-3}$ &  0.96 $\pm$  0.13 \\ 
 &  & 32.0 $\cdot 10^{-3}$ & 1.0 - 10.0 $\cdot 10^{-2}$ &  1.01 $\pm$  0.09 \\ 
\hline
480 & 320 - 640 & 32.0 $\cdot 10^{-3}$ & 1.0 - 10.0 $\cdot 10^{-2}$ &  1.10 $\pm$  0.14 \\ 
\hline
1000 & 640 - 10000 & 32.0 $\cdot 10^{-3}$ & 1.0 - 10.0 $\cdot 10^{-2}$ &  1.16 $\pm$  0.19 \\ 
\hline \hline 
 \multicolumn{5}{l}{Acceptance uncertainty: +5/-5 \%} \\\hline 
 \multicolumn{5}{l}{Energy scale uncertainty of $\pm$2 \%: -2/+2 \%} \\\hline 
 \multicolumn{5}{l}{Normalization error: 4 \%} \\\hline 
 \end{tabular} 
 \end{center} 
 \caption[The fraction of events with a leading neutron at $x_L=0.73$.]{\emcap The fraction of events with a leading neutron at $x_L=0.73$ in bins of $x$ and $Q^2$. \label{f2tabxl9} } 
 \end{table}

\newpage
\begin{table}
\centerline{\Large $ 0.76 < x_L < 0.82 $ }
\vspace{0.25cm}
\begin{center} 
 \begin{tabular}{|c|c|c|c|c|} \hline 
 $Q^2$ & $Q^2$ & $\xbj$ & $\xbj$ & $ratio$ (\%) \\ 
 GeV$^2$ & range & & range & $meas. \pm stat$ \\ 
\hline
7 & 4 - 10 & 1.1 $\cdot 10^{-4}$ & 8.0 - 15.0 $\cdot 10^{-5}$ &  0.98 $\pm$  0.06 \\ 
 &  & 2.1 $\cdot 10^{-4}$ & 1.5 - 3.0 $\cdot 10^{-4}$ &  0.83 $\pm$  0.04 \\ 
 &  & 4.2 $\cdot 10^{-4}$ & 3.0 - 6.0 $\cdot 10^{-4}$ &  0.90 $\pm$  0.04 \\ 
 &  & 8.5 $\cdot 10^{-4}$ & 6.0 - 12.0 $\cdot 10^{-4}$ &  0.86 $\pm$  0.05 \\ 
 &  & 17.0 $\cdot 10^{-4}$ & 1.2 - 2.4 $\cdot 10^{-3}$ &  0.93 $\pm$  0.06 \\ 
\hline
15 & 10 - 20 & 2.1 $\cdot 10^{-4}$ & 1.5 - 3.0 $\cdot 10^{-4}$ &  0.94 $\pm$  0.07 \\ 
 &  & 4.2 $\cdot 10^{-4}$ & 3.0 - 6.0 $\cdot 10^{-4}$ &  0.89 $\pm$  0.04 \\ 
 &  & 8.5 $\cdot 10^{-4}$ & 6.0 - 12.0 $\cdot 10^{-4}$ &  0.90 $\pm$  0.05 \\ 
 &  & 17.0 $\cdot 10^{-4}$ & 1.2 - 2.4 $\cdot 10^{-3}$ &  0.83 $\pm$  0.05 \\ 
 &  & 49.0 $\cdot 10^{-4}$ & 2.4 - 10.0 $\cdot 10^{-3}$ &  0.87 $\pm$  0.05 \\ 
\hline
30 & 20 - 40 & 4.2 $\cdot 10^{-4}$ & 3.0 - 6.0 $\cdot 10^{-4}$ &  0.77 $\pm$  0.07 \\ 
 &  & 8.5 $\cdot 10^{-4}$ & 6.0 - 12.0 $\cdot 10^{-4}$ &  0.89 $\pm$  0.05 \\ 
 &  & 17.0 $\cdot 10^{-4}$ & 1.2 - 2.4 $\cdot 10^{-3}$ &  0.82 $\pm$  0.05 \\ 
 &  & 49.0 $\cdot 10^{-4}$ & 2.4 - 10.0 $\cdot 10^{-3}$ &  0.89 $\pm$  0.04 \\ 
\hline
60 & 40 - 80 & 8.5 $\cdot 10^{-4}$ & 6.0 - 12.0 $\cdot 10^{-4}$ &  0.91 $\pm$  0.11 \\ 
 &  & 17.0 $\cdot 10^{-4}$ & 1.2 - 2.4 $\cdot 10^{-3}$ &  0.83 $\pm$  0.06 \\ 
 &  & 49.0 $\cdot 10^{-4}$ & 2.4 - 10.0 $\cdot 10^{-3}$ &  0.89 $\pm$  0.05 \\ 
 &  & 32.0 $\cdot 10^{-3}$ & 1.0 - 10.0 $\cdot 10^{-2}$ &  0.98 $\pm$  0.07 \\ 
\hline
120 & 80 - 160 & 17.0 $\cdot 10^{-4}$ & 1.2 - 2.4 $\cdot 10^{-3}$ &  1.22 $\pm$  0.18 \\ 
 &  & 49.0 $\cdot 10^{-4}$ & 2.4 - 10.0 $\cdot 10^{-3}$ &  0.67 $\pm$  0.06 \\ 
 &  & 32.0 $\cdot 10^{-3}$ & 1.0 - 10.0 $\cdot 10^{-2}$ &  0.89 $\pm$  0.07 \\ 
\hline
240 & 160 - 320 & 49.0 $\cdot 10^{-4}$ & 2.4 - 10.0 $\cdot 10^{-3}$ &  0.93 $\pm$  0.12 \\ 
 &  & 32.0 $\cdot 10^{-3}$ & 1.0 - 10.0 $\cdot 10^{-2}$ &  0.84 $\pm$  0.08 \\ 
\hline
480 & 320 - 640 & 32.0 $\cdot 10^{-3}$ & 1.0 - 10.0 $\cdot 10^{-2}$ &  0.60 $\pm$  0.10 \\ 
\hline
1000 & 640 - 10000 & 32.0 $\cdot 10^{-3}$ & 1.0 - 10.0 $\cdot 10^{-2}$ &  0.89 $\pm$  0.17 \\ 
\hline \hline 
 \multicolumn{5}{l}{Acceptance uncertainty: +5/-6 \%} \\\hline 
 \multicolumn{5}{l}{Energy scale uncertainty of $\pm$2 \%: +2/-1 \%} \\\hline 
 \multicolumn{5}{l}{Normalization error: 4 \%} \\\hline 
 \end{tabular} 
 \end{center} 
 \caption[The fraction of events with a leading neutron at $x_L=0.79$.]{\emcap The fraction of events with a leading neutron at $x_L=0.79$ in bins of $x$ and $Q^2$. \label{f2tabxl10} } 
 \end{table}

\newpage
\begin{table}
\centerline{\Large $ 0.82 < x_L < 0.88 $ }
\vspace{0.25cm}
\begin{center} 
 \begin{tabular}{|c|c|c|c|c|} \hline 
 $Q^2$ & $Q^2$ & $\xbj$ & $\xbj$ & $ratio$ (\%) \\ 
 GeV$^2$ & range & & range & $meas. \pm stat$ \\ 
\hline
7 & 4 - 10 & 1.1 $\cdot 10^{-4}$ & 8.0 - 15.0 $\cdot 10^{-5}$ &  0.73 $\pm$  0.05 \\ 
 &  & 2.1 $\cdot 10^{-4}$ & 1.5 - 3.0 $\cdot 10^{-4}$ &  0.71 $\pm$  0.03 \\ 
 &  & 4.2 $\cdot 10^{-4}$ & 3.0 - 6.0 $\cdot 10^{-4}$ &  0.65 $\pm$  0.04 \\ 
 &  & 8.5 $\cdot 10^{-4}$ & 6.0 - 12.0 $\cdot 10^{-4}$ &  0.74 $\pm$  0.05 \\ 
 &  & 17.0 $\cdot 10^{-4}$ & 1.2 - 2.4 $\cdot 10^{-3}$ &  0.74 $\pm$  0.05 \\ 
\hline
15 & 10 - 20 & 2.1 $\cdot 10^{-4}$ & 1.5 - 3.0 $\cdot 10^{-4}$ &  0.72 $\pm$  0.06 \\ 
 &  & 4.2 $\cdot 10^{-4}$ & 3.0 - 6.0 $\cdot 10^{-4}$ &  0.70 $\pm$  0.03 \\ 
 &  & 8.5 $\cdot 10^{-4}$ & 6.0 - 12.0 $\cdot 10^{-4}$ &  0.70 $\pm$  0.04 \\ 
 &  & 17.0 $\cdot 10^{-4}$ & 1.2 - 2.4 $\cdot 10^{-3}$ &  0.72 $\pm$  0.04 \\ 
 &  & 49.0 $\cdot 10^{-4}$ & 2.4 - 10.0 $\cdot 10^{-3}$ &  0.75 $\pm$  0.04 \\ 
\hline
30 & 20 - 40 & 4.2 $\cdot 10^{-4}$ & 3.0 - 6.0 $\cdot 10^{-4}$ &  0.73 $\pm$  0.07 \\ 
 &  & 8.5 $\cdot 10^{-4}$ & 6.0 - 12.0 $\cdot 10^{-4}$ &  0.65 $\pm$  0.04 \\ 
 &  & 17.0 $\cdot 10^{-4}$ & 1.2 - 2.4 $\cdot 10^{-3}$ &  0.64 $\pm$  0.04 \\ 
 &  & 49.0 $\cdot 10^{-4}$ & 2.4 - 10.0 $\cdot 10^{-3}$ &  0.68 $\pm$  0.03 \\ 
\hline
60 & 40 - 80 & 8.5 $\cdot 10^{-4}$ & 6.0 - 12.0 $\cdot 10^{-4}$ &  0.61 $\pm$  0.09 \\ 
 &  & 17.0 $\cdot 10^{-4}$ & 1.2 - 2.4 $\cdot 10^{-3}$ &  0.69 $\pm$  0.05 \\ 
 &  & 49.0 $\cdot 10^{-4}$ & 2.4 - 10.0 $\cdot 10^{-3}$ &  0.61 $\pm$  0.04 \\ 
 &  & 32.0 $\cdot 10^{-3}$ & 1.0 - 10.0 $\cdot 10^{-2}$ &  0.66 $\pm$  0.05 \\ 
\hline
120 & 80 - 160 & 17.0 $\cdot 10^{-4}$ & 1.2 - 2.4 $\cdot 10^{-3}$ &  0.72 $\pm$  0.13 \\ 
 &  & 49.0 $\cdot 10^{-4}$ & 2.4 - 10.0 $\cdot 10^{-3}$ &  0.63 $\pm$  0.05 \\ 
 &  & 32.0 $\cdot 10^{-3}$ & 1.0 - 10.0 $\cdot 10^{-2}$ &  0.60 $\pm$  0.05 \\ 
\hline
240 & 160 - 320 & 49.0 $\cdot 10^{-4}$ & 2.4 - 10.0 $\cdot 10^{-3}$ &  0.43 $\pm$  0.08 \\ 
 &  & 32.0 $\cdot 10^{-3}$ & 1.0 - 10.0 $\cdot 10^{-2}$ &  0.40 $\pm$  0.05 \\ 
\hline
480 & 320 - 640 & 32.0 $\cdot 10^{-3}$ & 1.0 - 10.0 $\cdot 10^{-2}$ &  0.40 $\pm$  0.08 \\ 
\hline
1000 & 640 - 10000 & 32.0 $\cdot 10^{-3}$ & 1.0 - 10.0 $\cdot 10^{-2}$ &  0.57 $\pm$  0.13 \\ 
\hline \hline 
 \multicolumn{5}{l}{Acceptance uncertainty: +6/-6 \%} \\\hline 
 \multicolumn{5}{l}{Energy scale uncertainty of $\pm$2 \%: +5/-8 \%} \\\hline 
 \multicolumn{5}{l}{Normalization error: 4 \%} \\\hline 
 \end{tabular} 
 \end{center} 
 \caption[The fraction of events with a leading neutron at $x_L=0.85$.]{\emcap The fraction of events with a leading neutron at $x_L=0.85$ in bins of $x$ and $Q^2$. \label{f2tabxl11} } 
 \end{table}

\newpage
\begin{table}
\centerline{\Large $ 0.88 < x_L < 1 $ }
\vspace{0.25cm}
\begin{center} 
 \begin{tabular}{|c|c|c|c|c|} \hline 
 $Q^2$ & $Q^2$ & $\xbj$ & $\xbj$ & $ratio$ (\%) \\ 
 GeV$^2$ & range & & range & $meas. \pm stat$ \\ 
\hline
7 & 4 - 10 & 1.1 $\cdot 10^{-4}$ & 8.0 - 15.0 $\cdot 10^{-5}$ &  0.52 $\pm$  0.04 \\ 
 &  & 2.1 $\cdot 10^{-4}$ & 1.5 - 3.0 $\cdot 10^{-4}$ &  0.53 $\pm$  0.03 \\ 
 &  & 4.2 $\cdot 10^{-4}$ & 3.0 - 6.0 $\cdot 10^{-4}$ &  0.48 $\pm$  0.03 \\ 
 &  & 8.5 $\cdot 10^{-4}$ & 6.0 - 12.0 $\cdot 10^{-4}$ &  0.41 $\pm$  0.03 \\ 
 &  & 17.0 $\cdot 10^{-4}$ & 1.2 - 2.4 $\cdot 10^{-3}$ &  0.45 $\pm$  0.04 \\ 
\hline
15 & 10 - 20 & 2.1 $\cdot 10^{-4}$ & 1.5 - 3.0 $\cdot 10^{-4}$ &  0.51 $\pm$  0.04 \\ 
 &  & 4.2 $\cdot 10^{-4}$ & 3.0 - 6.0 $\cdot 10^{-4}$ &  0.45 $\pm$  0.02 \\ 
 &  & 8.5 $\cdot 10^{-4}$ & 6.0 - 12.0 $\cdot 10^{-4}$ &  0.44 $\pm$  0.03 \\ 
 &  & 17.0 $\cdot 10^{-4}$ & 1.2 - 2.4 $\cdot 10^{-3}$ &  0.43 $\pm$  0.03 \\ 
 &  & 49.0 $\cdot 10^{-4}$ & 2.4 - 10.0 $\cdot 10^{-3}$ &  0.38 $\pm$  0.03 \\ 
\hline
30 & 20 - 40 & 4.2 $\cdot 10^{-4}$ & 3.0 - 6.0 $\cdot 10^{-4}$ &  0.40 $\pm$  0.04 \\ 
 &  & 8.5 $\cdot 10^{-4}$ & 6.0 - 12.0 $\cdot 10^{-4}$ &  0.39 $\pm$  0.03 \\ 
 &  & 17.0 $\cdot 10^{-4}$ & 1.2 - 2.4 $\cdot 10^{-3}$ &  0.41 $\pm$  0.03 \\ 
 &  & 49.0 $\cdot 10^{-4}$ & 2.4 - 10.0 $\cdot 10^{-3}$ &  0.43 $\pm$  0.02 \\ 
\hline
60 & 40 - 80 & 8.5 $\cdot 10^{-4}$ & 6.0 - 12.0 $\cdot 10^{-4}$ &  0.52 $\pm$  0.07 \\ 
 &  & 17.0 $\cdot 10^{-4}$ & 1.2 - 2.4 $\cdot 10^{-3}$ &  0.34 $\pm$  0.03 \\ 
 &  & 49.0 $\cdot 10^{-4}$ & 2.4 - 10.0 $\cdot 10^{-3}$ &  0.38 $\pm$  0.03 \\ 
 &  & 32.0 $\cdot 10^{-3}$ & 1.0 - 10.0 $\cdot 10^{-2}$ &  0.34 $\pm$  0.03 \\ 
\hline
120 & 80 - 160 & 17.0 $\cdot 10^{-4}$ & 1.2 - 2.4 $\cdot 10^{-3}$ &  0.45 $\pm$  0.09 \\ 
 &  & 49.0 $\cdot 10^{-4}$ & 2.4 - 10.0 $\cdot 10^{-3}$ &  0.36 $\pm$  0.04 \\ 
 &  & 32.0 $\cdot 10^{-3}$ & 1.0 - 10.0 $\cdot 10^{-2}$ &  0.35 $\pm$  0.04 \\ 
\hline
240 & 160 - 320 & 49.0 $\cdot 10^{-4}$ & 2.4 - 10.0 $\cdot 10^{-3}$ &  0.41 $\pm$  0.07 \\ 
 &  & 32.0 $\cdot 10^{-3}$ & 1.0 - 10.0 $\cdot 10^{-2}$ &  0.22 $\pm$  0.04 \\ 
\hline
480 & 320 - 640 & 32.0 $\cdot 10^{-3}$ & 1.0 - 10.0 $\cdot 10^{-2}$ &  0.25 $\pm$  0.06 \\ 
\hline
1000 & 640 - 10000 & 32.0 $\cdot 10^{-3}$ & 1.0 - 10.0 $\cdot 10^{-2}$ &  0.20 $\pm$  0.07 \\ 
\hline \hline 
 \multicolumn{5}{l}{Acceptance uncertainty: +6/-6 \%} \\\hline 
 \multicolumn{5}{l}{Energy scale uncertainty of $\pm$2 \%: +25/-26 \%} \\\hline 
 \multicolumn{5}{l}{Normalization error: 4 \%} \\\hline 
 \end{tabular} 
 \end{center} 
 \caption[The fraction of events with a leading neutron at $x_L=0.94$.]{\emcap The fraction of events with a leading neutron at $x_L=0.94$ in bins of $x$ and $Q^2$. \label{f2tabxl12} } 
 \end{table}

\newpage
\begin{table}
\centerline{\Large $ 0.2 < x_L < 0.64 $ }
\vspace{0.25cm}
\begin{center} 
 \begin{tabular}{|c|c|c|c|c|} \hline 
 $Q^2$ & $Q^2$ & $y$ & $y$ & $ratio$ (\%) \\ 
 GeV$^2$ & range & & range & $meas. \pm stat$ \\ 
\hline
0.11 & 0.10 - 0.13 & 0.60 & 0.54 - 0.64 &  3.15 $\pm$  0.43 \\ 
 &  & 0.70 & 0.64 - 0.74 &  2.85 $\pm$  0.39 \\ 
\hline
0.15 & 0.13 - 0.17 & 0.40 & 0.37 - 0.45 &  3.72 $\pm$  0.47 \\ 
 &  & 0.50 & 0.45 - 0.54 &  2.76 $\pm$  0.34 \\ 
 &  & 0.60 & 0.54 - 0.64 &  3.49 $\pm$  0.39 \\ 
 &  & 0.70 & 0.64 - 0.74 &  2.76 $\pm$  0.41 \\ 
\hline
0.20 & 0.17 - 0.21 & 0.26 & 0.23 - 0.30 &  3.23 $\pm$  0.41 \\ 
 &  & 0.33 & 0.30 - 0.37 &  3.09 $\pm$  0.38 \\ 
 &  & 0.40 & 0.37 - 0.45 &  3.41 $\pm$  0.38 \\ 
 &  & 0.50 & 0.45 - 0.54 &  3.70 $\pm$  0.43 \\ 
 &  & 0.60 & 0.54 - 0.64 &  2.96 $\pm$  0.43 \\ 
 &  & 0.70 & 0.64 - 0.74 &  3.98 $\pm$  0.63 \\ 
\hline
0.25 & 0.21 - 0.27 & 0.20 & 0.16 - 0.23 &  3.23 $\pm$  0.28 \\ 
 &  & 0.26 & 0.23 - 0.30 &  3.46 $\pm$  0.31 \\ 
 &  & 0.33 & 0.30 - 0.37 &  3.24 $\pm$  0.33 \\ 
 &  & 0.40 & 0.37 - 0.45 &  4.04 $\pm$  0.39 \\ 
 &  & 0.50 & 0.45 - 0.54 &  3.28 $\pm$  0.39 \\ 
 &  & 0.60 & 0.54 - 0.64 &  3.83 $\pm$  0.49 \\ 
\hline
0.30 & 0.27 - 0.35 & 0.12 & 0.08 - 0.16 &  3.88 $\pm$  0.29 \\ 
 &  & 0.20 & 0.16 - 0.23 &  3.82 $\pm$  0.30 \\ 
 &  & 0.26 & 0.23 - 0.30 &  3.03 $\pm$  0.30 \\ 
 &  & 0.33 & 0.30 - 0.37 &  2.86 $\pm$  0.32 \\ 
 &  & 0.40 & 0.37 - 0.45 &  3.36 $\pm$  0.40 \\ 
 &  & 0.50 & 0.45 - 0.54 &  2.97 $\pm$  0.41 \\ 
\hline
0.40 & 0.35 - 0.45 & 0.12 & 0.08 - 0.16 &  3.05 $\pm$  0.28 \\ 
 &  & 0.20 & 0.16 - 0.23 &  3.02 $\pm$  0.30 \\ 
 &  & 0.26 & 0.23 - 0.30 &  3.61 $\pm$  0.38 \\ 
 &  & 0.33 & 0.30 - 0.37 &  3.94 $\pm$  0.45 \\ 
 &  & 0.40 & 0.37 - 0.45 &  4.45 $\pm$  0.52 \\ 
\hline
0.50 & 0.45 - 0.58 & 0.12 & 0.08 - 0.16 &  3.35 $\pm$  0.33 \\ 
 &  & 0.20 & 0.16 - 0.23 &  3.02 $\pm$  0.34 \\ 
 &  & 0.26 & 0.23 - 0.30 &  2.50 $\pm$  0.36 \\ 
\hline
0.65 & 0.58 - 0.74 & 0.20 & 0.16 - 0.23 &  4.01 $\pm$  0.54 \\ 
 &  & 0.26 & 0.23 - 0.30 &  3.46 $\pm$  0.68 \\ 
\hline
\hline \hline 
 \multicolumn{5}{l}{Acceptance uncertainty: +2/-2 \%} \\\hline 
 \multicolumn{5}{l}{Energy scale uncertainty of $\pm$2 \%: -4/+4 \%} \\\hline 
 \multicolumn{5}{l}{Normalization error: 4 \%} \\\hline 
 \end{tabular} 
 \end{center} 
 \caption[The fraction of events with a leading neutron at $x_L=0.42$.]{\emcap The fraction of events with a leading neutron at $x_L=0.42$ in bins of $y$ and $Q^2$ for the BPC region. \label{bpctabxl31} } 
 \end{table}

\newpage
\begin{table}
\centerline{\Large $ 0.64 < x_L < 0.82 $ }
\vspace{0.25cm}
\begin{center} 
 \begin{tabular}{|c|c|c|c|c|} \hline 
 $Q^2$ & $Q^2$ & $y$ & $y$ & $ratio$ (\%) \\ 
 GeV$^2$ & range & & range & $meas. \pm stat$ \\ 
\hline
0.11 & 0.10 - 0.13 & 0.60 & 0.54 - 0.64 &  2.62 $\pm$  0.38 \\ 
 &  & 0.70 & 0.64 - 0.74 &  2.91 $\pm$  0.39 \\ 
\hline
0.15 & 0.13 - 0.17 & 0.40 & 0.37 - 0.45 &  2.30 $\pm$  0.36 \\ 
 &  & 0.50 & 0.45 - 0.54 &  2.58 $\pm$  0.32 \\ 
 &  & 0.60 & 0.54 - 0.64 &  2.35 $\pm$  0.31 \\ 
 &  & 0.70 & 0.64 - 0.74 &  2.17 $\pm$  0.36 \\ 
\hline
0.20 & 0.17 - 0.21 & 0.26 & 0.23 - 0.30 &  2.25 $\pm$  0.33 \\ 
 &  & 0.33 & 0.30 - 0.37 &  2.94 $\pm$  0.36 \\ 
 &  & 0.40 & 0.37 - 0.45 &  2.69 $\pm$  0.33 \\ 
 &  & 0.50 & 0.45 - 0.54 &  2.30 $\pm$  0.33 \\ 
 &  & 0.60 & 0.54 - 0.64 &  2.81 $\pm$  0.41 \\ 
 &  & 0.70 & 0.64 - 0.74 &  2.16 $\pm$  0.45 \\ 
\hline
0.25 & 0.21 - 0.27 & 0.20 & 0.16 - 0.23 &  2.53 $\pm$  0.24 \\ 
 &  & 0.26 & 0.23 - 0.30 &  2.46 $\pm$  0.25 \\ 
 &  & 0.33 & 0.30 - 0.37 &  2.63 $\pm$  0.28 \\ 
 &  & 0.40 & 0.37 - 0.45 &  3.04 $\pm$  0.33 \\ 
 &  & 0.50 & 0.45 - 0.54 &  2.23 $\pm$  0.31 \\ 
 &  & 0.60 & 0.54 - 0.64 &  2.13 $\pm$  0.36 \\ 
\hline
0.30 & 0.27 - 0.35 & 0.12 & 0.08 - 0.16 &  2.55 $\pm$  0.23 \\ 
 &  & 0.20 & 0.16 - 0.23 &  2.64 $\pm$  0.24 \\ 
 &  & 0.26 & 0.23 - 0.30 &  2.03 $\pm$  0.24 \\ 
 &  & 0.33 & 0.30 - 0.37 &  2.24 $\pm$  0.28 \\ 
 &  & 0.40 & 0.37 - 0.45 &  2.12 $\pm$  0.31 \\ 
 &  & 0.50 & 0.45 - 0.54 &  2.61 $\pm$  0.38 \\ 
\hline
0.40 & 0.35 - 0.45 & 0.12 & 0.08 - 0.16 &  2.73 $\pm$  0.25 \\ 
 &  & 0.20 & 0.16 - 0.23 &  2.70 $\pm$  0.28 \\ 
 &  & 0.26 & 0.23 - 0.30 &  2.64 $\pm$  0.31 \\ 
 &  & 0.33 & 0.30 - 0.37 &  2.53 $\pm$  0.35 \\ 
 &  & 0.40 & 0.37 - 0.45 &  2.78 $\pm$  0.40 \\ 
\hline
0.50 & 0.45 - 0.58 & 0.12 & 0.08 - 0.16 &  2.63 $\pm$  0.29 \\ 
 &  & 0.20 & 0.16 - 0.23 &  2.87 $\pm$  0.32 \\ 
 &  & 0.26 & 0.23 - 0.30 &  2.37 $\pm$  0.34 \\ 
\hline
0.65 & 0.58 - 0.74 & 0.20 & 0.16 - 0.23 &  1.77 $\pm$  0.34 \\ 
 &  & 0.26 & 0.23 - 0.30 &  2.56 $\pm$  0.57 \\ 
\hline
\hline \hline 
 \multicolumn{5}{l}{Acceptance uncertainty: +5/-5 \%} \\\hline 
 \multicolumn{5}{l}{Energy scale uncertainty of $\pm$2 \%: -2/+2 \%} \\\hline 
 \multicolumn{5}{l}{Normalization error: 4 \%} \\\hline 
 \end{tabular} 
 \end{center} 
 \caption[The fraction of events with a leading neutron at $x_L=0.73$.]{\emcap The fraction of events with a leading neutron at $x_L=0.73$ in bins of $y$ and $Q^2$ for the BPC region. \label{bpctabxl32} } 
 \end{table}

\newpage
\begin{table}
\centerline{\Large $ 0.82 < x_L < 1 $ }
\vspace{0.25cm}
\begin{center} 
 \begin{tabular}{|c|c|c|c|c|} \hline 
 $Q^2$ & $Q^2$ & $y$ & $y$ & $ratio$ (\%) \\ 
 GeV$^2$ & range & & range & $meas. \pm stat$ \\ 
\hline
0.11 & 0.10 - 0.13 & 0.60 & 0.54 - 0.64 &  1.15 $\pm$  0.23 \\ 
 &  & 0.70 & 0.64 - 0.74 &  0.87 $\pm$  0.19 \\ 
\hline
0.15 & 0.13 - 0.17 & 0.40 & 0.37 - 0.45 &  1.19 $\pm$  0.23 \\ 
 &  & 0.50 & 0.45 - 0.54 &  1.10 $\pm$  0.19 \\ 
 &  & 0.60 & 0.54 - 0.64 &  0.82 $\pm$  0.17 \\ 
 &  & 0.70 & 0.64 - 0.74 &  0.89 $\pm$  0.21 \\ 
\hline
0.20 & 0.17 - 0.21 & 0.26 & 0.23 - 0.30 &  1.03 $\pm$  0.20 \\ 
 &  & 0.33 & 0.30 - 0.37 &  1.18 $\pm$  0.20 \\ 
 &  & 0.40 & 0.37 - 0.45 &  1.06 $\pm$  0.19 \\ 
 &  & 0.50 & 0.45 - 0.54 &  1.00 $\pm$  0.19 \\ 
 &  & 0.60 & 0.54 - 0.64 &  1.16 $\pm$  0.24 \\ 
 &  & 0.70 & 0.64 - 0.74 &  0.82 $\pm$  0.25 \\ 
\hline
0.25 & 0.21 - 0.27 & 0.20 & 0.16 - 0.23 &  0.96 $\pm$  0.13 \\ 
 &  & 0.26 & 0.23 - 0.30 &  0.79 $\pm$  0.13 \\ 
 &  & 0.33 & 0.30 - 0.37 &  1.41 $\pm$  0.19 \\ 
 &  & 0.40 & 0.37 - 0.45 &  1.03 $\pm$  0.17 \\ 
 &  & 0.50 & 0.45 - 0.54 &  0.94 $\pm$  0.18 \\ 
 &  & 0.60 & 0.54 - 0.64 &  1.09 $\pm$  0.23 \\ 
\hline
0.30 & 0.27 - 0.35 & 0.12 & 0.08 - 0.16 &  1.47 $\pm$  0.16 \\ 
 &  & 0.20 & 0.16 - 0.23 &  1.33 $\pm$  0.15 \\ 
 &  & 0.26 & 0.23 - 0.30 &  0.95 $\pm$  0.15 \\ 
 &  & 0.33 & 0.30 - 0.37 &  0.93 $\pm$  0.16 \\ 
 &  & 0.40 & 0.37 - 0.45 &  0.86 $\pm$  0.18 \\ 
 &  & 0.50 & 0.45 - 0.54 &  1.14 $\pm$  0.23 \\ 
\hline
0.40 & 0.35 - 0.45 & 0.12 & 0.08 - 0.16 &  1.33 $\pm$  0.16 \\ 
 &  & 0.20 & 0.16 - 0.23 &  1.15 $\pm$  0.16 \\ 
 &  & 0.26 & 0.23 - 0.30 &  1.12 $\pm$  0.18 \\ 
 &  & 0.33 & 0.30 - 0.37 &  0.93 $\pm$  0.19 \\ 
 &  & 0.40 & 0.37 - 0.45 &  0.92 $\pm$  0.21 \\ 
\hline
0.50 & 0.45 - 0.58 & 0.12 & 0.08 - 0.16 &  1.11 $\pm$  0.17 \\ 
 &  & 0.20 & 0.16 - 0.23 &  0.94 $\pm$  0.17 \\ 
 &  & 0.26 & 0.23 - 0.30 &  1.40 $\pm$  0.24 \\ 
\hline
0.65 & 0.58 - 0.74 & 0.20 & 0.16 - 0.23 &  0.92 $\pm$  0.22 \\ 
 &  & 0.26 & 0.23 - 0.30 &  1.21 $\pm$  0.35 \\ 
\hline
\hline \hline 
 \multicolumn{5}{l}{Acceptance uncertainty: +6/-6 \%} \\\hline 
 \multicolumn{5}{l}{Energy scale uncertainty of $\pm$2 \%: +14/-16 \%} \\\hline 
 \multicolumn{5}{l}{Normalization error: 4 \%} \\\hline 
 \end{tabular} 
 \end{center} 
 \caption[The fraction of events with a leading neutron at $x_L=0.91$.]{\emcap The fraction of events with a leading neutron at $x_L=0.91$ in bins of $y$ and $Q^2$ for the BPC region. \label{bpctabxl33} } 
 \end{table}



\newpage

\clearpage

\clearpage

\clearpage

\clearpage

\clearpage

\clearpage

\vspace{-1cm}
\clearpage

\clearpage
%
\clearpage

\clearpage

\clearpage

\clearpage


\clearpage

\clearpage

\clearpage

\clearpage
\end{document}